\documentclass[cernpreprint,texmf,texlive=2014,UKenglish,txfonts,siunitx=false]{atlasdoc}
\pdfoutput=1
\usepackage[biblatex]{atlaspackage}
\usepackage{authblk}
\usepackage{atlasphysics}
\DeclareGraphicsExtensions{.pdf}
\usepackage{atlasbiblatex}
\usepackage{graphicx}
\usepackage{tabularx}
\usepackage{multirow}

\AtlasTitle{Determination of spin and parity of the Higgs boson in the $WW^*\rightarrow e \nu \mu \nu$ decay channel with the ATLAS detector}
\author{The ATLAS Collaboration}
\PreprintIdNumber{CERN-PH-EP-2015-037}
\AtlasJournal{Eur. Phys. J. C}

\AtlasAbstract{Studies of the spin and parity quantum numbers of the Higgs boson in the $WW^* \rightarrow e \nu \mu \nu$ final state  are presented, 
based on proton--proton collision data collected by the ATLAS detector at the Large Hadron Collider, corresponding to 
an integrated luminosity of 20.3~fb$^{-1}$ at a centre-of-mass energy of $\sqrt{s}=8$~TeV.
The Standard Model spin-parity $J^{CP} = 0^{++}$ hypothesis is compared with 
alternative hypotheses for both spin and CP. The case where the observed resonance is a mixture of the Standard-Model-like Higgs boson and CP-even 
($J^{CP} = 0^{++}$) or  CP-odd ($J^{CP} = 0^{+-}$) Higgs boson in scenarios beyond the Standard Model is also studied.  
The data are found to be consistent with the Standard Model prediction and limits are placed on alternative spin and CP hypotheses, including CP mixing in different scenarios. 
}

\begin{document}
\nolinenumbers
\newcommand{\ksm} {\ensuremath{\kappa_{\rm \sss SM}}\xspace}
\newcommand{\khww} {\ensuremath{\kappa_{\rm \sss HWW}}\xspace}
\newcommand{\kaww} {\ensuremath{\kappa_{\rm \sss AWW}}\xspace}
\newcommand{\khdw} {\ensuremath{\kappa_{\rm \sss H\partial{}W}}\xspace}
\newcommand{\nn} {\nonumber}
\newcommand{\bs}[1]{\boldsymbol{{#1}}}
\newcommand\sss{\scriptscriptstyle}
\newcommand{\kt}{k_T}

\newcommand{\lumi}{$\mathcal{L}$}

\newcommand{\madgraph} {\textsc{MadGraph}}
\newcommand{\mgaMC} {\textsc{MadGraph5\_}a\textsc{MC@NLO}}
\newcommand{\amcatnlo} {a\textsc{MC@NLO}}

\newcommand{\pTll}{\ensuremath{\pT^{\ell\ell}}}
\newcommand{\ptll}{\pTll}
\newcommand{\dpt}{\ensuremath{\Delta p_{\rm T}}}
\newcommand{\Efun}{\ensuremath{E_{\ell \ell \nu \nu}}}
\newcommand{\pTH}{\ensuremath{p_{\rm T}^{\rm H}}}

\newcommand{\fqqf}[1]{\ensuremath{f_{q\bar{q}}={#1}\% }}
\newcommand{\fqq}{\ensuremath{f_{q\bar{q}}}}
\newcommand{\spinzero}{\ensuremath{J^P=0^+}}
\newcommand{\spinzeromm}{\ensuremath{J^P=0^-}}
\newcommand{\spin}{\ensuremath{J^P}}
\newcommand{\spintwo}{\ensuremath{J^P=2^+}}
\newcommand{\spinonepp}{\ensuremath{J^P=1^+}}
\newcommand{\spinonemm}{\ensuremath{J^P=1^-}}
\newcommand{\spinalt}{\ensuremath{J^{P}_{ \mathrm {alt} }}}
\newcommand{\CLs}{\ensuremath{{\rm CL}_{\rm s}}}
\newcommand{\spinzerop}{\ensuremath{0^{+}}}
\newcommand{\spinzerom}{\ensuremath{0^{-}}}
\newcommand{\spinonep}{\ensuremath{1^{+}}}
\newcommand{\spinonem}{\ensuremath{1^{-}}}
\newcommand{\spintwopm}{\ensuremath{2^{+}_{m}}}
\newcommand{\spintwop}{\ensuremath{2^{+}}}
\newcommand{\spintwom}{\ensuremath{2^{-}}}
\newcommand{\spintwokg}{\ensuremath{ 2^+_{k_{g}=1,\, k_{q}=0} }}
\newcommand{\spintwokq}{\ensuremath{ 2^+_{k_{g}=0.5,\, k_{q}=1} }}
\newcommand{\qq}{\ensuremath{q\bar{q}}}

\newcommand{\bdtcp} {\ensuremath{\mbox{BDT}_{\mbox{\tiny CP}}}}
\newcommand{\pexpmusm} {\ensuremath{ p^{\mbox{\tiny SM}}_{\text{exp},\ \mu=1}}}
\newcommand{\pexpsm} {\ensuremath{ p^{\mbox{\tiny SM}}_{\text{exp},\ \mu=\hat{\mu}}}}
\newcommand{\pexpalt} {\ensuremath{ p^{\mbox{\tiny ALT}}_{\text{exp},\ \mu=\hat{\mu}}}}
\newcommand{\pobssm} {\ensuremath{ p^{\mbox{\tiny SM}}_\text{obs}}}
\newcommand{\pobsalt} {\ensuremath{ p^{\mbox{\tiny ALT}}_\text{obs}}}

\newcommand{\Amumu}{\ensuremath{h/H/A \rightarrow \mu^{+}\mu^{-}}}
\newcommand{\emu}{\ensuremath{e^{\pm}\mu^{\mp}}}
\newcommand{\mup}{\ensuremath{\mu^{+}}}
\newcommand{\mum}{\ensuremath{\mu^{-}}}
\newcommand{\bbA}{\ensuremath{b\bar{b}A}}

\newcommand{\bb}{\ensuremath{b\bar{b}}}
\newcommand{\ggF}{\ensuremath{gg\to H}}
\newcommand{\VBF}{\ensuremath{qq'\to qq'H}}
\newcommand{\VH}{\ensuremath{q\bar{q}\to WH/ZH}}
\newcommand{\ttH}{\ensuremath{q\bar{q}/gg \to t\bar{t}H}}
\newcommand{\ggttH}{\ensuremath{gg \to t\bar{t}H}}
\newcommand{\Zgjets}{$Z/\gamma^{*}$+jets}
\newcommand{\Zjets}{$Z$+jets}
\newcommand{\mll}{\ensuremath{m_{\ell\ell}}}
\newcommand{\dphill}{\ensuremath{\Delta \phi_{\ell\ell}}}
\newcommand{\metrel}{\ensuremath{E_{\rm T,rel}^{\rm miss}}}
\newcommand{\mT}{\ensuremath{m_{\rm T}}}

\newcommand{\Calometrel}{\ensuremath{\text{Calo}E_{\rm T,rel}^{\rm miss}}}
\newcommand{\Trkmetrel}{\ensuremath{\text{Track}E_{\rm T,rel}^{\rm miss}}}
\newcommand{\trackmet}{\ensuremath{E_{\rm T}^{\rm miss,J-TRK}}}
\newcommand{\trackmT}{\ensuremath{m_{\rm T}^{\rm track,jetCorr}}}
\newcommand{\vpT}{\ensuremath{\boldsymbol{p}_{\mathrm{T}}}}
\newcommand{\vpTll}{\ensuremath{\boldsymbol{p}_{\rm T}^{\ell\ell}}}

\newcommand{\ptt}{\ensuremath{p_{\mathrm{Tt}}}}
\newcommand{\hgg}{\ensuremath{H{\rightarrow\,}\gamma\gamma}}
\newcommand{\hww}{\ensuremath{H \rightarrow WW^{*}}}
\newcommand{\ggWW}{\ensuremath{gg{\rightarrow\,}WW}}
\newcommand{\hWWlnln}{\ensuremath{H{\rightarrow\,}WW^{*}{\rightarrow\,}\ell\nu\ell\nu}}
\newcommand{\hWWlvlv}{\ensuremath{H{\rightarrow\,}WW^{*}{\rightarrow\,}\ell\nu\ell\nu}}
\newcommand{\hwwlnln}{\ensuremath{H{\rightarrow\,}WW^{*}{\rightarrow\,}\ell\nu\ell\nu}}

\newcommand{\hWWenmun}{\ensuremath{H{\rightarrow\,}WW^{*}{\rightarrow\,}e\nu\mu\nu}}
\newcommand{\hWWevmuv}{\ensuremath{H{\rightarrow\,}WW^{*}{\rightarrow\,}e\nu\mu\nu}}
\newcommand{\hwwenmun}{\ensuremath{H{\rightarrow\,}WW^{*}{\rightarrow\,}e\nu\mu\nu}}

\newcommand{\lnln}{\ensuremath{\ell\nu\ell\nu}}
\newcommand{\enmun}{\ensuremath{e\nu\mu\nu}}

\newcommand{\hWWlnqq}{\ensuremath{H \rightarrow WW \rightarrow \ell\nu q\overline{q}'}}
\newcommand{\lnqq}{\ensuremath{\ell\nu q\overline{q}'}}
\newcommand{\hzz}{\ensuremath{H \rightarrow ZZ^{*}}}
\newcommand{\hZZ}{\ensuremath{H \rightarrow ZZ}}
\newcommand{\hZZllnn}{\ensuremath{H \rightarrow ZZ\rightarrow \ell^+\ell^-\nu\overline{\nu}}}
\newcommand{\hZZllqq}{\ensuremath{H \rightarrow ZZ\rightarrow \ell^+\ell^- q\overline{q}}}
\newcommand{\hZZllll}{\ensuremath{H{\rightarrow\,}ZZ^{*}{\rightarrow\,}4\ell}}
\newcommand{\hzzllnn}{\ensuremath{H \rightarrow ZZ\rightarrow \ell^+\ell^-\nu\overline{\nu}}}
\newcommand{\hzzllqq}{\ensuremath{H \rightarrow ZZ\rightarrow \ell^+\ell^- q\overline{q}}}
\newcommand{\hzzllll}{\ensuremath{H \rightarrow ZZ^{*}\rightarrow 4\ell}}
\newcommand{\htt}{\ensuremath{H \rightarrow \tau^+\tau^-}}
\newcommand{\hbb}{\ensuremath{H \rightarrow \bb}}
\newcommand{\llnn}{\ensuremath{\ell^+\ell^-\nu\bar{\nu}}}
\newcommand{\hllnn}{\ensuremath{H \rightarrow ZZ \rightarrow \llnn}}
\newcommand{\zzllnn}{\ensuremath{ZZ \rightarrow \llnn}}
\newcommand{\llll}{\ensuremath{\ell^+\ell^-\ell^+\ell^-}}
\newcommand{\hllll}{\ensuremath{H{\rightarrow\,}ZZ^{*}{\rightarrow\,}\llll}}
\newcommand{\llbb}{\ensuremath{\ell^+\ell^-b\bar{b}}}
\newcommand{\lvbb}{\ensuremath{\ell\nu b\bar{b}}}
\newcommand{\wbb}{\ensuremath{Wb\bar{b}}}
\newcommand{\Zbb}{\ensuremath{Zb\bar{b}}}
\newcommand{\Zjj}{\ensuremath{Zq\bar{q}}}
\newcommand{\hllbb}{\ensuremath{H \rightarrow ZZ \rightarrow \llbb}}
\newcommand{\heenn}{\ensuremath{H \rightarrow ZZ \rightarrow e^+e^-\nu\bar{\nu}}}
\newcommand{\hmmnn}{\ensuremath{H \rightarrow ZZ \rightarrow \mu^+\mu^-\nu\bar{\nu}}}
\newcommand{\ztoll}{\ensuremath{Z{\rightarrow\,}\ell^+\ell^-}}
\newcommand{\ztoee}{\ensuremath{Z{\rightarrow\,}e^+e^-}}
\newcommand\tthbb{\ensuremath{t\overline{t}}\hbb}
\newcommand\htollbb{\hllbb}
\newcommand\htollnunu{\hllnn}
\newcommand\zztollnunu{\zzllnn}
\newcommand\htollll{$H{\rightarrow\,}ZZ^{*}{\rightarrow\,}4\ell$}
\newcommand{\htoeenunu}{\heenn}
\newcommand{\htomumununu}{\hmmnn}
\newcommand{\lumiuncertainty}{$\pm3.7$\%}
\newcommand{\lhood}{\ensuremath{{\cal L}}}
\newcommand{\gaussprob}{\ensuremath{\textrm{Norm}(\delta_i|\bar{\delta}_i)}}
\newcommand{\nuisance}{\ensuremath{m_{\delta_i}}}
\newcommand{\ZZbkg}{\ensuremath{ZZ^{(*)}}}
\newcommand{\zz}{\ensuremath{ZZ}}
\newcommand{\ww}{\ensuremath{WW}}
\newcommand{\wz}{\ensuremath{WZ}}
\newcommand{\Wjets}{\ensuremath{W}+jets}
\newcommand{\SM}{Standard Model}
\newcommand{\httlh}{\ensuremath{H\rightarrow \tau\tau \rightarrow \ell \tau_{had} 3 \nu }}
\newcommand{\httllj}{\ensuremath{H\rightarrow \tau\tau \rightarrow \ell^+\ell^-4\nu }}
\newcommand{\lh}{\ensuremath{\ell\tau_{had} 3 \nu}}
\newcommand{\llj}{\ensuremath{\tau_{\ell} \tau_{\ell} + jet}}
\newcommand{\ttoWb}{t{\rightarrow\,}Wb}
\newcommand{\Ztt}{\ensuremath{Z/\gamma^*\rightarrow \tau\tau}}

\newcommand\htollvv{$H\to ZZ\to\ell^+\ell^-\nu\bar{\nu}$\ }
\newcommand\htollqq{$H\to ZZ\to\ell^+\ell^- q \bar{q}$\ }
\newcommand\htoeevv{$H\to ZZ\to ee\nu\nu$\ }
\newcommand\htoeeqq{$H\to ZZ\to eeq \bar{q}$\ }
\newcommand\htouuvv{$H\to ZZ\to\mu\mu\nu\nu$\ }
\newcommand\htouuqq{$H\to ZZ\to\mu\mu q \bar{q}$\ }
\newcommand{\inpb}{pb$^{-1}$}
\newcommand{\infb}{fb$^{-1}$}
\newcommand\Wlv          {W{\TO}\ell\nu}
\newcommand\ZZ           {ZZ}
\newcommand\HWW          {H{\TO}\WWs}
\newcommand\WWs          {WW^\ast}
\newcommand\WW           {WW}
\newcommand\Wg           {W\gamma}
\newcommand\Wgs          {W\gstar}
\newcommand\gstar        {\gamma^{\ast}}
\newcommand\WZ           {W\!Z}
\newcommand\Zg           {Z\gamma}
\newcommand\Zgs          {Z\gstar}
\newcommand\ZDY          {Z/\gstar}
\newcommand\alphaS       {\alpha_{\rm S}}

\def\vec#1{{\mbox{$\boldsymbol{#1}$}}}

\renewcommand{\ttbar}{\ensuremath{t\overline{t}}}

\def\thetast{\ensuremath{\theta^*}}
\def\costs{\ensuremath{|\cos \theta^*|}}
\newcommand{\mgg}{\ensuremath{m_{\gamma\gamma}}}
\newcommand{\mm}{\ensuremath{\,\mbox{mm}}}
\newcommand{\kg}{\ensuremath{\kappa_{g}}}
\newcommand{\kq} {\ensuremath{\kappa_{q}}}

\newcommand\ABS  [1]{|\,{#1}\,|}            
\newcommand\CDOT   {\SPACE{\cdot}}
\newcommand\SPACE[1]{\,{#1}\,}              
\newcommand\EQ     {\SPACE{=}}
\newcommand\TO     {\SPACE{\rightarrow}}
\newcommand\MINUS  {\SPACE{-}}
\newcommand\PLUS   {\SPACE{+}}
\newcommand\OTIMES {\SPACE{\otimes}}
\newcommand\CAP    {\SPACE{\cap}}
\newcommand\CUP    {\SPACE{\cup}}
\newcommand\PM     {\SPACE{\pm}}
\newcommand\EQUIV  {\SPACE{\equiv}}
\newcommand\APPROX {\SPACE{\approx}}
\newcommand\GT     {\SPACE{>}}
\newcommand\LT     {\SPACE{<}}
\newcommand\GE     {\SPACE{\ge}}
\newcommand\LE     {\SPACE{\le}}
\newcommand\GG     {\SPACE{\gg}}
\newcommand\LL     {\SPACE{\mathchar"321C}}
\newcommand\GTRSIM {\SPACE{\gtrsim}}
\newcommand\TIMES  {\SPACE{\times}}
\newcommand\data   {\rm data}

\newcommand\no           {\!\!}
\newcommand\np           {\no\no}
\newcommand\nq           {\np\np}
\newcommand\nqq          {\nq\nq\nq}
\newcommand\nqqq         {\nq\nq\nq\nq}
\newcommand\nr           {\phantom{i}\nq}
\newcommand\z            {\phantom{0}}
\newcommand\Z            {\phantom{.0}}
\newcommand\QUAD         {~\,}

\newcommand\JETVHETO     {{\sc JetVHeto}}
\newcommand\PROPHECY     {{\sc prophecy}}
\newcommand\MCFM         {{\sc MCFM}}
\newcommand\HAWK         {{\sc hawk}}
\newcommand\MINLO        {{\sc minlo}}
\newcommand\VBFATNLO     {{\sc vbf@nlo}}
\newcommand\VBFATNNLO    {{\sc vbf@nnlo}}
\newcommand\MCATNLO      {{\sc mc@nlo}}
\newcommand\aMCATNLO     {a{\sc mc@nlo}}
\newcommand\GGTOWW       {{\sc gg2ww}}
\newcommand\GGTOVV       {{\sc gg2VV}}
\newcommand\ACERMC       {{\sc AcerMC}}
\newcommand\HDECAY       {{\sc hdecay}}
\newcommand\HRES         {{\sc HRes}}
\newcommand\TOPPP        {{\sc Top${++}$}}

\newcommand\dbline{\noalign{\vskip 0.10truecm\hrule\vskip 0.05truecm\hrule\vskip 0.10truecm}}
\newcommand\sgline{\noalign{\vskip 0.10truecm\hrule\vskip 0.10truecm}}
\newcommand\clineskip{\noalign{\vskip 0.10truecm}}

\newcommand\iab    {\mbox{\,ab$^{-1}$}}

\newcommand\DR           {{\Delta}R}
\newcommand\DeltaR       {\DR}
\newcommand\myeta  {\protect\raisebox{0.5px}{\ensuremath{\eta}}}

\hyphenation{ATLAS}
\renewcommand{\arraystretch}{1.2}

\section{Introduction} 
\label{sec:Introduction}


This paper presents studies of the spin and parity quantum numbers of the newly discovered Higgs particle~\cite{HiggsObservationATLAS,HiggsObservationCMS} 
in the $WW^* \rightarrow e \nu \mu \nu$ final state, where only final states with opposite-charge, different-flavour 
leptons ($e,\mu$) are considered. Determining the spin of the newly discovered resonance and 
its properties under charge-parity (CP) conjugation is of primary importance to firmly establish its nature, and in particular whether it is the Standard Model (SM) Higgs boson or not. Compared to the previous ATLAS publication~\cite{HiggsSpin2013}, this paper contains significant updates and improvements: the SM Higgs-boson hypothesis is compared with improved spin-2 scenarios.  The case where the observed resonance\footnote{In the following the abbreviated notation $J^{P}$ is used instead of $J^{CP}$.} has  $J^{P} = 1^{+}$ or $1^{-}$ is not studied in this 
paper as it is already excluded by previous publications both by the ATLAS~\cite{HiggsSpin2013} and CMS collaborations~\cite{CMSSpinComb}.

To simulate the alternative Higgs-boson hypotheses, the \mgaMC~\cite{MG5} generator is adopted. It includes terms of higher order ($\alphaS^3$) in the Lagrangian, in contrast to the JHU~\cite{jhu1,jhu2} event generator used in the previous publication~\cite{HiggsSpin2013}. In the context of this 
study, the 1-jet final state, which is more sensitive to contributions from the  higher-order terms, is analysed, in addition to the 0-jet final state.
 
Furthermore, the parity of the Higgs resonance is studied by testing the compatibility of the data with a beyond-the-Standard-Model (BSM) CP-even or CP-odd Higgs boson~\cite{HC}.
Finally, the case where the observed resonance is a mixed CP-state, namely  a mixture of a SM Higgs boson and a BSM CP-even or CP-odd Higgs boson, is investigated. 

This study follows the recently published $H\rightarrow W W^*$ analysis~\cite{ATLAS-CONF-2014-060} in the 0- and 1-jet channels with one 
major difference:  the spin and parity analysis uses multivariate techniques to disentangle the various signal hypotheses and the backgrounds 
from each other, namely Boosted Decision Trees (BDT)~\cite{TMVA}. The reconstruction and identification of physics objects in the event, the 
simulation and normalisation of backgrounds, and the main systematic uncertainties are the same as described in Ref.~\cite{ATLAS-CONF-2014-060}. 
This paper focuses in detail on the aspects of the spin and parity analysis that differ from that publication.


The outline of this paper is as follows: Sect.~\ref{sec:Theory} describes the theoretical framework for the spin and parity analysis, Sect.~\ref{sec:detector_samples} discusses 
the ATLAS detector, the data and Monte Carlo simulation samples used. The event selection and the background estimates are described in Sects.~\ref{sec:event_selection} 
and~\ref{sec:backgrounds}, respectively. The BDT analysis is presented in Sect.~\ref{sec:bdt_analysis}, followed by a description of the statistical tools used and of the various uncertainties in 
Sects.~\ref{sec:fit_uncertainties} and~\ref{systematics}, respectively. Finally, the results are presented in Sect.~\ref{sec:results}.

\label{sec:theory}

\section{Theoretical framework for the spin and parity analyses}
\label{sec:Theory}
In this section, the theoretical framework for the study of the spin and parity of the newly discovered 
resonance is discussed. The effective field theory (EFT) approach is adopted in this paper, within the 
Higgs characterisation model~\cite{HC} implemented in the 
\mgaMC~\cite{MG5} generator. 
Different hypotheses for the Higgs-boson spin and parity are studied. Three main categories  can be 
distinguished: the hypothesis that the observed resonance is a spin-2 resonance, a pure CP-even or 
CP-odd BSM Higgs boson, or a mixture of an SM Higgs and CP-even or CP-odd BSM Higgs bosons. 
The latter case would imply CP violation in the Higgs sector. 
In all cases, only the  Higgs boson with a  mass of 125~\GeV\ is considered.
In case of CP mixing, the Higgs boson would be a mass eigenstate, but not a CP eigenstate. 


The approach used by this model relies on an EFT, which by definition is only valid up to a certain energy scale 
$\Lambda$. This Higgs characterisation model considers that the resonance structure
recently observed corresponds to one new boson with $J^{P}=0^{\pm}, 1^{\pm}$ or $2^+$ and with 
mass of 125~\GeV, assuming that any other BSM particle exists at an energy scale larger than $\Lambda$.
The EFT approach has the advantage of being easily and systematically improvable by adding higher-dimensional 
operators  in the Lagrangian, which effectively corresponds to adding higher-order corrections, following the same 
approach as that used in perturbation theory. The  $\Lambda$ cutoff scale is set to 1~\TeV\ in this paper, to account 
for  the experimental results from the LHC and previous collider experiments that show no evidence of new physics at lower energy 
scales. More details can be found in Ref.~\cite {HC}.
In the EFT approach adopted, the Higgs-boson couplings to particles other than $W$ bosons are ignored as they would impact the signal 
yield with no effects on the $H\rightarrow WW^*$ decay kinematics, which is not studied in this analysis.


This section is organised as follows. Higgs-like resonances in the framework of the Higgs characterisation 
model are introduced in Sects.~\ref{sec:spin2theory} and ~\ref{sec:CPtheory}, for spin-2 and spin-0 particles, 
respectively. The specific benchmark models under study are described in Sects.~\ref{sec:choicespin2} and ~\ref{sec:choiceCP}.


 \subsection{Spin-2 theoretical model and benchmarks}

\subsubsection{ Spin-2 theoretical model}
\label{sec:spin2theory}

Given the large number of possible spin-2 benchmark models, a specific one is chosen, corresponding to a graviton-inspired 
tensor with minimal couplings to the SM particles~\cite{randall}. In the spin-2 boson rest frame,
its polarisation states projected onto the parton collision axis can take only the values of $\pm 2$ for the gluon fusion (ggF)
process and $\pm1$ for the \qq\ production process. For the spin-2 model studied in this analysis, only these two production 
mechanisms are considered. The Lagrangian $\mathcal{L}^{p}_{2}$ for a spin-2 minimal coupling model is defined as:

\begin{equation} \label{eq:spin2}
   \mathcal{L}^{p}_{2} = \sum_{p=V,f} -\frac{1}{\Lambda}\kappa_{p}T^{p}_{\mu\nu}\, X^{\mu\nu}_{2},
\end{equation}
where $T^{p}_{\mu\nu}$ is the energy-momentum tensor, $X^{\mu\nu}_{2}$ is the spin-2 particle field and $V$ and $f$ denote vector bosons ($Z$, $W$, $\gamma$ and gluons)  
and fermions (leptons and quarks), respectively. The $\kappa_{p}$ are the couplings of the Higgs-like resonance to particles, e.g. \kq\ and \kg\ label the couplings to quarks and gluons, respectively.

With respect to the previous publication~\cite{HiggsSpin2013}, the spin-2 analysis presented in this paper uses the \mgaMC~\cite{MG5} 
generator, which includes higher-order tree-level QCD calculations. As discussed in the following, these calculations have an important 
impact on the Higgs-boson transverse momentum \pTH\ distribution, compared to the studies already performed using a Monte Carlo (MC) 
generator at leading order~\cite{jhu1,jhu2}. In fact, when \kq\ is not equal to \kg\ (non-universal couplings), due to order-$\alphaS^3$ terms, 
a tail in the \pTH\ spectrum appears.

\begin{figure}[htb]
  \centering
\subfloat{\includegraphics[width=0.5\textwidth]{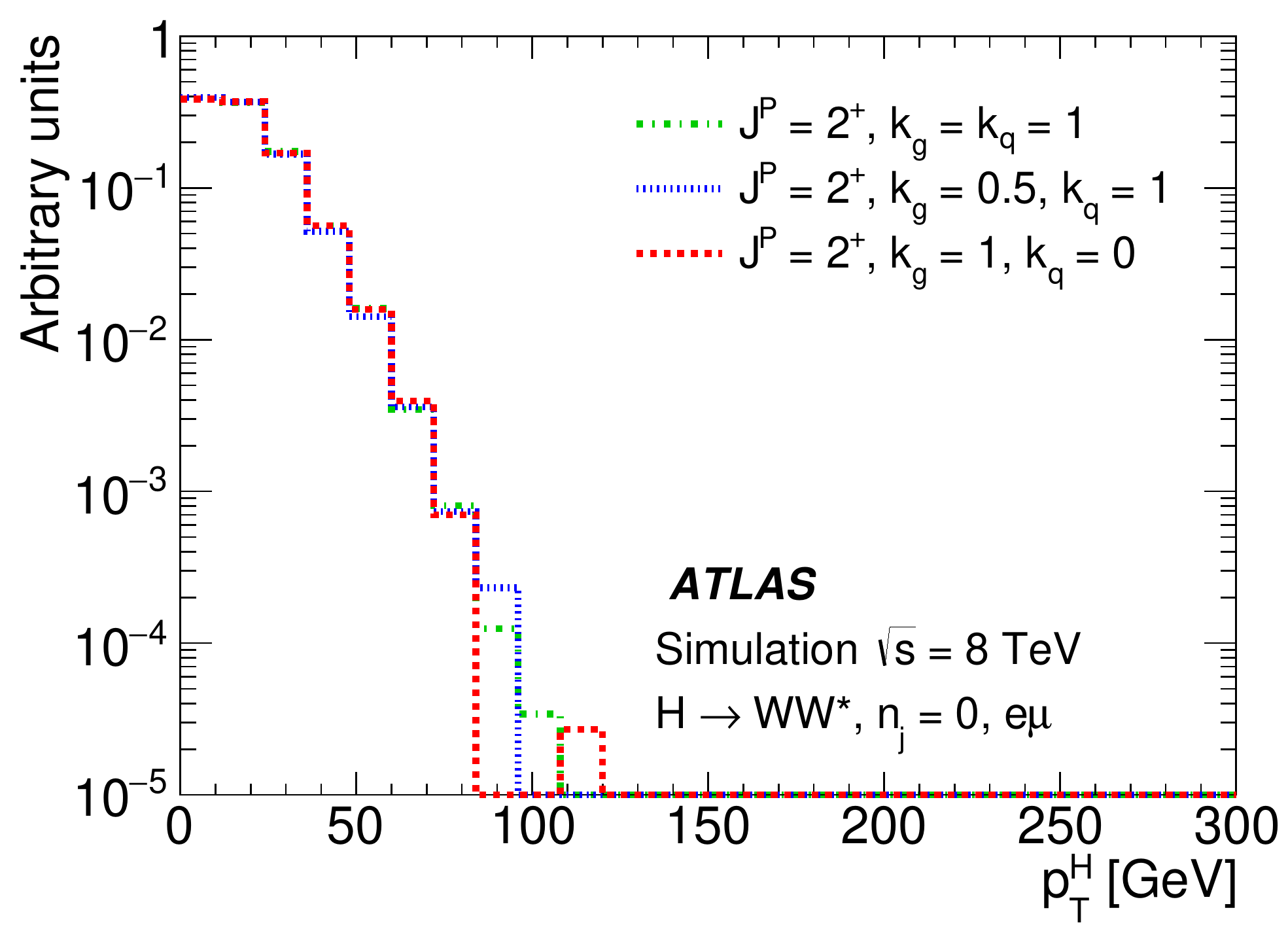}}
\subfloat{\includegraphics[width=0.5\textwidth]{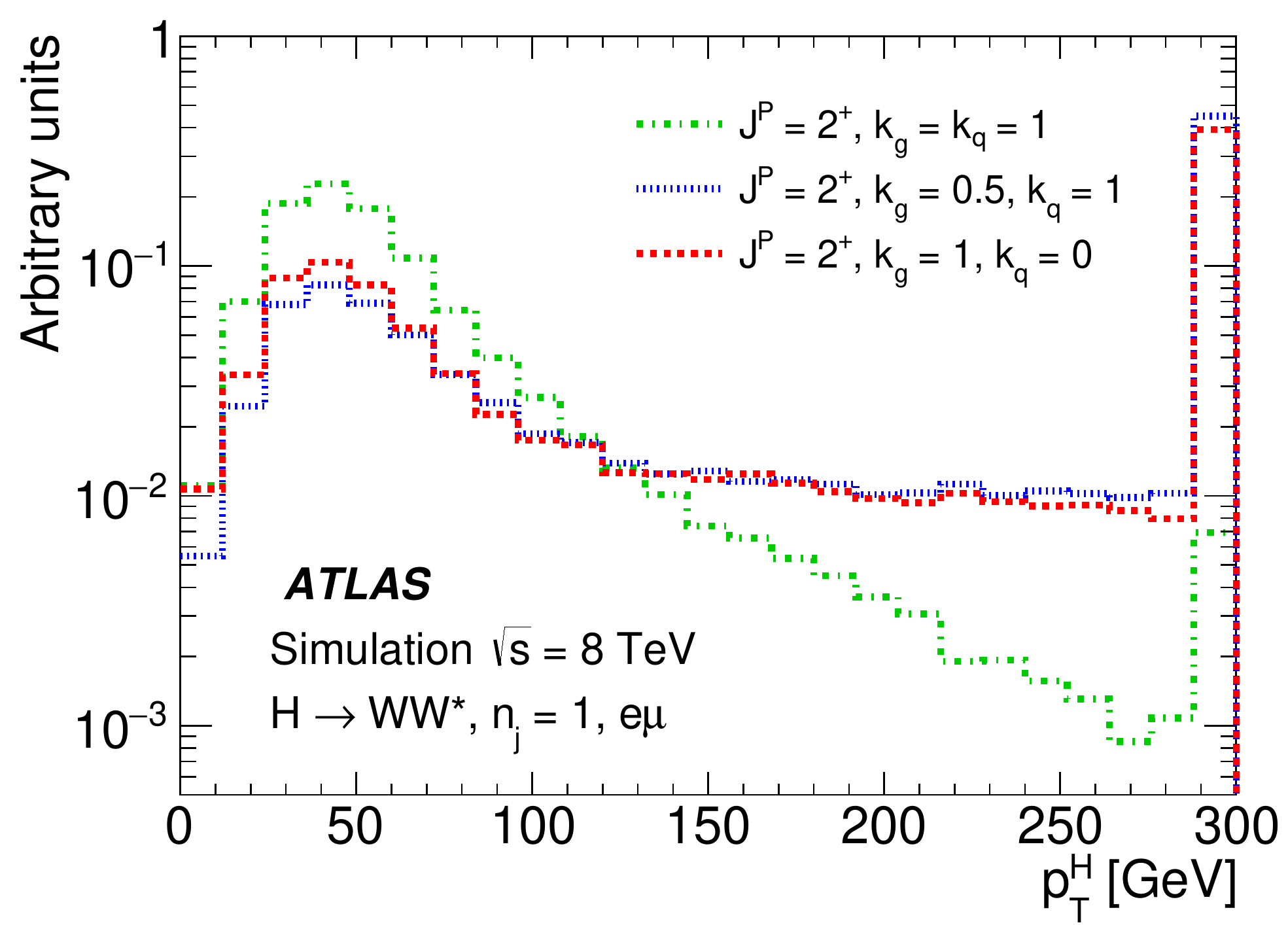}}
    \caption{The distribution of the transverse momentum of the Higgs boson, \pTH, at the Monte Carlo event-generator level for 0-jet (left) and 1-jet (right) final states. Three spin-2 signal hypotheses are shown:  $\kg\  = \kq\ = 1$, $\kg\  =0.5$,  $\kq\ = 1$ and $\kg\  = 1$, $\kq\ = 0$. The last bin in each plot includes the overflow.} 
  \label{fig:pttail}
\end{figure}

For leading-order (LO) effects, the \qq\ and ggF production processes are completely independent, but the beyond-LO processes contain diagrams with extra partons that give rise to a term proportional to $(\kq - \kg)^{2}$, which grows with the 
centre-of-mass energy squared of the hard process $(s)$ as $s^{3}/(m^{4}\Lambda^{2})$ (where $m$ is the mass of the spin-2 particle), and leads to a large tail at high values of \pTH. The distributions of some spin-sensitive observables are affected by this tail. For a more detailed discussion, see Ref.~\cite{HC}. This feature appears in final states with at least one jet, which indeed signals the presence of effects beyond leading order. 
Therefore, the 1-jet category is analysed in addition to the 0-jet category in this paper, in order to increase the sensitivity for these production
modes. Figure~\ref{fig:pttail} shows the \pTH\ distribution for the 0- and 1-jet final states at generator level after basic selection requirements 
(the minimum \pT\ required for the jets used for this study is 25~\GeV). Three different signal hypotheses are shown: one corresponding to universal 
couplings, $\kg  = \kq$, and two examples of non-universal couplings. The tail at high values of \pTH\ is clearly visible in the 1-jet category for the cases of non-universal couplings.

This \pTH\ tail would lead to unitarity violation if there were no cutoff scale for the validity of the theory. By definition, in the context 
of the EFT approach, at a certain scale $\Lambda$, new physics should appear and correct the unitarity-violating behaviour, even below the 
scale $\Lambda$. There is a model-dependent theoretical uncertainty on the \pT\ scale at which the EFT would be corrected by new physics: this uncertainty dictates the need to study benchmarks that use different \pTH\ cutoffs, as discussed in the following subsection.

\subsubsection{Choice of spin-2 benchmarks}
\label{sec:choicespin2}

Within the spin-2 model described in the previous section, a few benchmarks, corresponding to a range of possible scenarios, are studied in 
this paper. In order to  make sensible predictions for the spin-sensitive observables in the case of 
non-universal couplings, a cutoff on the Higgs-boson transverse momentum is introduced at a scale where the EFT is assumed to still be valid: this is chosen to be one-third of the scale $\Lambda$, corresponding to $\pT<300\GeV$. On the other hand, the lowest possible value up to which the EFT is valid by construction is the mass of the resonance itself; therefore it is important to study the effect of a threshold on \pTH\ at 125~\GeV.

Five different hypotheses are tested against the data:
\begin{itemize}
\item universal couplings: $\kg=\kq$;
\item $\kg=1$ and $\kq=0$, with two \pTH\ cutoffs at 125 and 300~\GeV;
\item   $\kg=0.5$ and $\kappa_q=1$, with two \pTH\ cutoffs at 125 and 300~\GeV.
\end{itemize}

The case $\kappa_g=0$ and $\kappa_q=1$ is not considered here, because it leads to a \pTH\ distribution which disagrees with the data, as shown in the $H\rightarrow \gamma \gamma$ and $H\rightarrow ZZ$ differential cross-section measurements~\cite{gamgam_diff,ZZ_diff}. 

\subsection{Spin-0 and CP-mixing theoretical models and benchmarks}
\subsubsection{Spin-0 and CP-mixing theoretical models}

\label{sec:CPtheory}

In the case where the spin of the Higgs-like resonance is zero, there are several BSM scenarios that predict the parity of the Higgs particle to be either even or odd~\cite{CPworkshop}. Another interesting possibility is that the Higgs-like resonance is not a CP eigenstate, but a mixture of CP-even and CP-odd states. This would imply CP violation in the Higgs sector, which is possible in the context of the Minimal Supersymmetric Standard Model~\cite{MSSMCP} or of two Higgs-doublet models~\cite{2HDM}. This CP violation might be large enough to explain the prevalence of matter over antimatter in the universe.

In the adopted EFT description, the scalar boson has the same properties as the SM Higgs boson, and its interactions with the SM particles are described by the appropriate operators. The BSM effects are expressed in terms of interactions with SM particles via higher-dimensional operators.

The  effective Lagrangian ${\cal L}_0^W$ adopted for this study, in order to describe the interactions of $W$ bosons with scalar and pseudoscalar states, is expressed as:

\begin{align}
 {\cal L}_0^W =\bigg\{&
  c_{\alpha}\kappa_{\rm \sss SM}\big[g_{\rm \sss HWW}\, W^+_\mu W^{-\mu}\big] \nn
  -\frac{1}{2}\frac{1}{\Lambda}\big[c_{\alpha}\kappa_{\rm \sss HWW} \, W^+_{\mu\nu}W^{-\mu\nu}
        +s_{\alpha}\kappa_{\rm \sss AWW}\,W^+_{\mu\nu}\widetilde W^{-\mu\nu}\big] \nn\\ 
  &-\frac{1}{\Lambda}c_{\alpha} 
    \big[
         \big(\kappa_{\rm \sss H\partial W} \, W_{\nu}^+\partial_{\mu}W^{-\mu\nu}+h.c.\big)
 \big]
 \bigg\} X_0\,,
 \label{eq:CPlag}
\end{align}
where $ W_{\mu\nu} =\partial_{\mu}W^{\pm}_{\nu}-\partial_{\nu}W^{\pm}_{\mu}$, 
$\widetilde W_{\mu\nu} =1/2 \cdot \epsilon_{\mu\nu\rho\sigma}W^{\rho\sigma}$
and $\epsilon_{\mu\nu\rho\sigma}$ is the Levi-Civita tensor, while $X_0$ represents the spin-0 Higgs-boson field~\cite{HC}.
In the SM, the coupling of the Higgs boson to the $W$ bosons is given by $g_{\rm \sss HWW}$, while the angle 
$\alpha$ describes the mixing between CP-even and CP-odd states. The notation $c_\alpha\equiv \cos \alpha\,, s_\alpha\equiv \sin \alpha\,$
is used in the Lagrangian. The dimensionless coupling parameters  $\kappa_i$ are real and describe CP violation in the most general way. The parameter
\ksm\ describes the deviations of the Higgs-boson coupling to the vector boson $W$ from those predicted by the SM, while \kaww\ and \khww\ are the BSM CP-odd and CP-even coupling parameters, respectively.\footnote{The Lagrangian terms associated to the higher-dimensional operators 
are called in this paper {\it BSM CP-even} and {\it BSM CP-odd Higgs bosons}. }
The mixing between the CP-even SM Higgs boson and the CP-even BSM Higgs boson can be achieved by changing the relative strength of the couplings \ksm\ and  \khww. The $\cos \alpha$ term multiplies both the SM and BSM CP-even terms in the Lagrangian and therefore its value does not change the relative importance of those contributions. This is different from the mixing of CP-even and  CP-odd states, as a $\sin \alpha$ term multiplies  the CP-odd state in the Lagrangian. The last term of the Lagrangian is due to derivative operators which are relevant in the case one of the two vector bosons is off-shell.

The higher-dimensional operator terms in the Lagrangian are the terms that contain \kaww\ and \khww\ and  are 
suppressed by a factor $1/\Lambda$. The  SM Higgs boson is described by the first term of the Lagrangian, corresponding to the following 
choice of parameters:
$\ksm=1$, $\kaww = \khww=0$ and 
$| c_\alpha|=1$. 
The derivative operator (the \khdw\ term) described in the Lagrangian of Eq.~(\ref{eq:CPlag}) would modify the results below the sensitivity achievable with the available data statistics. In fact, the effects on the kinematic distributions introduced by the derivative operator in the same range of variation of  \khww\ are at most 10--20\% of the ones produced by \khww\ itself. Since the present analysis is barely sensitive to \khww, the even smaller \khdw\ variations are not studied further, and the
corresponding term in the Lagrangian is neglected.


\subsubsection{Choice of CP benchmarks}
\label{sec:choiceCP}

The following approach to study different CP hypotheses under the assumption of a spin-0 hypothesis is taken in this paper. First of all, in the 
fixed-hypothesis scenario, the cases where the observed resonance is a pure BSM CP-even or CP-odd Higgs boson are considered. In addition, 
the mixing between the CP-even SM and BSM CP-odd or CP-even Higgs bosons is studied. In the CP-odd case, the mixing depends on 
the value of \kaww\ and on the mixing angle $\alpha$. As can be deduced from Eq.~(\ref{eq:CPlag}), varying  $\alpha$ or 
\kaww\ has an equivalent effect on the kinematic variable distributions; therefore in this paper only the  $\alpha$ parameter is 
varied while \kaww\ is kept constant. The scan range of $\alpha$ covers the entire range from $-\pi/2$~to~$\pi/2$ as the final state kinematic distributions differ for positive and negative values of $\alpha$. On the other hand, the mixing between the CP-even SM and CP-even BSM Higgs bosons depends exclusively  on the 
value of \khww\ and not on the value of $\alpha$.

To summarise, four hypotheses are tested against the data in this paper (for the cutoff value $\Lambda  = 1$ TeV):
\begin{itemize}
\item  Compare the SM Higgs-boson case with the pure BSM CP-even case, defined as $\ksm = 0$, $\kaww= 0$, $\khww = 1$, $ c_{\alpha} = 1$.

\item Compare the SM Higgs-boson case with the BSM CP-odd case, defined as $\ksm= 0$, $\kaww = 1$, $\khww= 0$, $c_{\alpha} = 0$.

\item Scan over $\tan \alpha$: under the assumption of a mixing between a CP-even SM Higgs boson and a CP-odd BSM Higgs boson. The mixing parameter is defined as 
 $\left(\tilde{\kappa}_{\rm \sss AWW}/\ksm\right) \cdot \tan \alpha$, where $\tilde{\kappa}_{\rm \sss AWW}=(1/4) \cdot \left(v/\Lambda\right)\cdot \kaww$, $v$ is the vacuum expectation value and $\tan \alpha$ is the only variable term (corresponding to variations
of $c_{\alpha}$ between --1 and 1). The other parameters are set as follows: $\ksm = 1$, $\kaww = 1$, $\khww = 0$.

\item Scan over \khww: under the assumption of a mixing between a CP-even SM Higgs boson and a CP-even BSM Higgs boson. The mixing parameter is defined as $\tilde{\kappa}_{\rm \sss HWW}/\ksm$, where $\tilde{\kappa}_{\rm \sss HWW}=(1/4) \cdot \left(v/\Lambda\right) \cdot \khww$ and the only variable
term is \khww\ (corresponding to variations of $\tilde{\kappa}_{\rm \sss HWW}/\ksm$ between --2.5 and 2.5). 
For larger values of this ratio, the kinematic distributions of the final-state particles asymptotically tend to the ones obtained in presence of a pure CP-even BSM Higgs boson. The latter is used as the last point of the scan. The other parameters are set as follows:  $\ksm = 1$, $\kaww = 0$, $c_{\alpha} = 1$.

\end{itemize}

In the case of CP-mixing, only one MC sample is generated (see Sect.~\ref{sec:detector_samples}), and all other samples are obtained from it by reweighting the events on the basis of the matrix element amplitudes derived from Eq.~(\ref{eq:CPlag}). The precision of this procedure is verified to be better than the percent level. The mixing parameters used to produce this sample are chosen such that the kinematic phase space for all CP-mixing scenarios considered here was fully populated, and the values of the parameters are: $\ksm=1$, $\kaww=2$, $\khww=2$, $c_\alpha=0.3$.

In addition, it is interesting to study the case where the SM, the BSM CP-even and the CP-odd Higgs bosons all mix. Unfortunately, in the $H\rightarrow WW^*$ channel, the present data sample size limits the possibility to constrain such a scenario, which would imply a simultaneous scan of two parameters $\tan \alpha$ and \khww. 
In particular this is due to the lack of sensitivity in the \khww\ scan, consequently, as already stated, both the two and the three parameter scans, including in addition the derivative operators, are not pursued further. These studies are envisaged for the future.

\section{ATLAS detector, data and MC simulation samples} 
\label{sec:detector_samples}

This section describes  the ATLAS detector, along with the data and MC simulation samples used for this analysis. 

\subsection { The ATLAS detector}

The ATLAS detector \cite{atlas-det} is a multipurpose particle detector with approximately forward-backward 
symmetric cylindrical geometry and a near $4\pi$ coverage in solid angle.\footnote{The experiment uses a right-handed coordinate system with the origin at the nominal $pp$ interaction point at the centre of the detector. The
positive $x$-axis is defined by the direction from the origin to the centre of the LHC ring, the positive $y$-axis
points upwards, and the $z$-axis is along the beam direction. Cylindrical coordinates $(r,\phi)$ are used in the
plane transverse to the beam, with $\phi$ the azimuthal angle around the beam axis. Transverse components of 
vectors are indicated by the subscript T. The pseudorapidity is defined in terms of the polar angle 
$\theta$ as $\eta=-\mathrm{ ln~tan}(\theta/2)$. The angular distance between two objects is defined 
as $\DeltaR= \sqrt{(\Delta \eta) ^2 + (\Delta \phi)^2}$.}

The inner tracking detector (ID) consists of a silicon-pixel detector, 
which is closest to the interaction point, a silicon-strip detector surrounding the pixel detector, both covering up to 
$| \eta|= 2.5$, and an outer transition-radiation straw-tube tracker (TRT) covering $| \eta|< 2$. The ID is surrounded 
by a thin superconducting solenoid providing a 2~T axial magnetic field. 

A highly segmented lead/liquid-argon (LAr) sampling electromagnetic calorimeter measures the energy and the 
position of electromagnetic showers over $| \eta |< 3.2$. The LAr calorimeter includes a presampler (for $| \eta |< 1.8$)
and three sampling layers, longitudinal in shower depth, up to $| \eta |<2.5$. LAr sampling calorimeters are
also used to measure hadronic showers in the end-cap ($1.5<| \eta |<3.2$) and both the electromagnetic and hadronic
showers in the forward ($3.1< | \eta |< 4.9$) regions, while an iron/scintillator tile sampling calorimeter measures hadronic 
showers in the central region ($ | \eta |< 1.7$).

The muon spectrometer (MS) surrounds the calorimeters and is designed to detect muons in the pseudorapidity 
range $ | \eta |< 2.7$. The MS consists of one barrel ($ | \eta |<1.05$) and two end-cap regions. A system of
three large superconducting air-core toroid magnets provides a magnetic field with a bending integral of about 
2.5~T$\cdot$m (6~T$\cdot$m) in the barrel (end-cap) region. Monitored drift-tube chambers in both the barrel and 
end-cap regions and cathode strip chambers covering $2.0<| \eta |<2.7$ are used as precision measurement chambers, 
whereas resistive plate chambers in the barrel and thin gap chambers in the end-caps are used as trigger chambers, 
covering up to $ | \eta |$  = 2.4. 

A three-level trigger system selects events to be recorded for offline analysis. The first-level trigger is hardware-based, while 
the higher-level triggers are software-based. 

\subsection{Data and Monte Carlo simulation samples}

The data and MC simulation samples used in this analysis are a subset of those used in Ref.~\cite{ATLAS-CONF-2014-060}
with the exception of the specific spin/CP signal samples produced for this paper.

The data were recorded by the ATLAS detector during the 2012 LHC run with proton--proton collisions at a centre-of-mass 
energy of 8~\TeV, and correspond to an integrated luminosity of 20.3~fb$^{-1}$. This analysis uses events selected by 
triggers that required either a single high-\pT\ lepton or two leptons. Data quality requirements are applied to reject events 
recorded when the relevant detector components were not operating correctly. 

Dedicated MC samples are generated to evaluate all but the \Wjets\ and multi-jet backgrounds, which are estimated using data 
as discussed in Sect.~\ref{sec:backgrounds}.  Most samples use the \POWHEG~\cite{Nason:2004rx} generator, which includes
corrections at next-to-leading order (NLO) in $\alphaS$ for the processes of interest. In cases where higher parton multiplicities 
are important, \ALPGEN~\cite{alpgen} or \SHERPA~\cite{Gleisberg:2008ta}~provide merged calculations at tree level for up to five 
additional partons.  In a few cases, only leading-order generators (such as \ACERMC~\cite{Kersevan:2004yg}~or \GGTOVV~\cite{Kauer:2013qba}) 
are available.  Table~\ref{tab:mc} shows the event generator and production cross-section times branching fraction used for each 
of the signal and background processes considered in this analysis.

\begin{table}[t!]
\caption{
  Monte Carlo samples used to model the signal and background processes.  The 
  corresponding cross-sections times branching fractions, $\sigma{\CDOT}\mathcal{B}$,
  are quoted at $\sqrt{s}{\EQ}8\TeV$.  The branching fractions include the decays
  $t{\TO}Wb$, $\Wlv$, and $Z{\TO}\ell\ell$ (except for the process 
  $\ZZ{\TO}\ell\ell\,\nu\nu$).  Here $\ell$ 
  refers to $e$, $\mu$, or $\tau$.  The neutral 
  current $Z/\gamma^\ast{\TO}\ell\ell$ process is denoted $Z$ or $\gamma^\ast$, 
  depending on the mass 
  of the produced lepton pair. The parameters \kg, \kq\ are defined in Sect.~\ref{sec:spin2theory}, while \ksm, \khww, \kaww, $
  c_\alpha$ are defined in Sect.~\ref{sec:CPtheory}.
}
\centering
\scalebox{1}{

\begin{tabularx}{\textwidth}{llll} 

\dbline
{Process}  &  MC generator$\nq$ & Filter & {$\sigma{\CDOT}\mathcal{B}$\,(pb) }
\\
\sgline
 \multicolumn{4}{l}{ Signal samples used in $J^{P}=2^+$ analysis } \\

\quad SM {$\HWW$}   &  \POWHEG+\PYTHIA8  &    & 0.435 \\
\quad  {$\kg=\kq$ } &   {\mgaMC+\PYTHIA6} & & - \\
\quad  {$\kg=1$, $\kq=0$} &  {\mgaMC+\PYTHIA6} & & - \\
 \quad  {$\kg=0.5$, $\kq=1$} &  {\mgaMC+\PYTHIA6} & & - \\
\sgline
 \multicolumn{4}{l}{ Signal samples used in CP-mixing analysis} \\
\quad {$c_\alpha=0.3$, $\ksm=1$} & \multirow{2}{*} {\mgaMC+\PYTHIA6} & & \multirow{2}{*}- \\
   \quad {$\khww=2$, $\kaww=2$} &     & &  \\

\dbline
 \multicolumn{4}{l}{ Background samples } \\
\quad $\WW$ & & \\
\qquad $\qq{\TO}\WW$ and $qg{\TO}\WW$  &  \POWHEG+\PYTHIA6 &     & 5.68 \\ 
\qquad $gg{\TO}\WW$                   &   \GGTOVV+\HERWIG &      & 0.196 \\
\sgline
\quad Top quarks & & & \\
\qquad $\ttbar$ &  \POWHEG+\PYTHIA6   &   & 26.6 \\
\qquad $Wt$ &  \POWHEG+\PYTHIA6 &     & 2.35 \\
\qquad $tq\bar{b}$ &  \ACERMC+\PYTHIA6 &     & 28.4 \\
\qquad $t\bar{b}$ &  \POWHEG+\PYTHIA6  &    & 1.82 \\
\sgline
\quad Other dibosons ($VV$)& & & \\
\qquad $\Wg$  & \ALPGEN+\HERWIG & $\pT^{\gamma}{\GT}8\GeV$      & 369 \\
\qquad $\Wgs$     & \SHERPA      & $\mll{\LE}7\GeV$           & 12.2 \\ 
\qquad $\WZ$        & \POWHEG+\PYTHIA8 & $\mll{\GT}7\GeV$      & 12.7 \\ 
\qquad $\Zg$  & \SHERPA      & $\pT^{\gamma}{\GT}8\GeV$         & 163 \\
\qquad $\Zgs$  & \SHERPA      & min.\ $\mll{\LE}4\GeV$           & 7.31 \\
\qquad $\ZZ$     & \POWHEG+\PYTHIA8  & $\mll{\GT}4\GeV$     & 0.733 \\
\qquad $\ZZ {\TO}\ell\ell\,\nu\nu$ &\POWHEG+\PYTHIA8   &  $\mll{\GT}4\GeV$    & 0.504 \\
\sgline
\quad Drell --Yan & & & \\
\qquad $Z/\gamma^*$       & \ALPGEN+\HERWIG  $\np$ & $\mll{\GT}10\GeV$ & 16500 \\
\dbline
\end{tabularx}
}

\label{tab:mc}
\end{table}


The matrix-element-level Monte Carlo calculations are matched to a model of the parton shower, underlying event and 
hadronisation, using either \PYTHIA6~\cite{Sjostrand:2006za}, \PYTHIA8~\cite{Sjostrand:2007gs}, \HERWIG~\cite{Corcella:2000bw} 
(with the underlying event modelled by {\JIMMY}~\cite{jimmy}), or \SHERPA.  Input parton distribution functions (PDFs) 
are taken from CT10~\cite{Lai:2010vv} for the \POWHEG~and \SHERPA~samples and CTEQ6L1~\cite{cteq6} 
for the \ALPGEN+\HERWIG~and \ACERMC~samples.  The Drell--Yan (DY) sample ($\ZDY$+jets) is reweighted to the MRST PDF set~\cite{mrst}.

The effects of the underlying event and of additional minimum-bias interactions occurring in the same or neighbouring bunch 
crossings, referred to as pile-up in the following,  are modelled with \PYTHIA8, and the ATLAS detector response is 
simulated~\cite{atlassim} using either \GEANT4~\cite{GEANT4} or \GEANT4 combined with a parametrised \GEANT4-based 
calorimeter simulation~\cite{AFII}. 

For the signal, the ggF production mode for the $\HWW$ signal is modelled with \POWHEG+\PYTHIA8~\cite{Bagnaschi:2011tu, Nason:2009ai} at 
$m_\text{H} = 125\GeV$ for the SM Higgs-boson signal in the spin-2 analysis, whereas {\textsc{MadGraph5\_aMC@}
{\textsc{NLO}}~\cite{MG5} is 
used for the CP analysis. The $H + 0,1,2$~partons samples are generated with LO accuracy, and subsequently showered with \PYTHIA6. 
For the BSM signal, the \mgaMC\ generator is used in all cases. For the CP analysis, all samples (SM and BSM) 
are obtained by using the matrix-element reweighting method applied to a CP-mixed sample, as mentioned in Sect.~\ref{sec:CPtheory}, 
to provide a description of different CP-mixing configurations. The PDF set used is CTEQ6L1. To improve the modelling of the SM 
Higgs-boson \pT, a reweighting scheme is applied that reproduces the prediction of the next-to-next-to-leading-order (NNLO)  and next-to-next-to-leading-logarithms (NNLL) dynamic-scale calculation 
given by the \HRES2.1 program~\cite{deFlorian:2011xf,Grazzini:2013mca}. The BSM spin-0 Higgs-boson \pT\ is reweighted to 
the same distribution.

Cross-sections are calculated for the dominant diboson and top-quark processes as follows: the inclusive $\WW$ cross-section 
is calculated to NLO with \MCFM~\cite{mcfm6}; non-resonant gluon fusion is calculated and modelled to LO in 
$\alphaS$ with~\GGTOVV, including both $\WW$ and $\ZZ$ production and their interference; $\ttbar$ production is 
normalised to the calculation at NNLO in $\alphaS$, with resummation of higher-order terms to NNLL accuracy, evaluated with \TOPPP2.0~\cite{Czakon:2011xx}; single-top-quark processes 
are normalised to NNLL, following the calculations from Refs.~\cite{Kidonakis:2010tc,Kidonakis:2011wy}~and~\cite{Kidonakis:2010ux} 
for the $s$-channel, $t$-channel, and $Wt$ processes, respectively.  

The $WW$ background and the dominant backgrounds involving top-quark production ($\ttbar$ and $Wt$) are modelled using 
the {\POWHEG+\PYTHIA6}  event generator~\cite{Melia:2011tj,Frixione:2007nw,Re:2010bp,Alioli:2009je}. For $\WW$, $\WZ$, and $\ZZ$ production via non-resonant vector boson scattering, 
the \SHERPA~generator provides the 
LO cross-section and is used for event modelling.  The negligible vector-boson-scattering (VBS) $\ZZ$ process is not shown in Table~\ref{tab:mc} but is included in 
the background modelling for completeness.  
The process $\Wgs$ is defined as associated $W$+$\ZDY$ production, containing an opposite-charge same-flavour 
lepton pair with invariant mass $\mll$ less than $7\GeV$.  This process is modelled using \SHERPA~with up to one additional 
parton.  The range $\mll{\GT}7\GeV$ is simulated with \POWHEG+\PYTHIA8 and normalised to the {\POWHEG} cross-section. 
The use of {\SHERPA} for $\Wgs$ is due to the inability of \POWHEG+\PYTHIA8 to model invariant masses down to 
the production threshold.  The {\SHERPA} sample requires two leptons with $\pT{\GT}5\GeV$ and $\ABS{\myeta}{\LT}3$.  
The jet multiplicity is corrected using a {\SHERPA} sample generated with $0.5{\LT}\mll{\LT}7\GeV$ and up to two additional partons,
while the total cross-section is corrected using the ratio of the {\MCFM} NLO to {\SHERPA} LO calculations in the same restricted 
mass range.  A similar procedure is used to model $Z\gamma^*$, defined as $\ZDY$ pair production with one same-flavour 
opposite-charge lepton pair having $\mll{\LE}4\GeV$ and the other having $\mll{\GT}4\GeV$.

The $\Wg$ and DY processes are modelled using \ALPGEN +\HERWIG~with merged tree-level calculations of up to five jets.  The merged 
samples are normalised to the NLO calculation of \MCFM~(for $\Wg$) or the NNLO calculation of DYNNLO~\cite{Catani:2009sm} 
(for DY).  The $\Wg$ sample is generated with the requirements $\pT^{\gamma}{\GT}8\GeV$ and $\DeltaR(\gamma, \ell){\GT}0.25$. 

A \SHERPA~sample is used to accurately model the $Z(\to\ell\ell) \gamma$ background.  The photon is required to have $\pT^{\gamma}{\GT}8\GeV$ 
and $\DeltaR(\gamma, \ell){\GT}0.1$; the lepton pair must satisfy $\mll{\GT}10\GeV$. The cross-section is normalised to NLO using \MCFM.  
Events are removed from the {\ALPGEN +\HERWIG} DY samples if they overlap with the kinematics defining the {\SHERPA} $Z(\to\ell\ell) \gamma$ sample.

\section{Event selection} 
\label{sec:event_selection}

The object reconstruction in terms of leptons, jets, and missing transverse momentum, as well as the lepton identification and isolation criteria, which 
were optimised to minimise the impact of the background from misidentified isolated prompt leptons, are the same as described in detail in 
Ref.~\cite{ATLAS-CONF-2014-060}: these aspects are therefore not discussed in this paper. The selection criteria and the analysis methodology used for 
the spin/CP studies described here are different however, since they are motivated not only by the need to distinguish the background processes from the 
Higgs-boson signal, but also by the requirement to optimise the separation power between different signal hypotheses. Thus, several selection requirements 
used in Ref.~\cite{ATLAS-CONF-2014-060} are loosened or removed in the selection described below.

This section is organised in four parts. First, the event preselection is described, followed by the discussion of
the spin- and parity-sensitive variables. These variables motivate the choice of topological selection requirements in the 0-jet
and 1-jet categories described in the last two sections. All selection criteria are summarised in 
Table~\ref{tab:comparecuts} and the corresponding expected and observed event yields are presented in Table~\ref{tab:Cutflow}.

\subsection{Event preselection}
\label{sec:preselection}

The $WW \rightarrow e\nu \mu \nu$ final state chosen for this analysis consists of $e\mu$ pairs, namely pairs of opposite-charge, different-flavour, identified and isolated 
prompt leptons. This choice is based on the expected better sensitivity of this channel compared to the same-flavour channel, which involves a large potential 
background from $Z/\gamma^* \rightarrow ee/\mu\mu$ processes. The preselection requirements are designed to reduce substantially the dominant 
background processes to the Higgs-boson signal (see Sect.~\ref{sec:backgrounds}) and can be summarised briefly as follows:
\begin{itemize}
\item The leading lepton is required to have $\pT > 22$~\GeV\ to match the trigger requirements.
\item The subleading lepton is required to have $\pT > 15$~\GeV.
\item The mass of the lepton pair is required to be above 10~\GeV.
\item The missing transverse momentum in the event is required to be $\ptmiss >20$~\GeV.
\item The event must contain at most one jet with $\pT > 25$~\GeV\ and $|\eta| < 4.5$. The jet \pT\ is required to be higher than 30~\GeV\ in the forward region, $2.4 < |\eta| < 4.5$, to minimise the impact of pile-up.
\end{itemize}

This analysis considers only $e\mu$ pairs in the 0-jet and 1-jet categories for the reasons explained in Sect.~\ref{sec:theory}. Each category is analysed independently since they display rather different background compositions and signal-to-background ratios.

\subsection{Spin- and CP-sensitive variables}
\label{sec:variables}

The shapes of spin- and CP-sensitive variable distributions are discussed in this section for the preselected events. 

Figures \ref{fig:signal_cutvars_0jet} and~\ref{fig:signal_cutvars_1jet}  show the variables used to discriminate different spin-2 signal hypotheses 
from the SM Higgs-boson hypothesis for the 0-jet and the 1-jet category, respectively. For both the 0-jet and the 1-jet categories,
the most sensitive variables are \ptll\ (transverse momentum of the dilepton system), \mll, \dphill\ ($\phi$ 
angle between the two leptons) and \mT\ (transverse mass of the dilepton and missing momentum system). These  variables are 
the same as those used for the spin-2 analysis in the previous publication~\cite{HiggsSpin2013}.

Similarly, Figs.~\ref{fig:inputvars_shapes_training_CP_even} and~\ref{fig:inputvars_shapes_training_CP_odd} show the 
the variables that best discriminate between an SM Higgs boson and a BSM CP-even or CP-odd signal, respectively. The BSM 
CP-even variables are the same as those used in the spin-2 analysis, apart from the \ptmiss\ variable which is substituted for \mT. 
The variables for the CP-odd analysis are \mll, \Efun, \dpt, \dphill, where $ E_{\ell \ell \nu \nu} =p_{\text T}^{{\ell}_1} -0.5p_{\text T}^{{\ell}_2}+0.5\ptmiss$,
$p_{\text T}^{{\ell}_1}$ and $p_{\text T}^{{\ell}_2}$ are respectively the transverse momenta of the leading and subleading leptons, 
and ~\dpt\ is the absolute value of their difference.

The CP-mixing analysis studies both the positive and negative values of the mixing parameter, as explained in Sect.~\ref{sec:choiceCP}.
In the BSM CP-even benchmark scan, for negative values of the mixing parameter,  interference  between the SM and BSM CP-even 
Higgs-boson couplings causes a cancellation that drastically changes the shape of the discriminating variable distributions. As an example, 
Fig.~\ref{fig:DPhineg} shows the distribution of \dphill\ for the SM Higgs boson together with the distributions for several different values of the CP-mixing parameter.

While for positive values of $\tilde{\kappa}_{\rm \sss HWW}/\ksm$ (Fig.~\ref{fig:DPhineg}, left) and for the SM Higgs-boson hypothesis, the \dphill\ 
distribution peaks towards low values, when reaching the maximum of the interference (at about $\tilde{\kappa}_{\rm \sss HWW}/\ksm \sim{-1}$), 
the mean of the \dphill\ distribution slowly moves towards higher values. This significantly improves the separation power between the SM and the 
BSM CP-even Higgs-boson hypotheses (Fig.~\ref{fig:DPhineg}, right). For values of $\tilde{\kappa}_{\rm \sss HWW}/\ksm<-1$, the peak of distribution gradually moves back to low values of \dphill, as in the case of the SM Higgs-boson hypothesis. The sum of the backgrounds is also shown on 
the same figure. The other CP-sensitive variables exhibit a similar behaviour in this specific region of parameter space. The impact of this feature on 
the results is discussed in Sect.~\ref{sec:CPunbresults}.

\begin{figure}[h]
\centering
\subfloat{\includegraphics[width=0.49\textwidth]{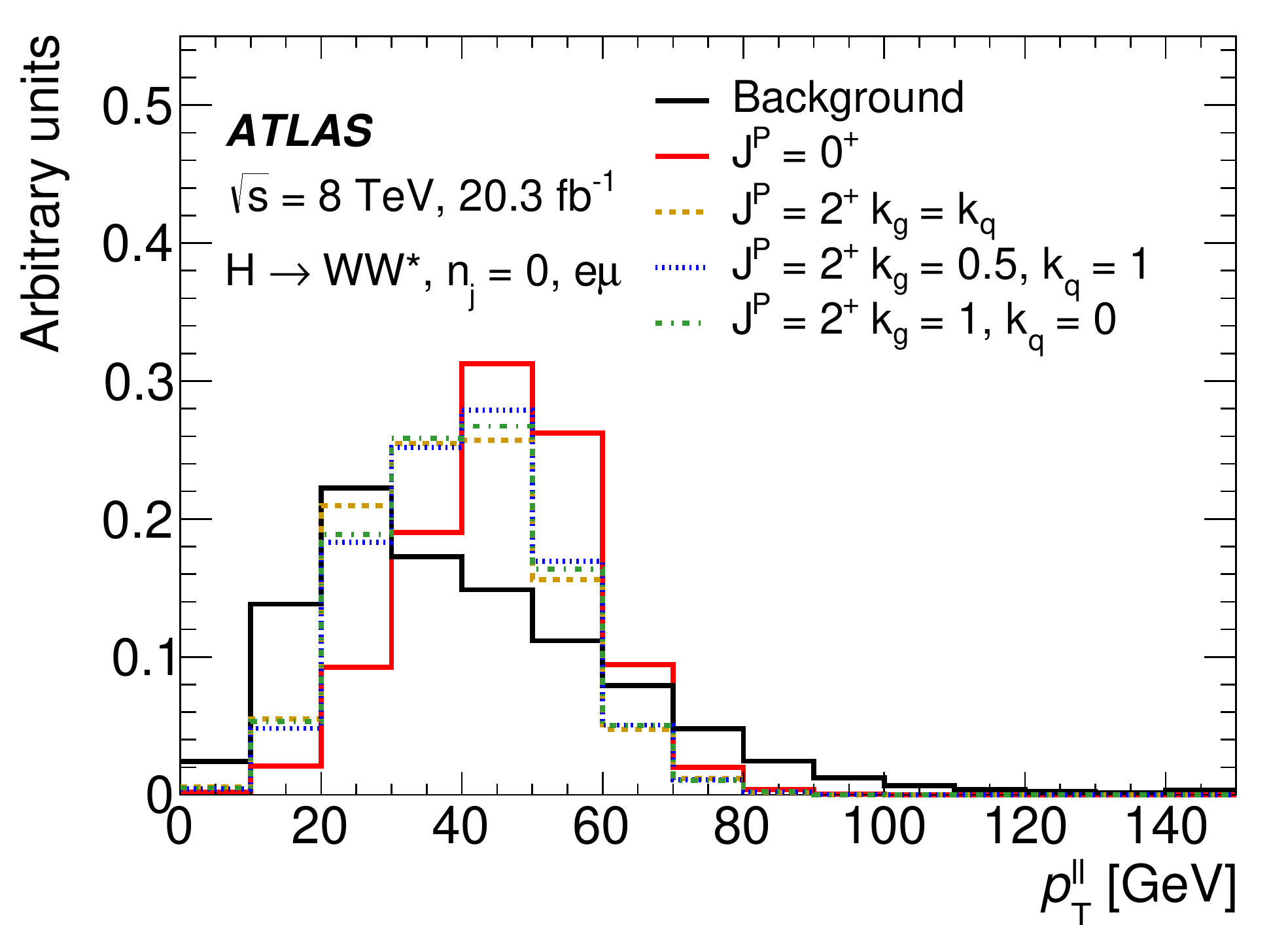}}
\subfloat{\includegraphics[width=0.49\textwidth]{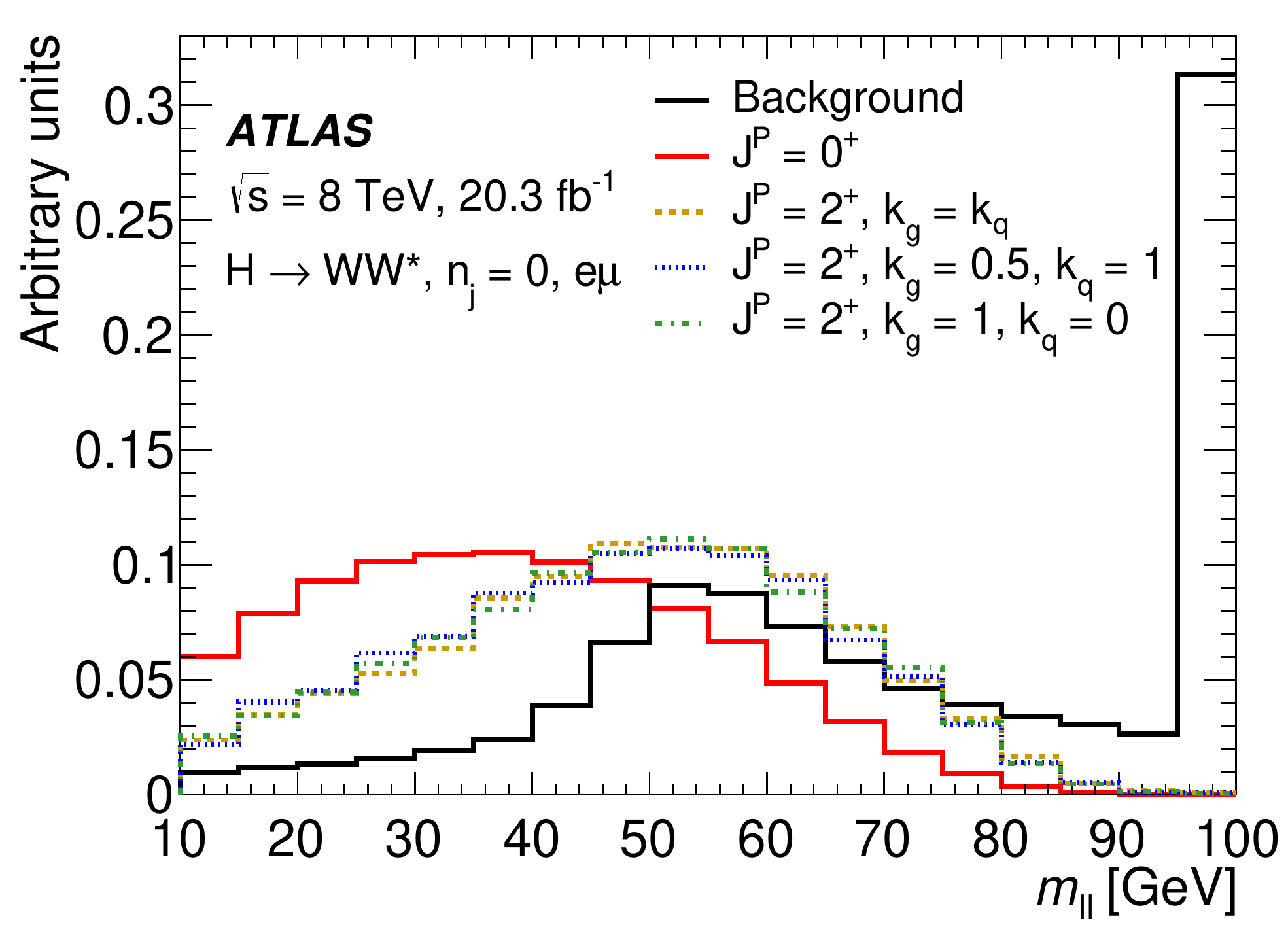}}\\
\subfloat{\includegraphics[width=0.49\textwidth]{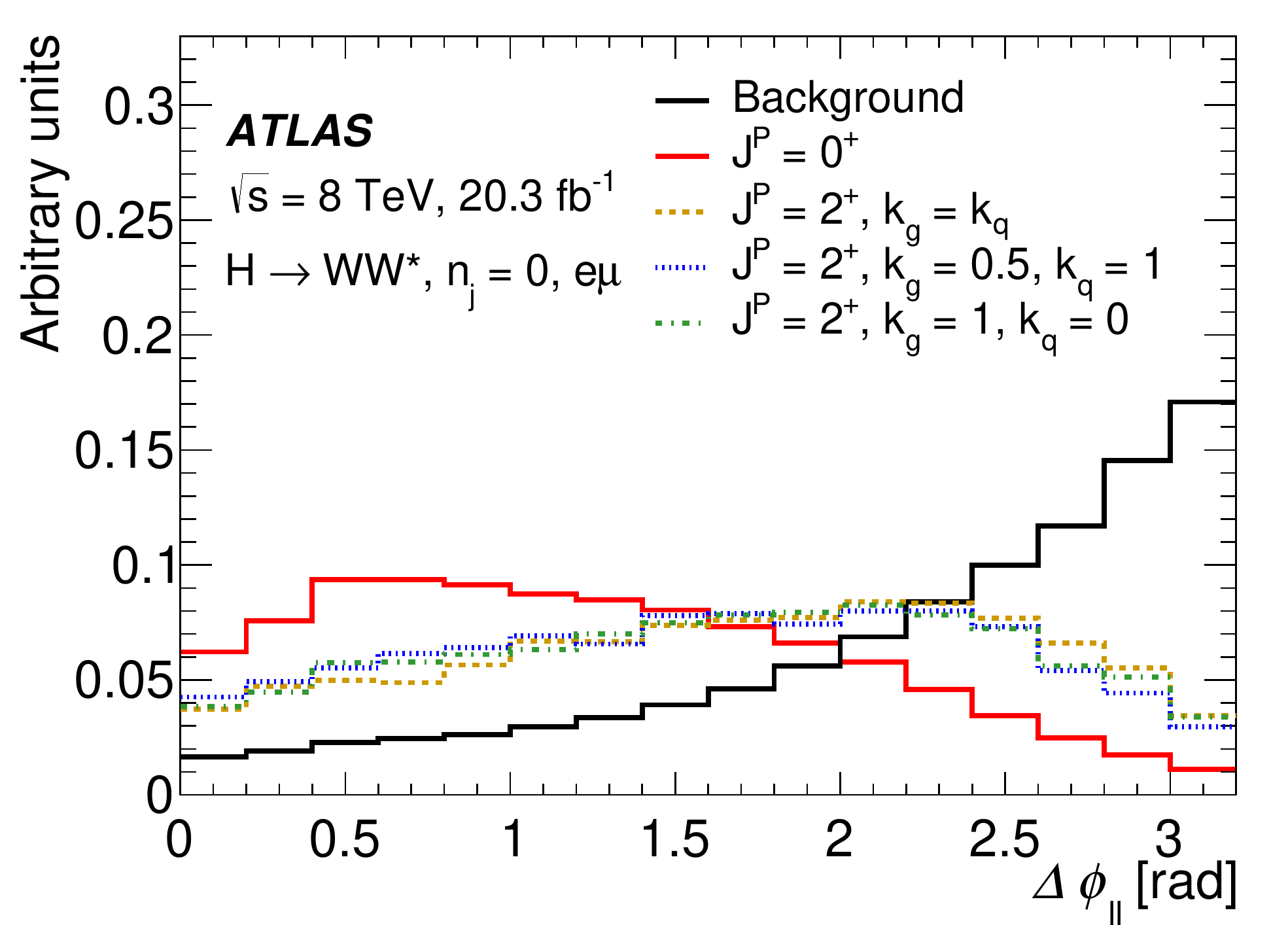}}
\subfloat{\includegraphics[width=0.49\textwidth]{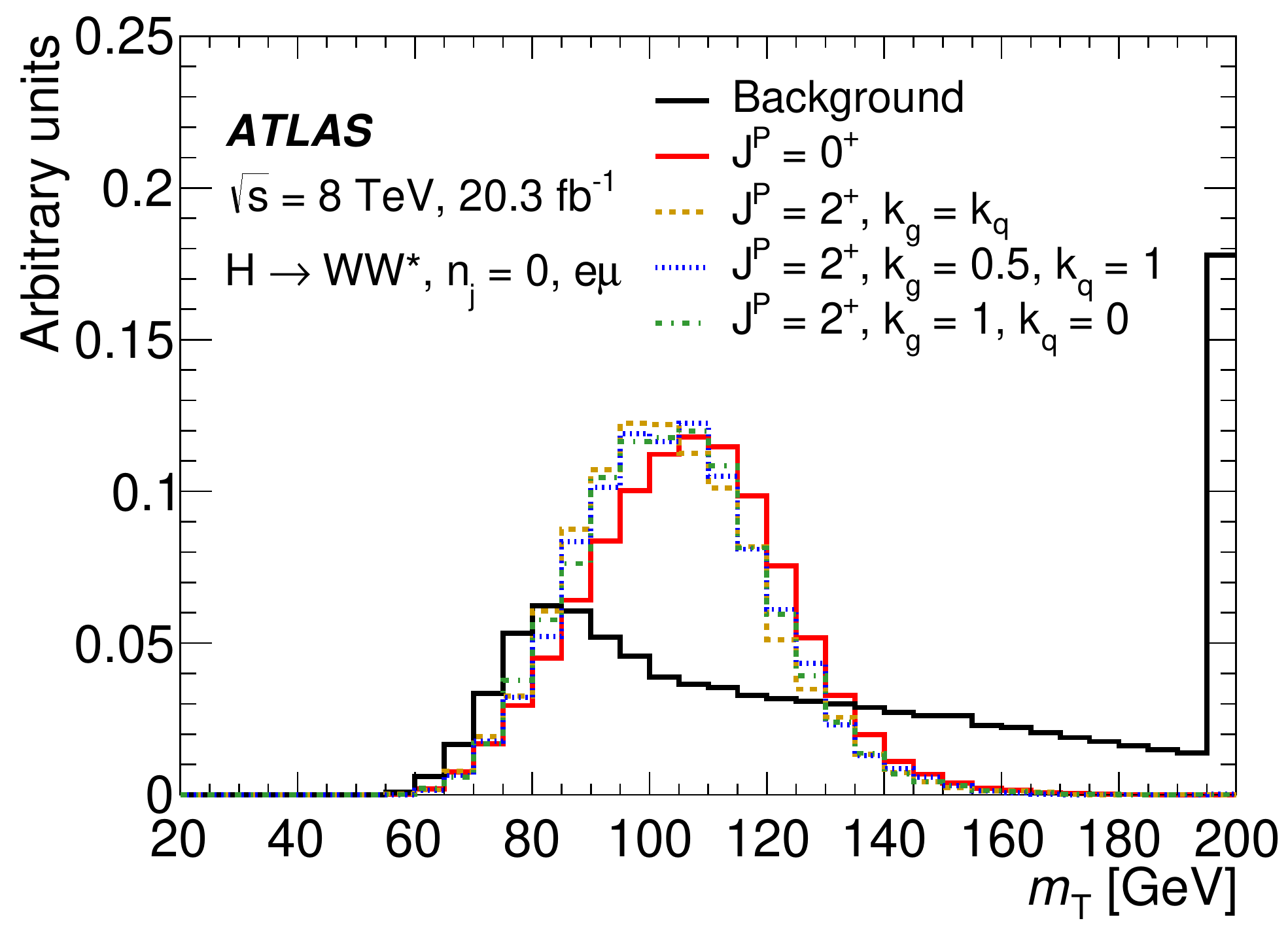}}
\caption{Expected normalised Higgs-boson distributions of  the transverse momentum of the dilepton system \ptll, the dilepton mass \mll, the azimuthal angular difference between the leptons \dphill\ and and 
the transverse mass \mT\ for the $e\mu$+0-jet category. The distributions are shown for the SM signal hypothesis (solid red line) and for three spin-2 hypotheses, namely $J^P = 2^+$, \kg\ = 0.5, \kq\ = 1 (dashed yellow line), $J^P = 2^+$, \kg\ = 1, \kq\ = 0 (blue dashed line) and $J^P = 2^+$, \kg\ = \kq\  (green dashed line). The expected shapes for the sum of all backgrounds, including the data-derived $W$+jets background, is also shown (solid black line). The last bin in each plot includes the overflow.}   
\label{fig:signal_cutvars_0jet} 
\end{figure}

\begin{figure}[h]
\centering{
\subfloat{\includegraphics[width=0.49\textwidth]{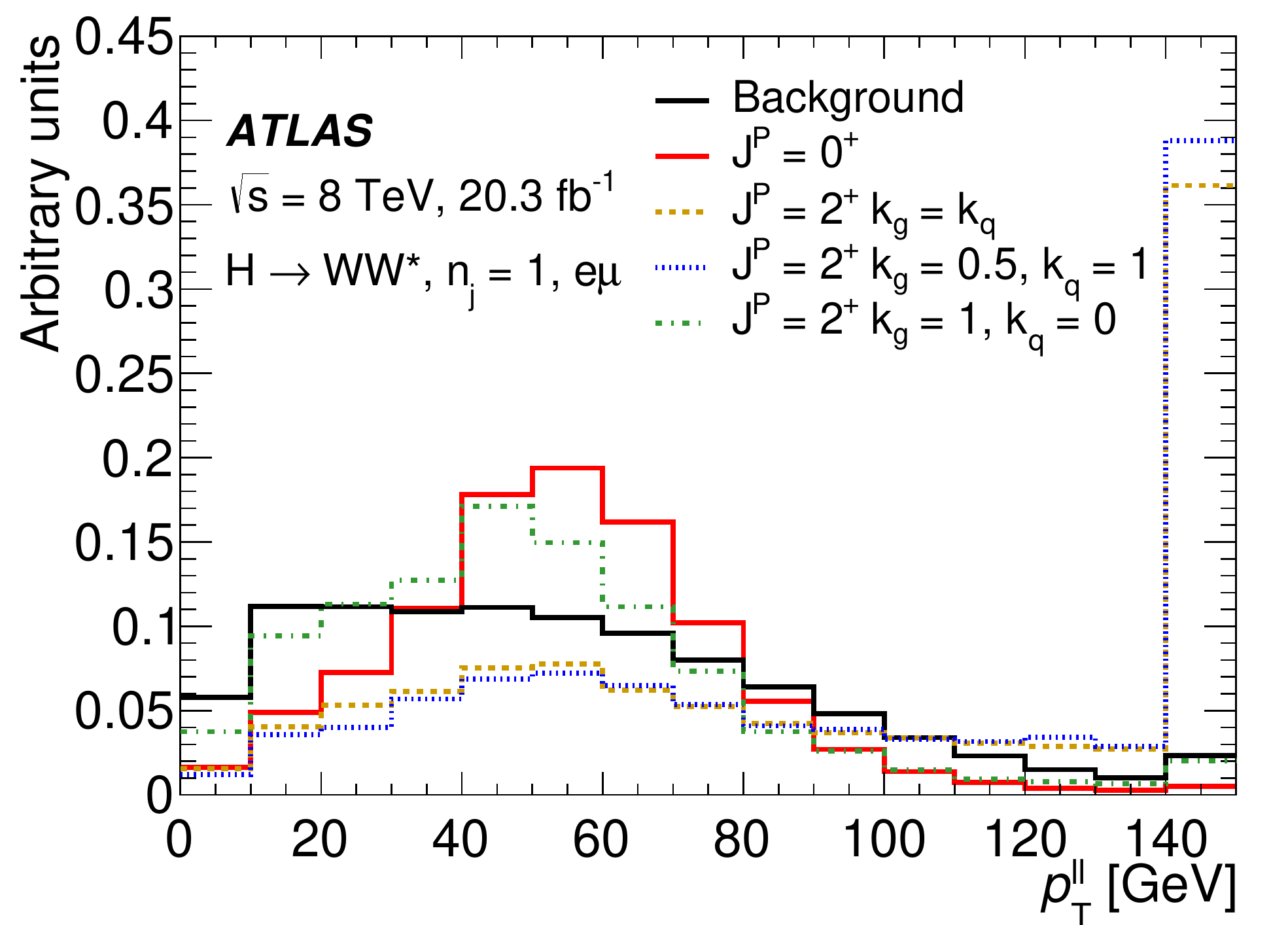}}}
\subfloat{\includegraphics[width=0.49\textwidth]{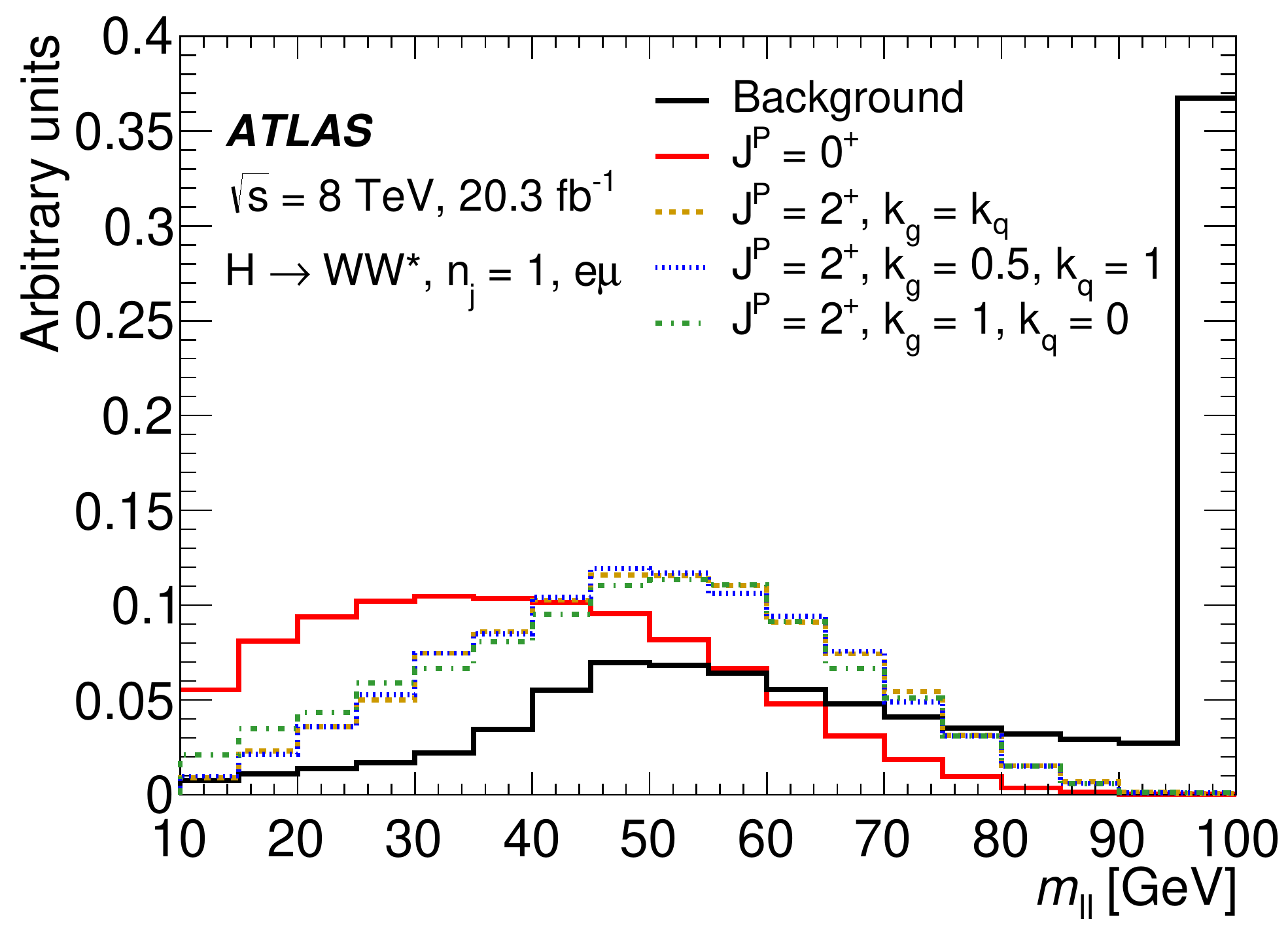}}\\
\subfloat{\includegraphics[width=0.49\textwidth]{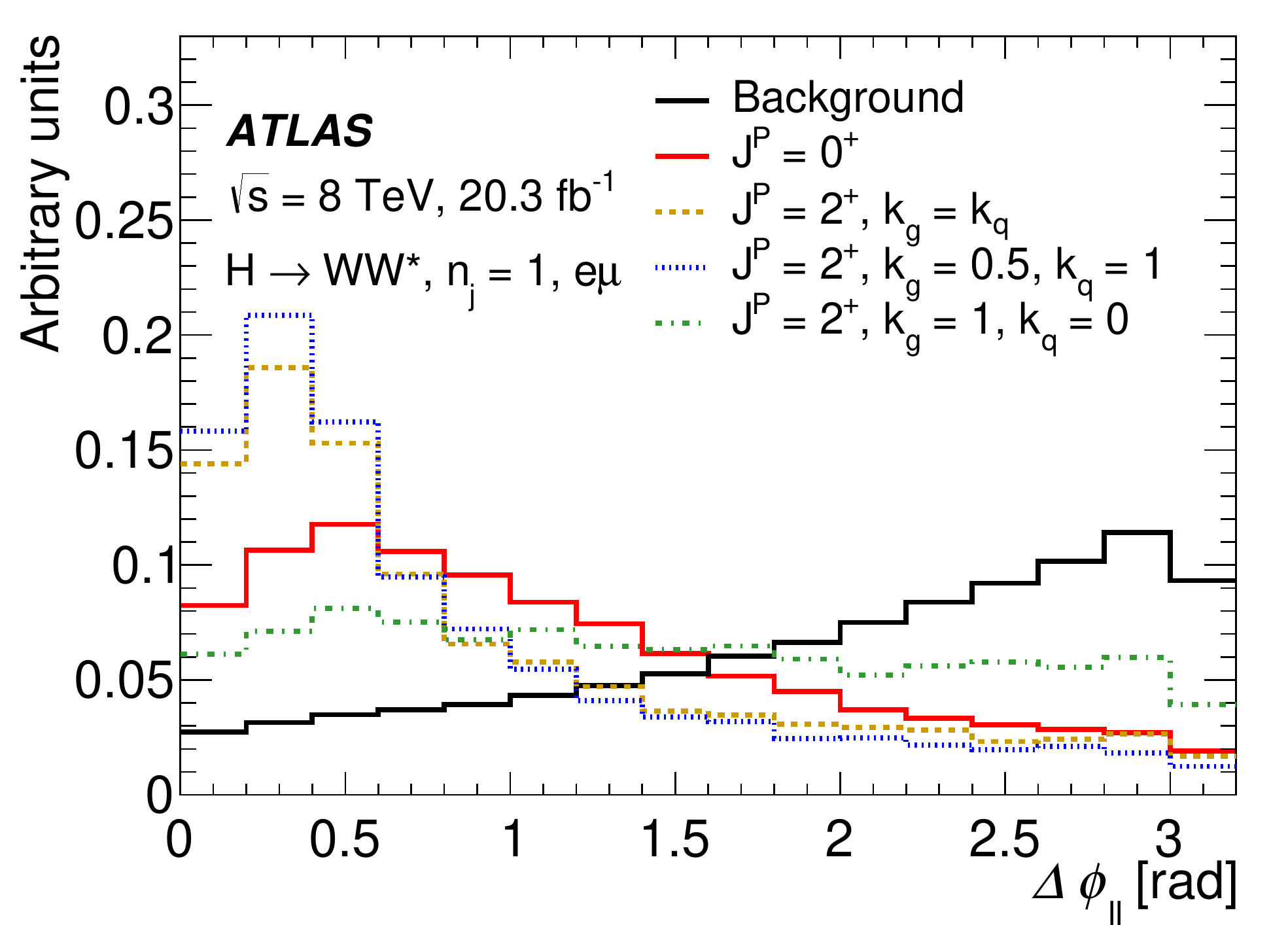}}
\subfloat{\includegraphics[width=0.49\textwidth]{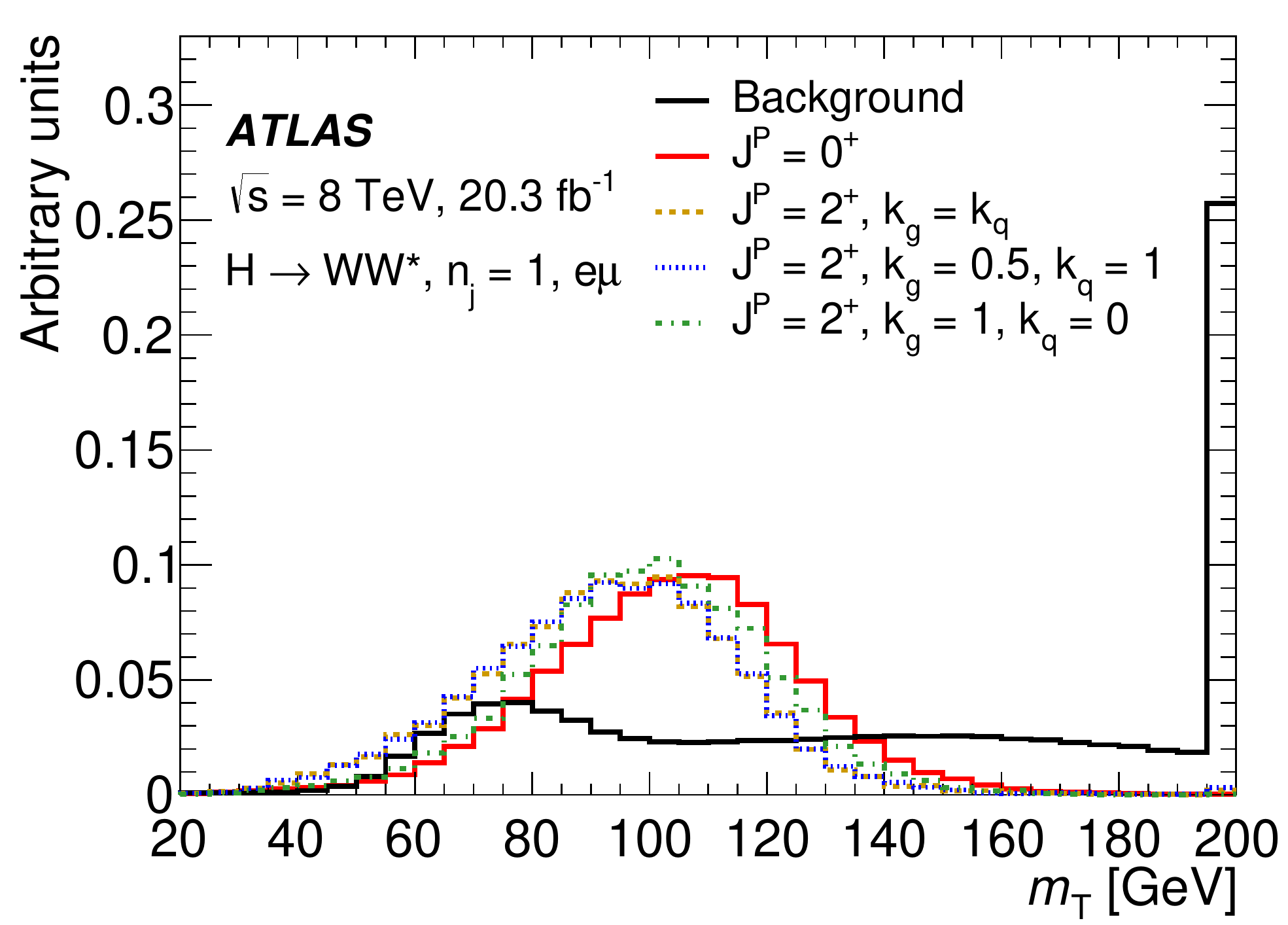}}
\caption{Expected normalised Higgs-boson distributions of  \ptll, \mll, \dphill\ and \mT\ for the $e\mu$+1-jet category. The distributions are shown for the SM signal hypothesis (solid red line) and for three spin-2 hypotheses, namely $J^P = 2^+$, \kg\ = 0.5, \kq\ = 1 (dashed yellow line), $J^P = 2^+$, \kg\ = 1, \kq\ = 0 (blue dashed line) and $J^P = 2^+$, \kg\ = \kq\  (green dashed line). The expected shapes for the sum of all backgrounds, including the data-derived $W$+jets background, is also shown (solid black line). The last bin in each plot includes the overflow.}
\label{fig:signal_cutvars_1jet}
\end{figure}

\begin{figure}[htb]
\begin{center}
\subfloat{\includegraphics[width=0.49\textwidth]{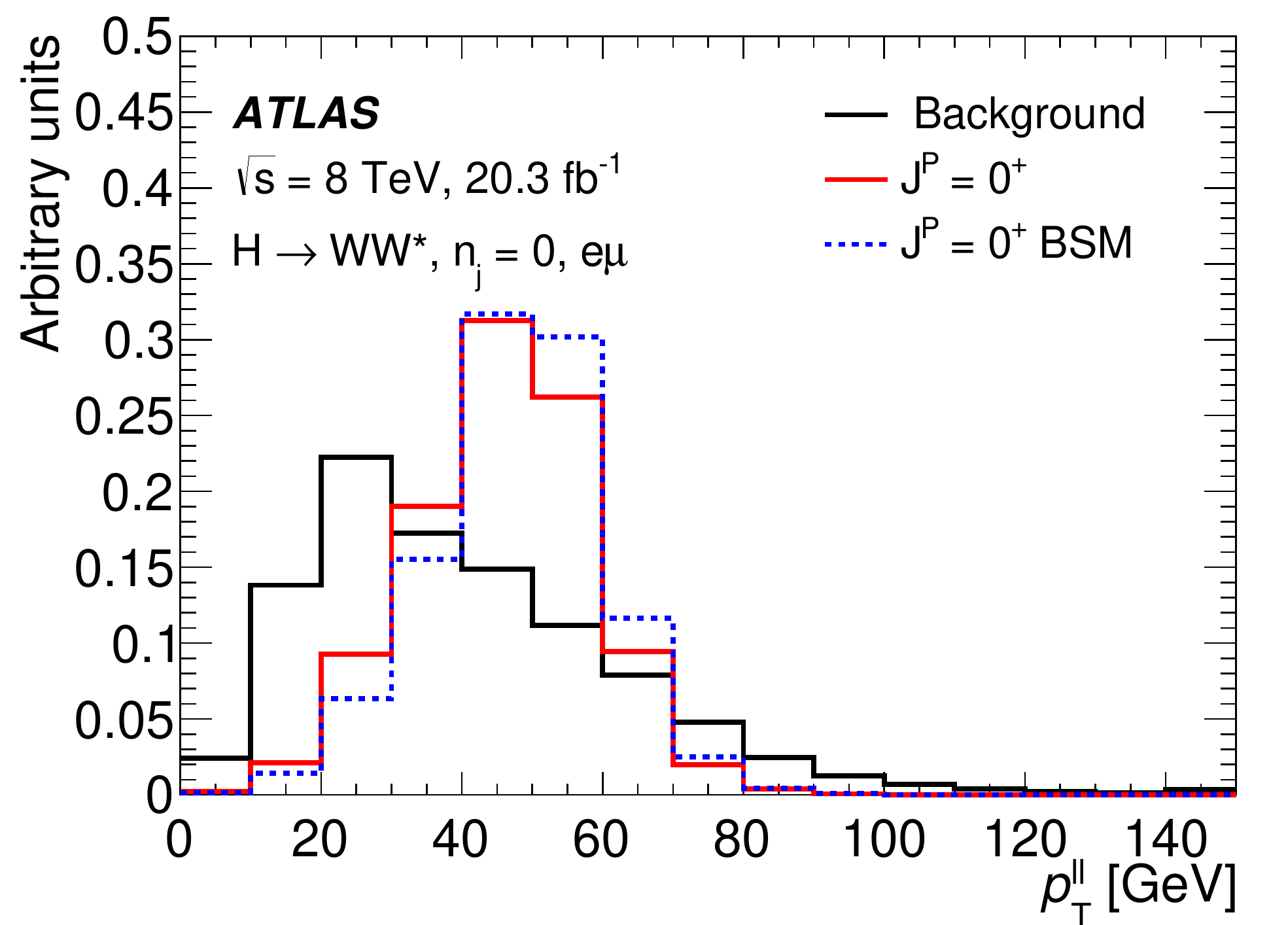}}
\subfloat{\includegraphics[width=0.49\textwidth]{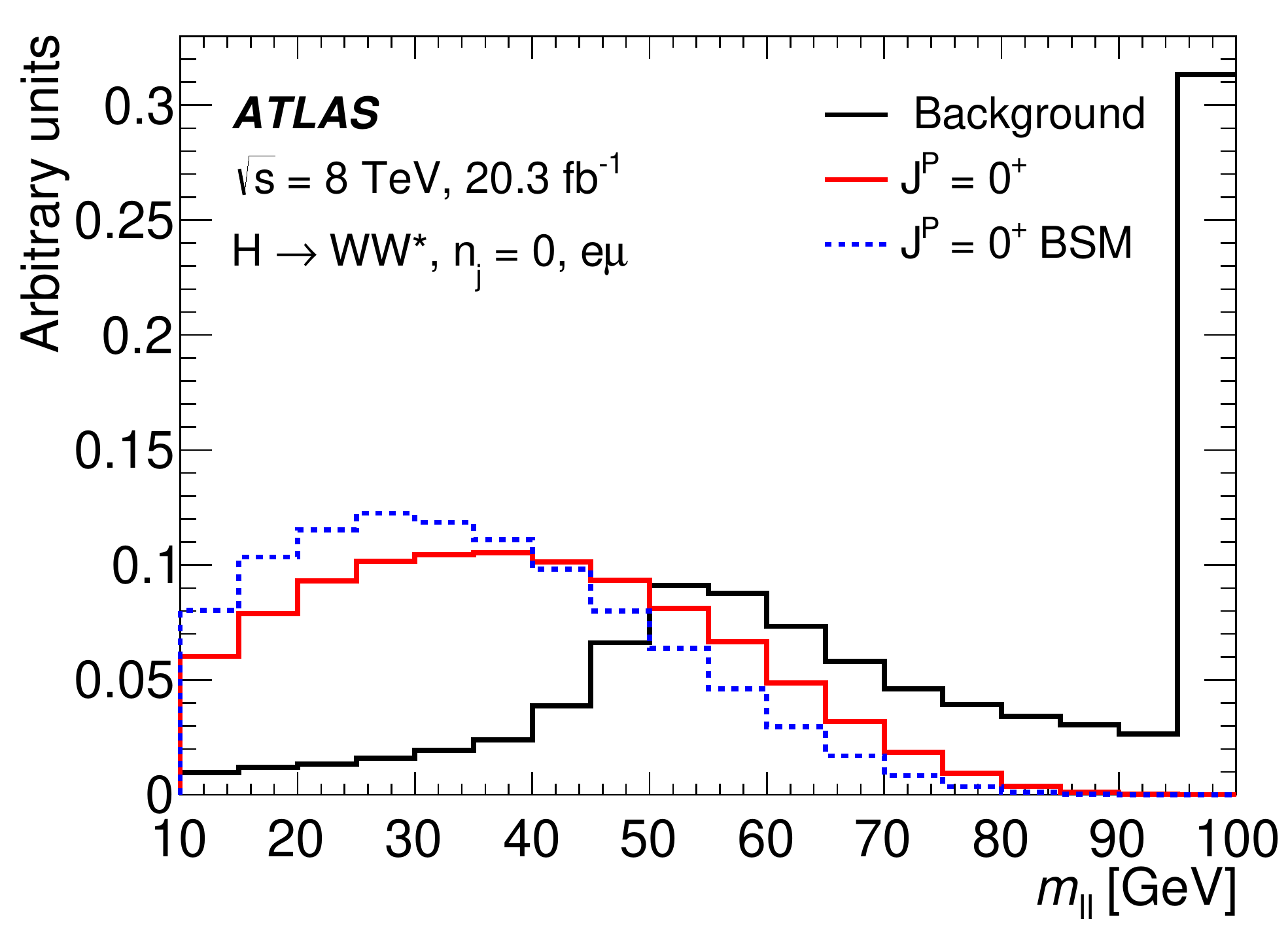}}\\
\subfloat{\includegraphics[width=0.49\textwidth]{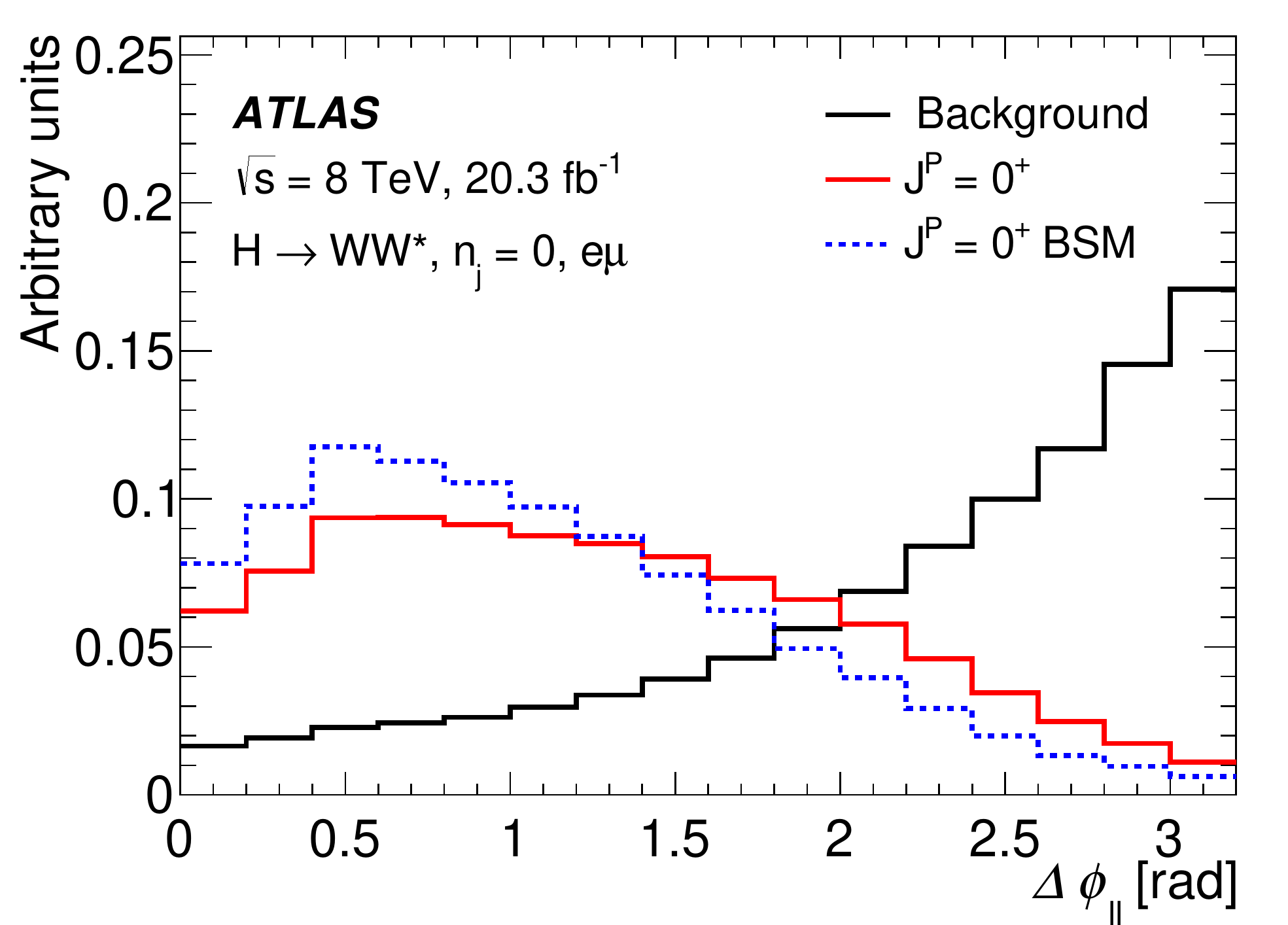}}
\subfloat{\includegraphics[width=0.49\textwidth]{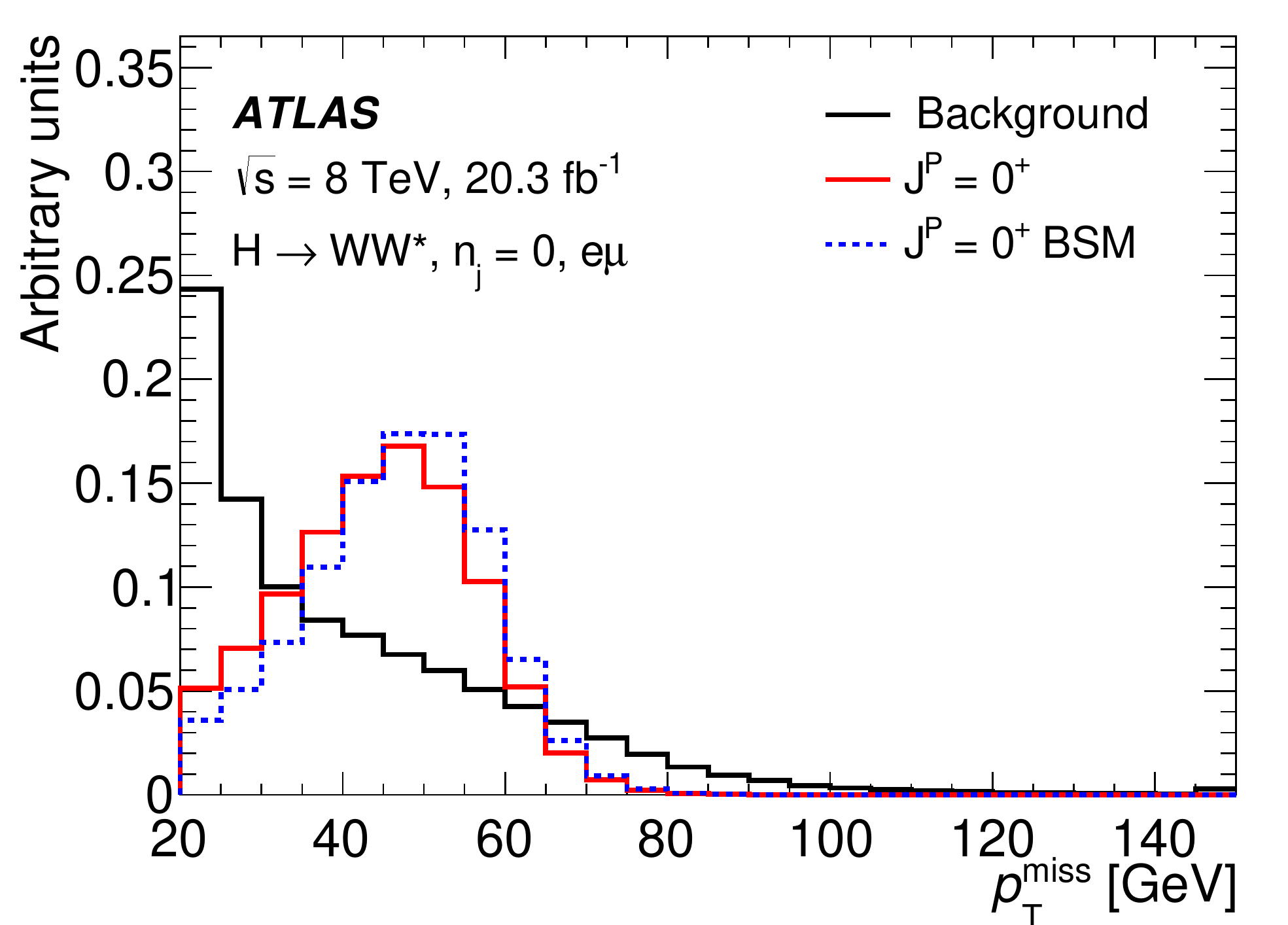}}

\caption{Expected normalised Higgs-boson distributions of  \ptll, \mll, \dphill\ and the missing transverse momentum \ptmiss\  for the $e\mu$+0-jet category. The distributions are shown for the SM signal hypothesis (solid red line) and for the BSM CP-even signal (dashed line). The expected shapes for the sum of all backgrounds, including the data-derived $W$+jets background, is also shown (solid black line). The last bin in each plot includes the overflow.}
\label{fig:inputvars_shapes_training_CP_even}
\end{center}
\end{figure}

\begin{figure}[htb]
\begin{center}
\subfloat{\includegraphics[width=0.49\textwidth]{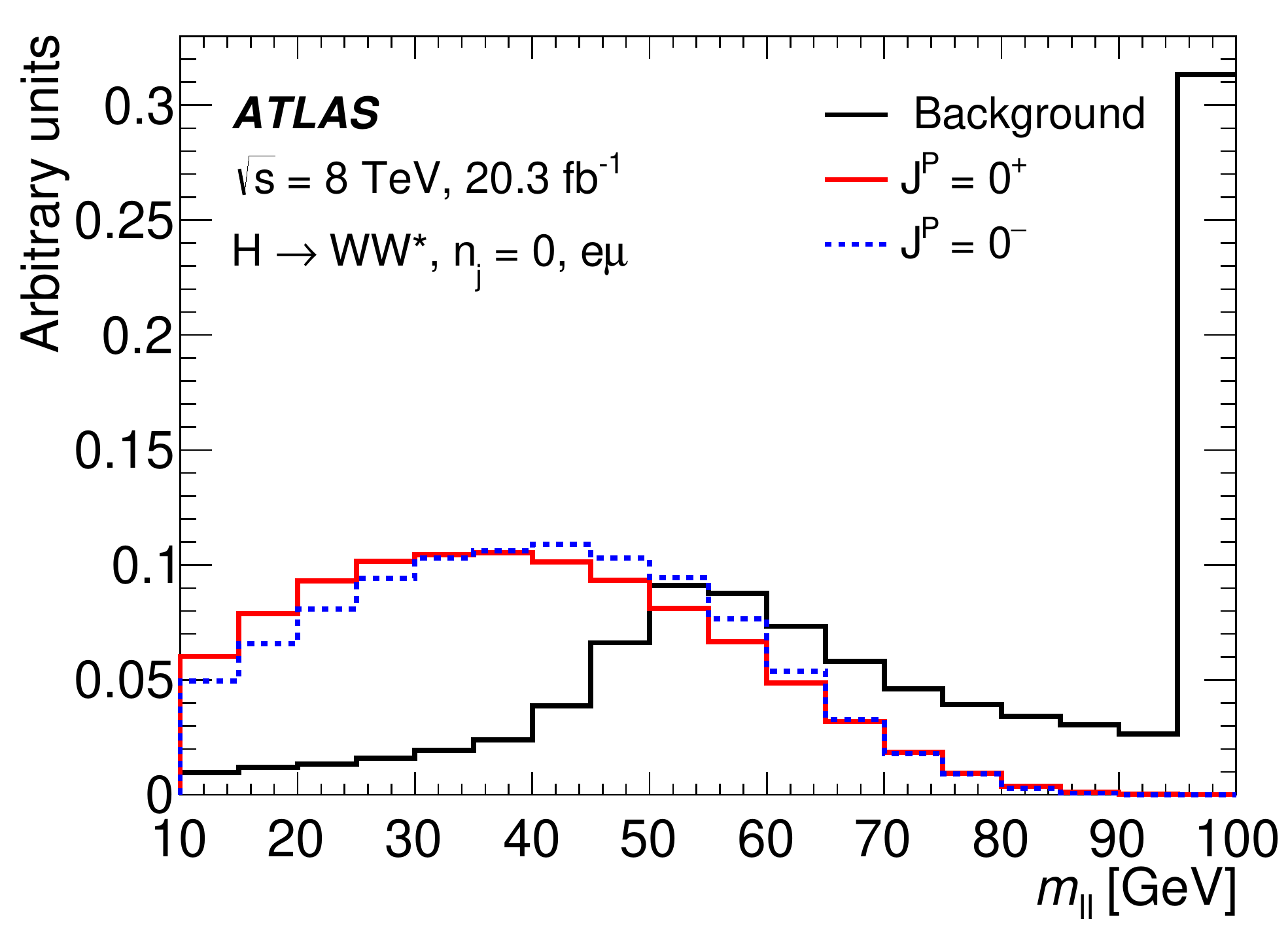}}
\subfloat{\includegraphics[width=0.49\textwidth]{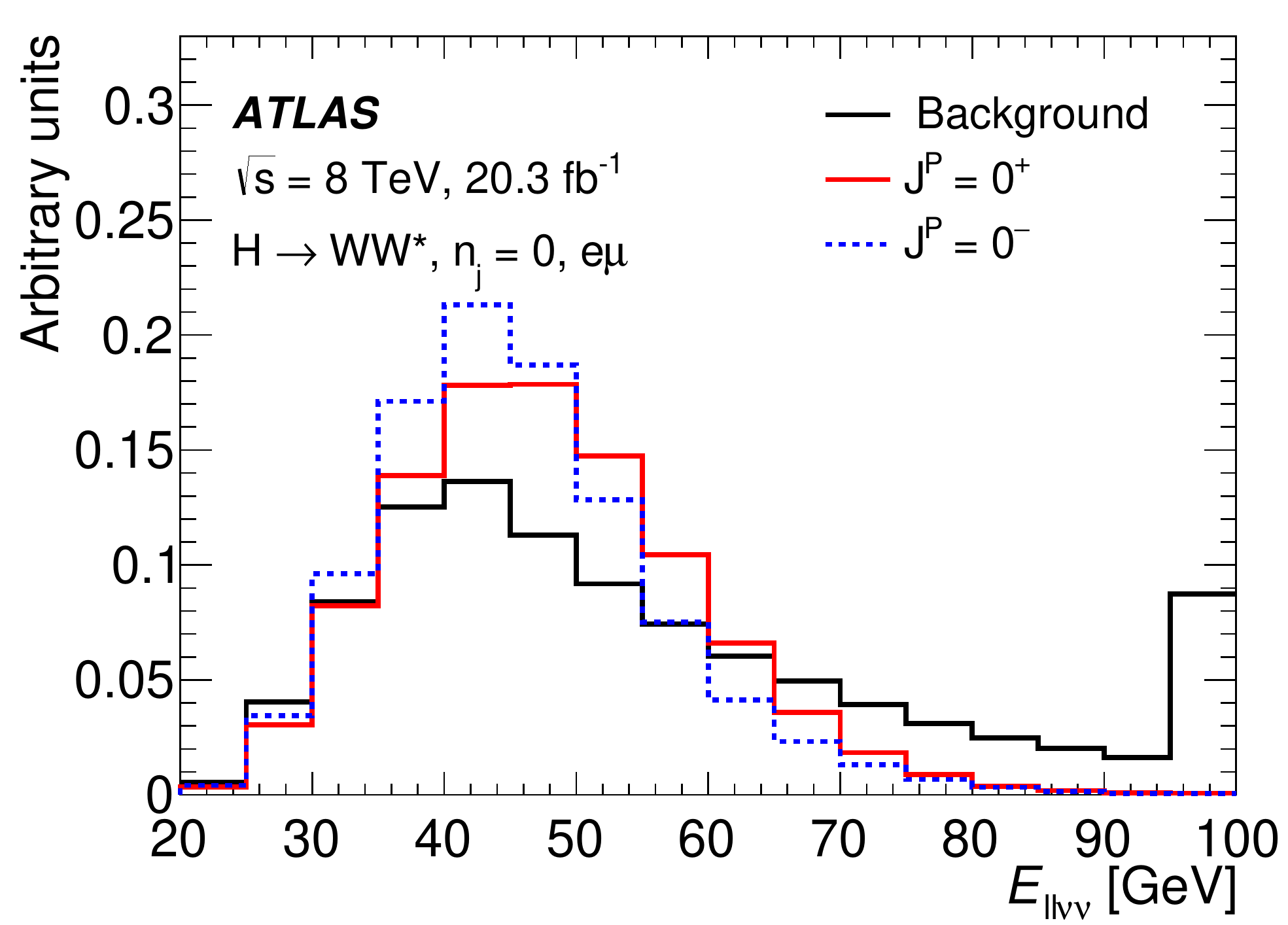}}\\
\subfloat{\includegraphics[width=0.49\textwidth]{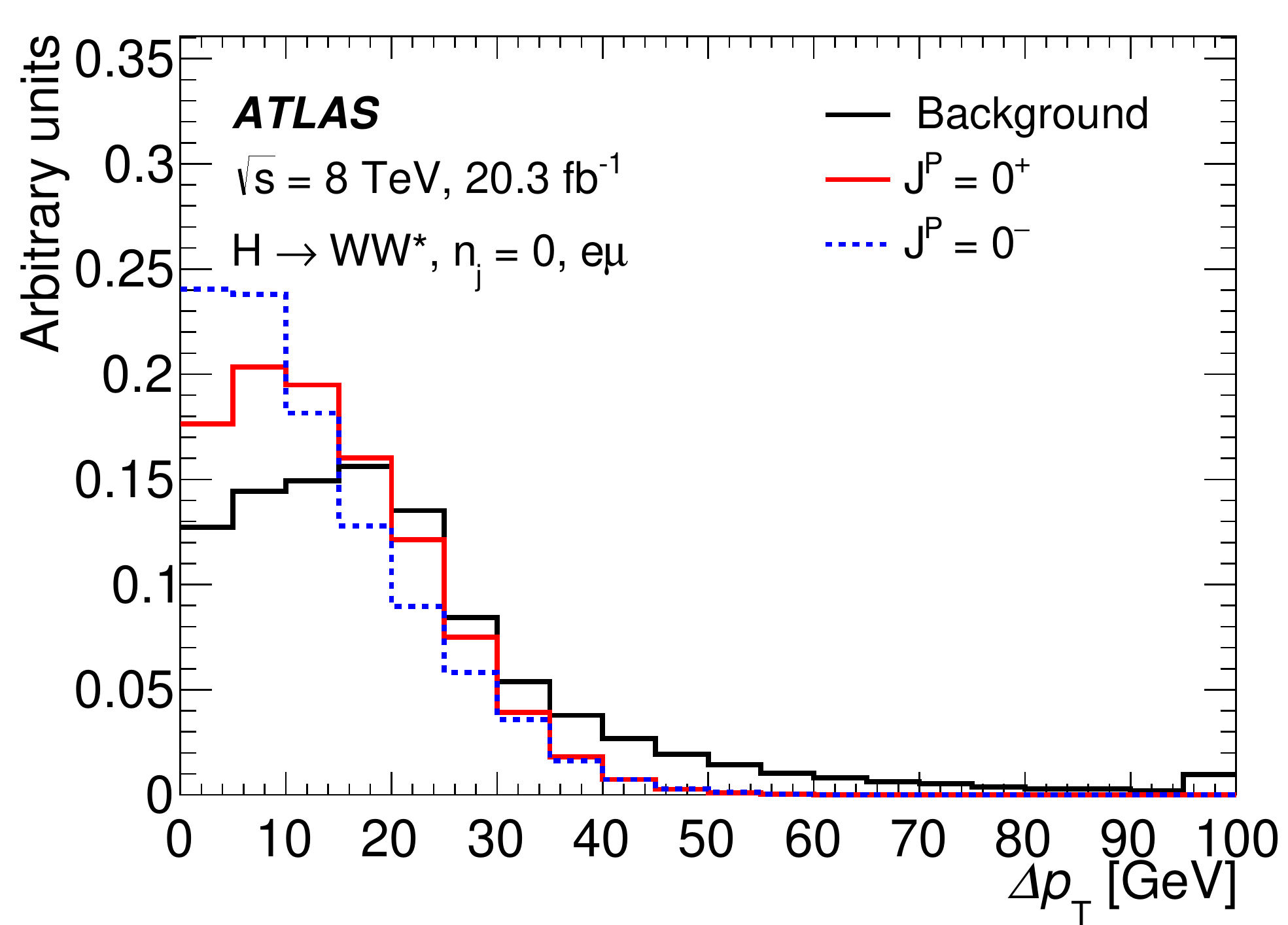}}
\subfloat{\includegraphics[width=0.49\textwidth]{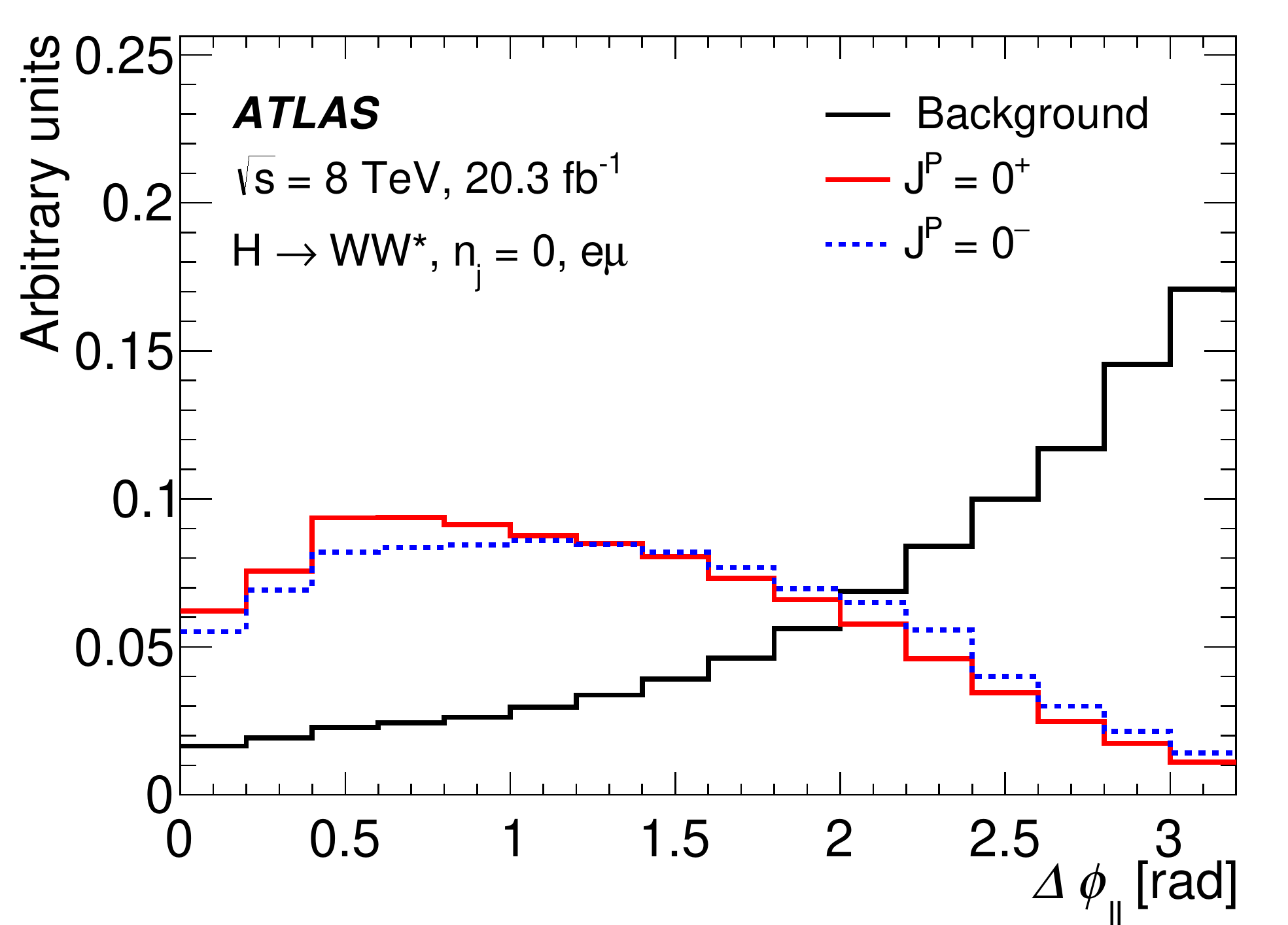}}
\caption{Expected normalised Higgs-boson distributions of  \mll, the \Efun\ variable defined in Sect.~\ref{sec:variables}, the difference between the transverse momenta of the leading and subleading leptons \dpt\ and \dphill\   for the $e\mu$+0-jet category. The distributions are shown for the SM signal hypothesis (solid red line) and for the BSM CP-odd signal (dashed line). The expected shapes for the sum of all backgrounds, including the data-derived $W$+jets background, is also shown (solid black line). The last bin in each plot includes the overflow.}
\label{fig:inputvars_shapes_training_CP_odd}
\end{center}
\end{figure}

\begin{figure}[h]
\centering
\subfloat{\includegraphics[width=0.49\textwidth]{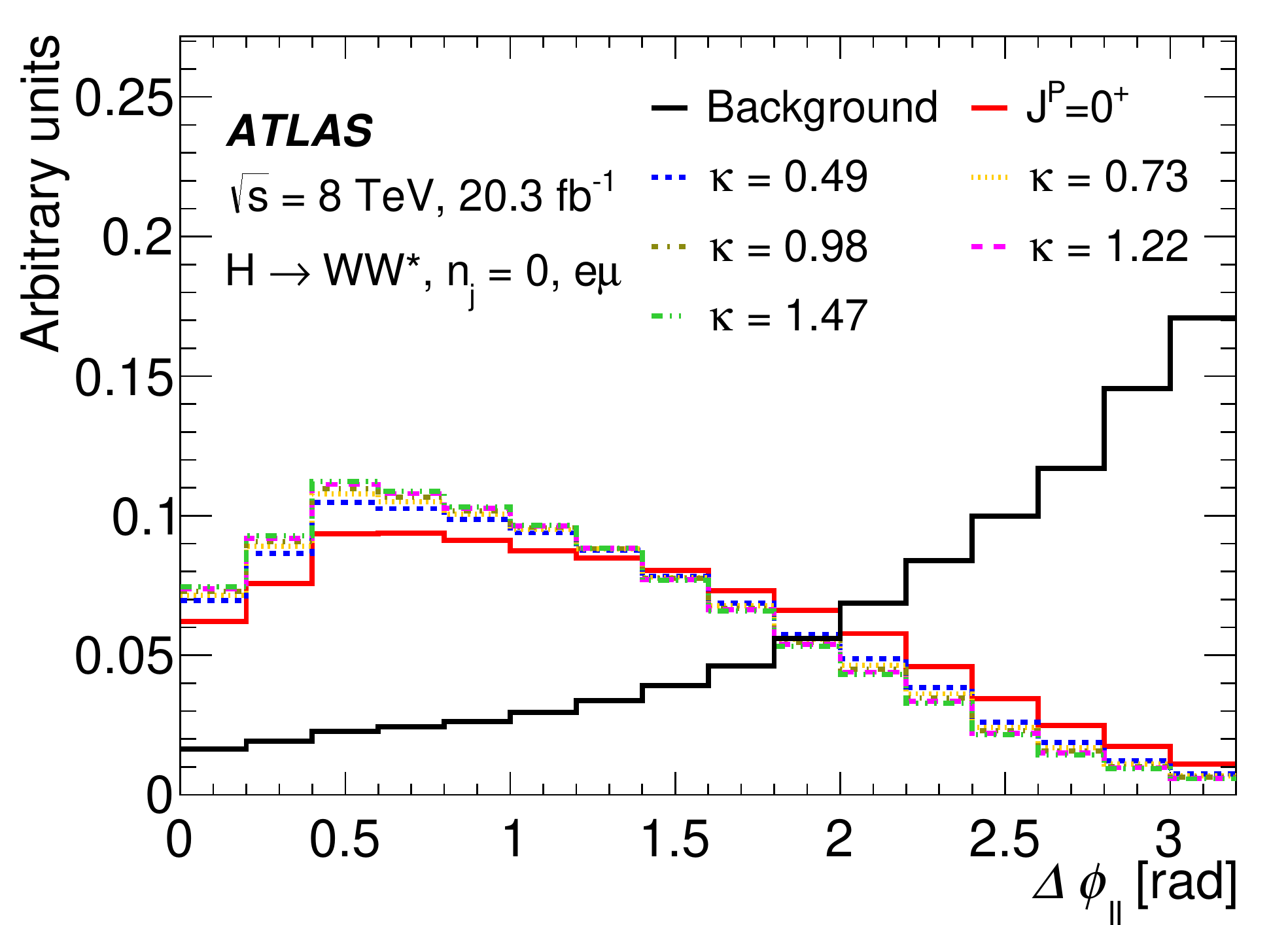}}
\subfloat{\includegraphics[width=0.49\textwidth]{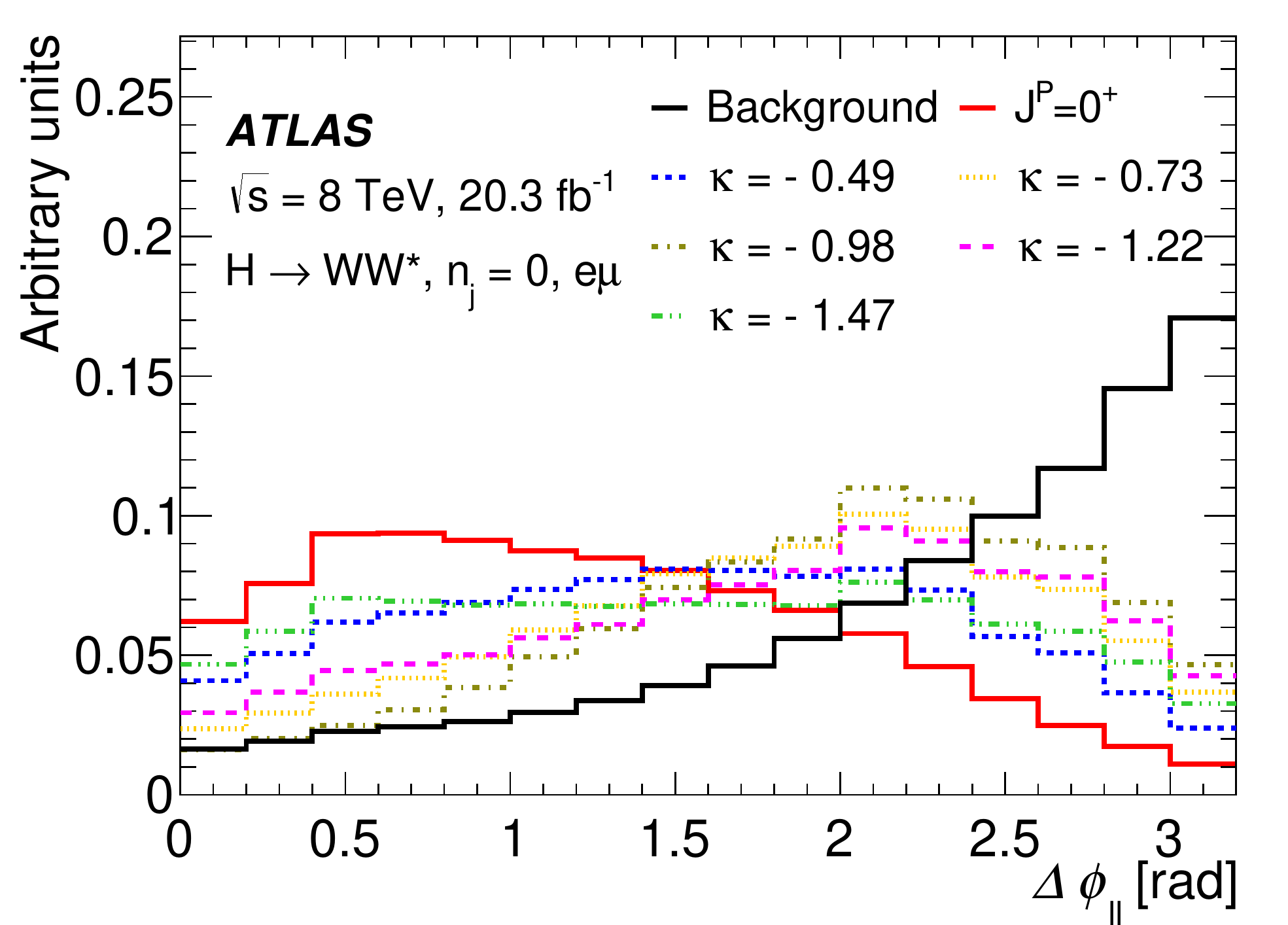}}
\caption{ Expected normalised Higgs-boson distributions of \dphill\ for the $e\mu$+0-jet category. The distributions are shown for the SM signal hypothesis (solid red line) and for different mixing hypotheses of the SM Higgs and CP-even BSM Higgs bosons, corresponding to positive (left) and negative (right) values of the mixing parameter $\tilde{\kappa}_{HWW}/{\kappa}_{\rm SM}$ (abbreviated to $\kappa$ in the legend). The expected shapes for the sum of all backgrounds, including the data-derived $W$+jets background, is also shown (solid black line). The last bin in each plot includes the overflow.}   
\label{fig:DPhineg} 
\end{figure}

\subsection{Event selection in the 0-jet and 1-jet categories}
\label{sec:selection}

Table~\ref{tab:comparecuts} summarises the preselection requirements discussed in Sect.~\ref{sec:preselection}, together with the selections applied specifically to the 0-jet and 1-jet categories. These selection requirements are optimised in terms of sensitivity for the different spin and CP hypotheses studied while maintaining the required rejection against the dominant backgrounds. In general, they are looser than those described in Ref.~\cite{ATLAS-CONF-2014-060}, which were optimised for the SM Higgs boson.

\begin{table}[htb]
\caption{\label{tab:comparecuts} List of selection requirements in the signal region adopted for both the
spin and CP analyses. ~The \pTH\ selection requirement~(*) is applied to all samples when testing 
the spin-2 benchmarks with non-universal couplings. } 
\begin{center}
  \begin{tabular}{c|c}
   \dbline
    Variable & Requirements  \\
   \sgline
    \multicolumn{2}{c}{Preselection}\\
    \sgline
    $N_{\mathrm{leptons}}$ & Exactly 2 with $\pt > 10 \GeV$, $e\mu$, opposite sign \\
    $p_{\text T}^{\ell_1}$  & $> 22\GeV$ \\
     $p_{\text T}^{\ell_2}$  & $> 15\GeV$ \\
     $\mll$ & $> 10 \GeV$ \\
    \ptmiss  & $>20 \GeV$ \\
    \sgline
    \multicolumn{2}{c}{0-jet selection}\\
    \sgline
    \ptll     & $>20 \GeV$ \\
    \mll      & $<80 \GeV$ \\
    \dphill   & $<2.8$    \\
    \pTH     & $< 125$ or $300\GeV$ (*) \\
     \sgline
    \multicolumn{2}{c}{1-jet selection}\\
    \sgline   
    $b-$veto & No $b$-jets with $\pt > 20 \GeV$ \\
    $m_{\tau\tau}$ & $< m_Z - 25\GeV$ \\
    $m_{\rm T}^{\ell}$ & $> 50\GeV$ \\
    \mll      & $<80 \GeV$ \\
    \dphill   & $<2.8$    \\
    \mT   & $<150\GeV$    \\
        \pTH     & $< 125$ or $300\GeV$ (*) \\
       \dbline
\end{tabular}

\end{center}
\end{table}

Some of these looser selection requirements are applied to both the 0-jet and 1-jet categories:
\begin{itemize}
\item The mass of the lepton pair, \mll, must satisfy $\mll\ < 80$~\GeV, a selection which strongly reduces the dominant WW continuum background.
\item The azimuthal angle, \dphill, between the two leptons, must satisfy $\dphill < 2.8$.
\end{itemize}

Events in the 0-jet category are required to also satisfy $\ptll > 20$~\GeV, while events in the 1-jet category, which suffer potentially from a much larger background from top-quark production, must also satisfy the following requirements:
\begin{itemize}
\item No $b$-tagged jet~\cite{ATLAS-CONF-2014-046} $\pt > 20$~\GeV\ is present in the event.

\item Using the direction of the missing transverse momentum a $\tau$-lepton pair can be reconstructed with a mass $m_{\tau \tau}$ by applying the collinear approximation~\cite{Ellis:1987xu}; $m_{\tau \tau}$ is required to pass the $m_{\tau \tau} < m_Z - 25$~\GeV\ requirement to reject $\Ztt$ events.
\item The transverse mass, $m^{\ell}_{\rm T}$, chosen to be the largest transverse mass of single leptons defined as $m^{\ell_{\text i}}_{\rm T} = \sqrt{2p_{\text T}^{\ell_{\text i}} \ptmiss (1-\cos\Delta\phi)}$, where $\Delta\phi$ is the angle between the lepton transverse momentum and \ptmiss, is required to satisfy $m^{\ell}_{\rm T} > 50$~\GeV\ to reject the \Wjets\ background.
\item The total transverse mass of the dilepton and missing transverse momentum system, \mT, is required to satisfy $\mT < 150$~\GeV.
\end{itemize}

For alternative spin-2 benchmarks with non-universal couplings, as listed in Sect.~\ref{sec:choicespin2}, an additional requirement on the reconstructed 
Higgs-boson transverse momentum \pTH\ is applied in the signal and control regions for all MC samples and data. The \pTH\ variable is 
reconstructed as the transverse component of the vector sum of the four-momenta of both leptons and the missing transverse energy.

Table~\ref{tab:Cutflow} shows the number of events for data, expected SM signal and the various background components after event selection. The background estimation methods are described in detail in Sect.~\ref{sec:backgrounds}. Good agreement is seen between the observed numbers of events in each of the two categories and the sum of the total background and the expected signal from an SM Higgs boson. The 0-jet category is the most sensitive one with almost three times larger yields than the 1-jet category. As expected, however, the requirements on \pTH\ affect mostly the 1-jet category, which is sensitive to possible tails at high values of \pTH, as explained in Sect.~\ref{sec:choicespin2}. Figures~\ref{fig:vars_unblind_sr0jet} and~\ref{fig:vars_unblind_sr1jet} show the distributions of discriminating variables used in the analysis after the full selection for the 0-jet and 1-jet categories, respectively. These figures show reasonable agreement between the data and the sum of all expected contributions, including that from the SM Higgs boson.

\begin{table}[ht]
\begin{center}
 {
    \caption{\label{tab:Cutflow}
    Expected event yields in the signal regions (SR) for the 0- and 1-jet categories (labelled as 0j and 1j, respectively). For the dominant backgrounds, the expected yields are normalised using the control regions defined in Sect.~\ref{sec:backgrounds}.  The expected contributions from various processes are listed, namely the ggF SM Higgs-boson production ($N_{\rm ggF}$), and the background contribution from $WW$ ($N_{WW}$), top quark (top-quark pairs $N_{t\bar{t}}$, and single-top quark $N_t$), Drell--Yan $Z/\gamma^*$ to $\tau\tau$ ($N_{\rm DY,\tau\tau}$), misidentified leptons ($N_{W+\rm jets}$), $WZ/ZZ/W\gamma$ ($N_{\mathrm{VV}}$) and Drell--Yan $Z/\gamma^*$ to $ee/\mu\mu$ ($N_{\rm DY,SF}$).
The total sum of the backgrounds ($N_{\rm bkg}$) is also shown together with the data. Applying the \pTH\ requirement in the 0-jet category does not change substantially the event yields, while it has an effect in the 1-jet category, as expected. The errors on the ratios of the data over total background, $N_{\rm bkg}$, only take into account the statistical uncertainties on the observed and expected yields.
    }
  }
\scalebox{0.94}{
{\small  
\begin{tabular}{l|c|cccccccc|cc}
\dbline
 & $N_{\rm ggF}$ & $N_{WW}$  & $N_{t\bar{t}}$ & $N_{t}$ & $N_{\rm DY,\tau\tau}$ & $N_{W+\rm jets}$ & $N_{\mathrm{VV}}$ & $N_{\rm DY,SF}$ & $N_{\rm bkg}$ & Data &  Data/$N_{\rm bkg}$  \\
\sgline
0j SR                                     & 218  & 2796    & 235  & 135   & 515  & 366  & 311 & 32 & 4390  & 4730 & 1.08 $\pm$ 0.02 \\

\sgline
1j SR:                                    & 77  & 555    & 267  & 103  & 228  & 123  & 131 & 5.8  & 1413 & 1569  & 1.11 $\pm$ 0.03 \\
1j SR: $\pTH < 300$~\GeV & 77  & 553    & 267  & 103  & 228  & 123  & 131 & 5.8  & 1411  & 1567 & 1.11 $\pm$ 0.03 \\
1j SR: $\pTH < 125$~\GeV & 76  & 530    & 259  & 101  & 224  & 121  & 128 & 5.8  & 1367  & 1511 & 1.11 $\pm$ 0.03 \\
\dbline
\end{tabular}
 }}
\end{center}
\end{table}

\begin{figure}
  \centering
\subfloat{\includegraphics[width=0.43\textwidth]{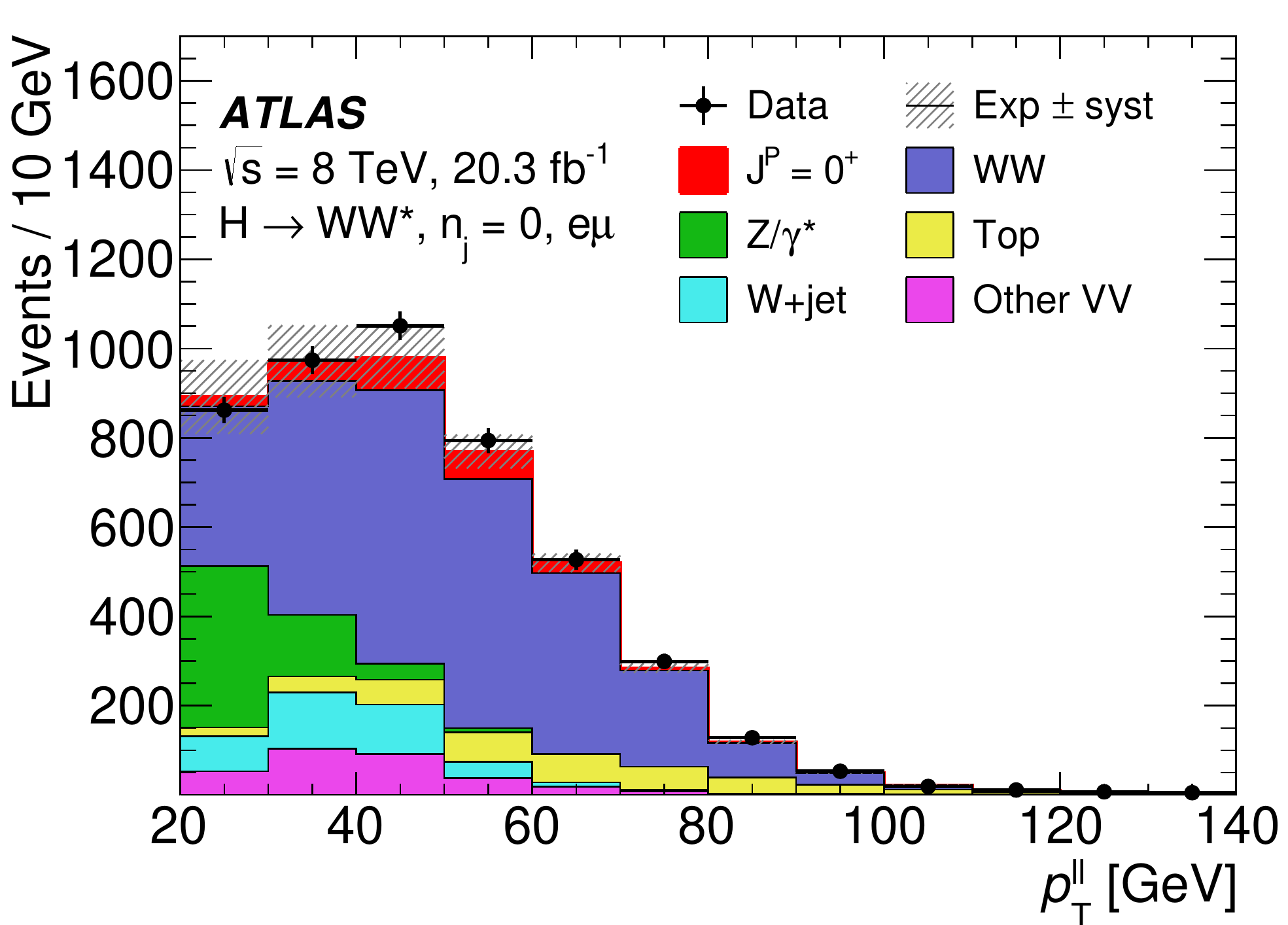}}
\subfloat{\includegraphics[width=0.43\textwidth]{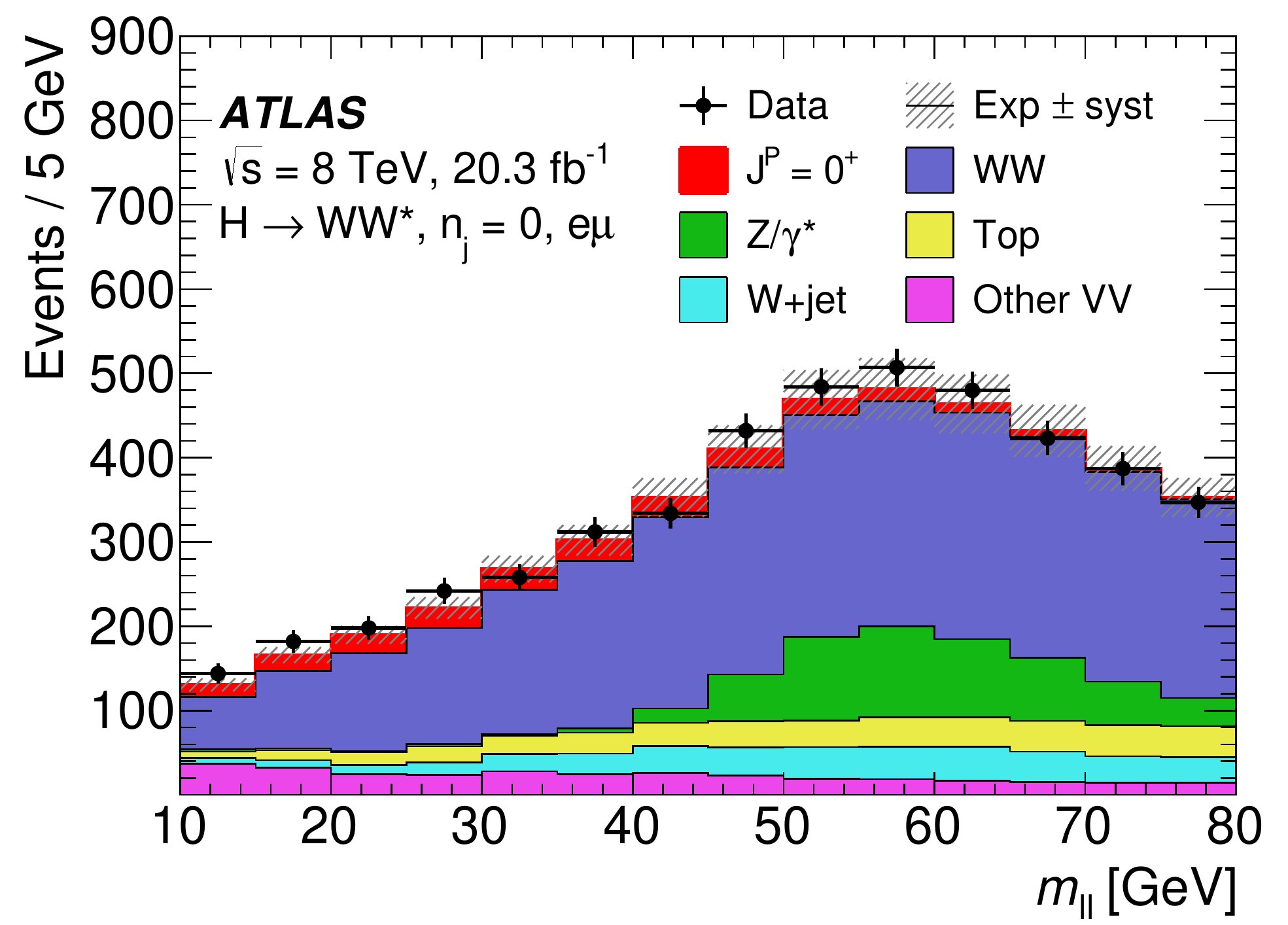}} \\
\subfloat{\includegraphics[width=0.43\textwidth]{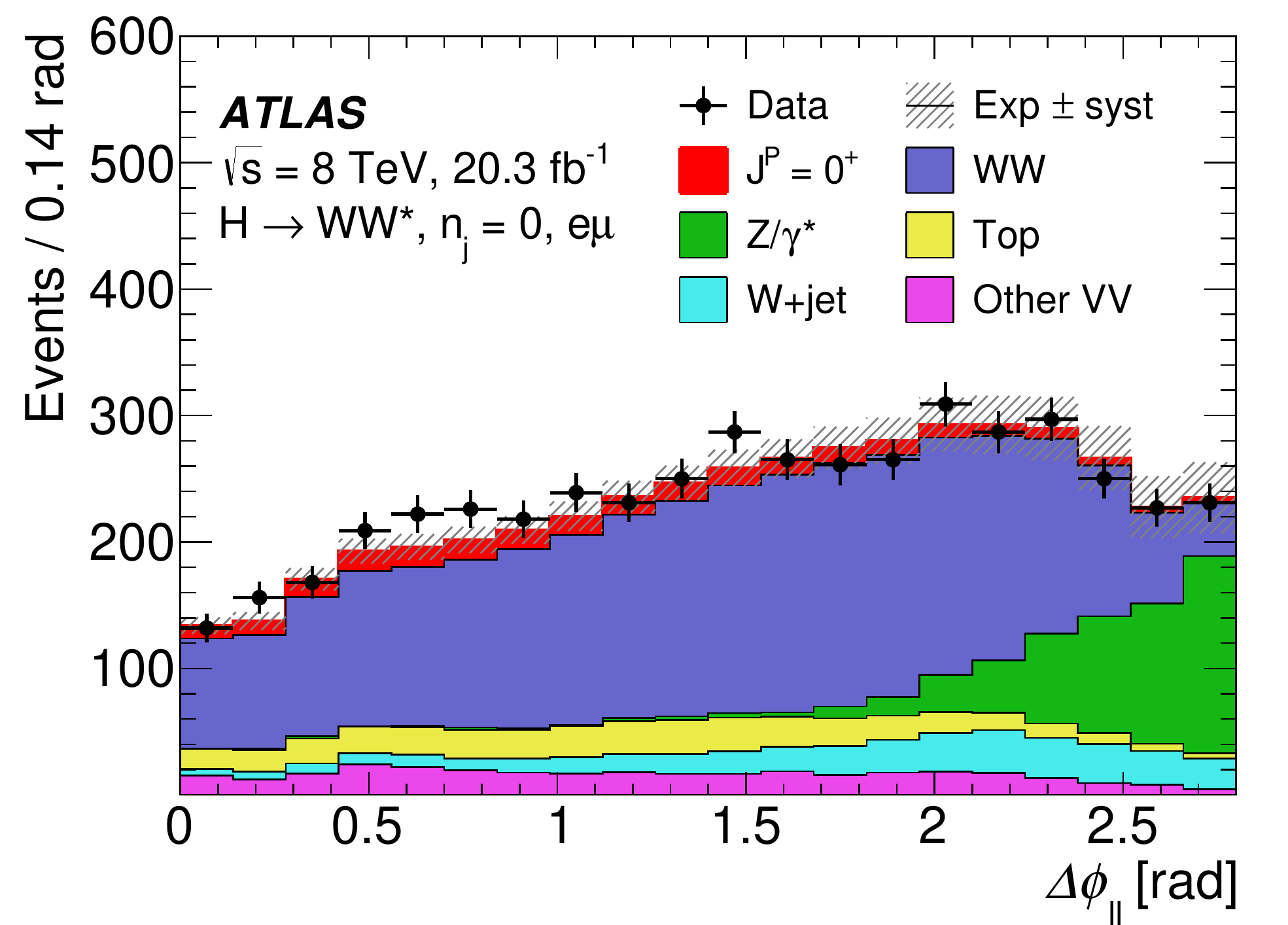}} 
\subfloat{\includegraphics[width=0.43\textwidth]{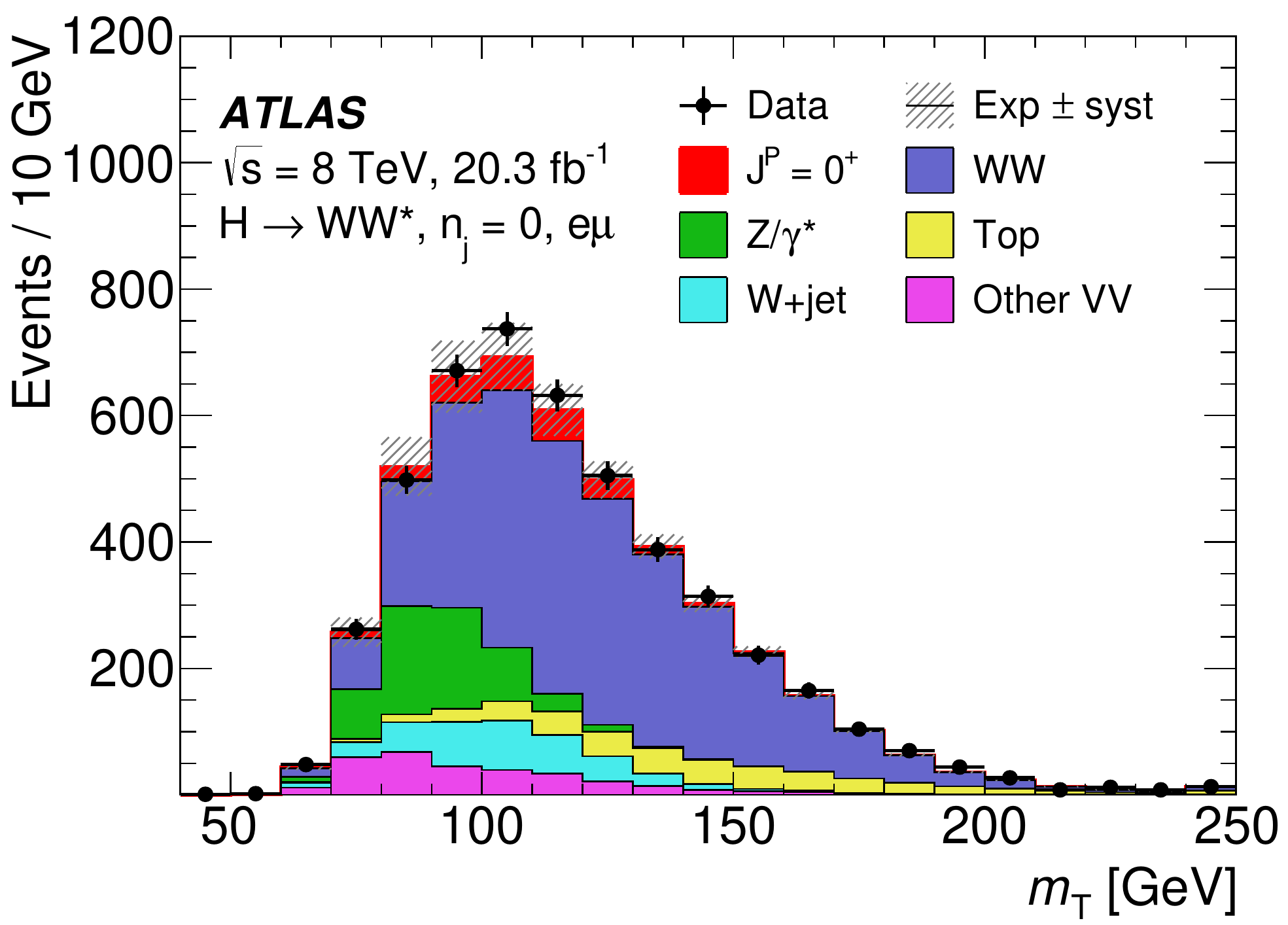}} \\
\subfloat{\includegraphics[width=0.43\textwidth]{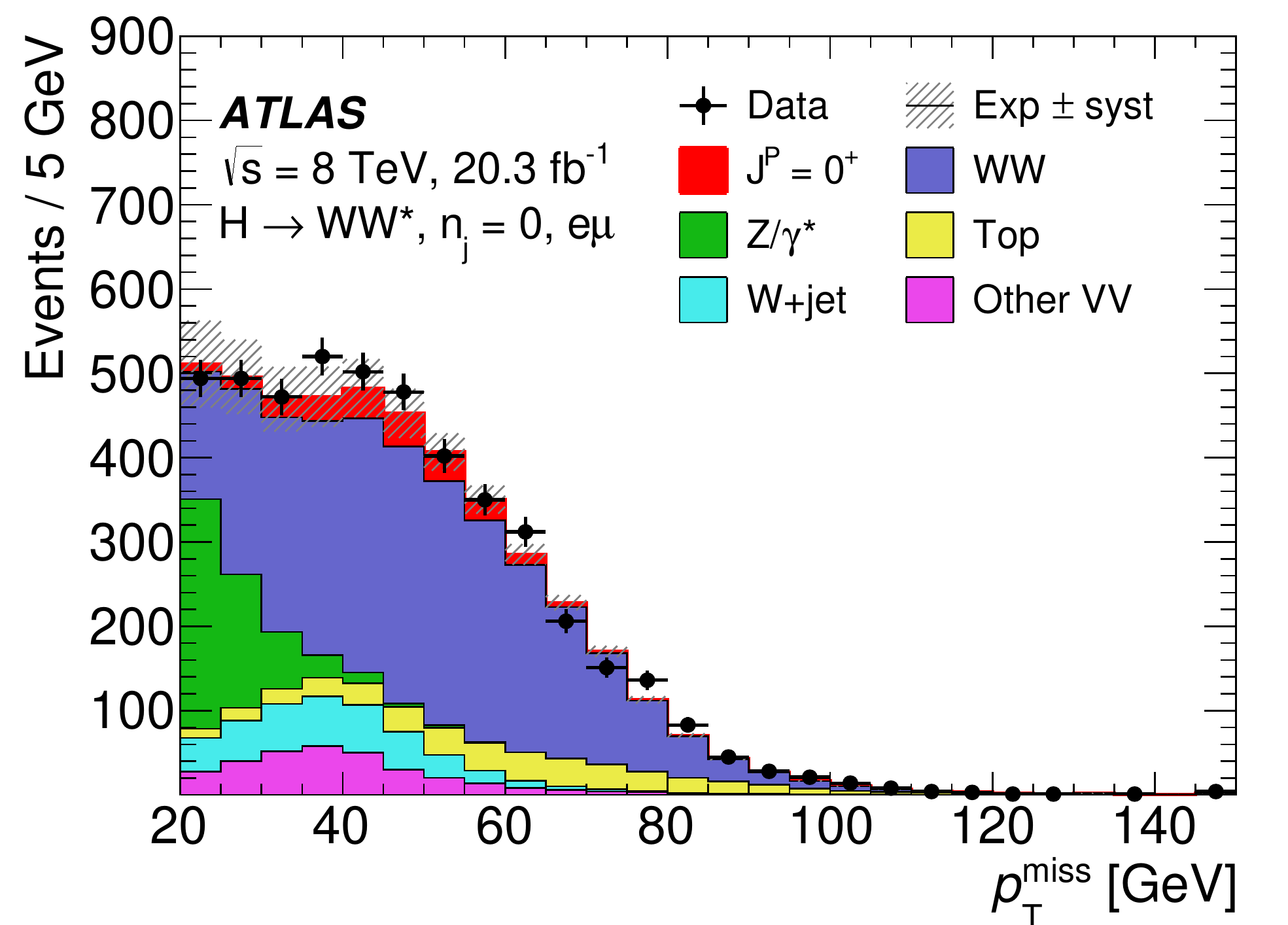}}
\subfloat{\includegraphics[width=0.43\textwidth]{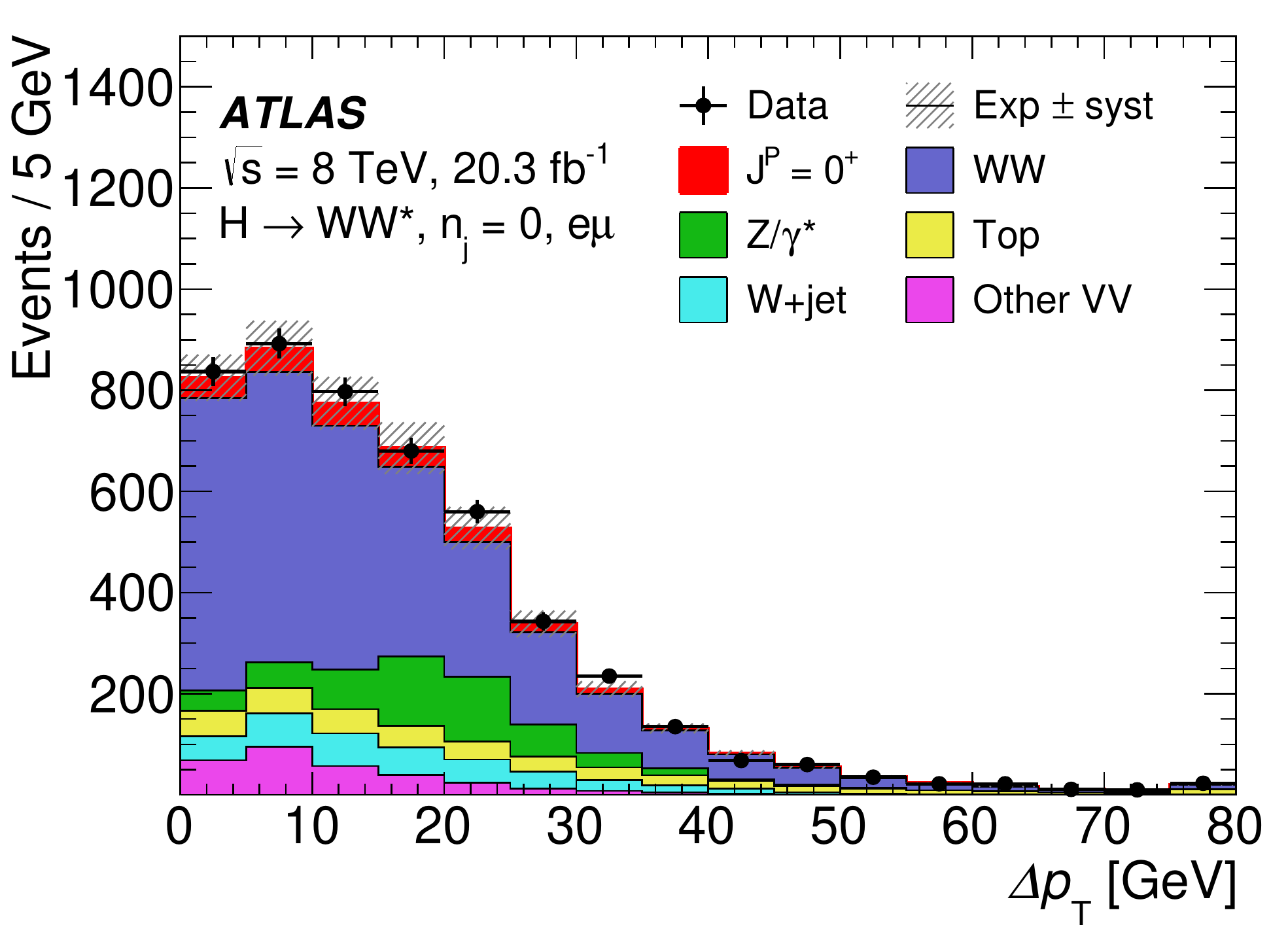}}\\
\subfloat{\includegraphics[width=0.43\textwidth]{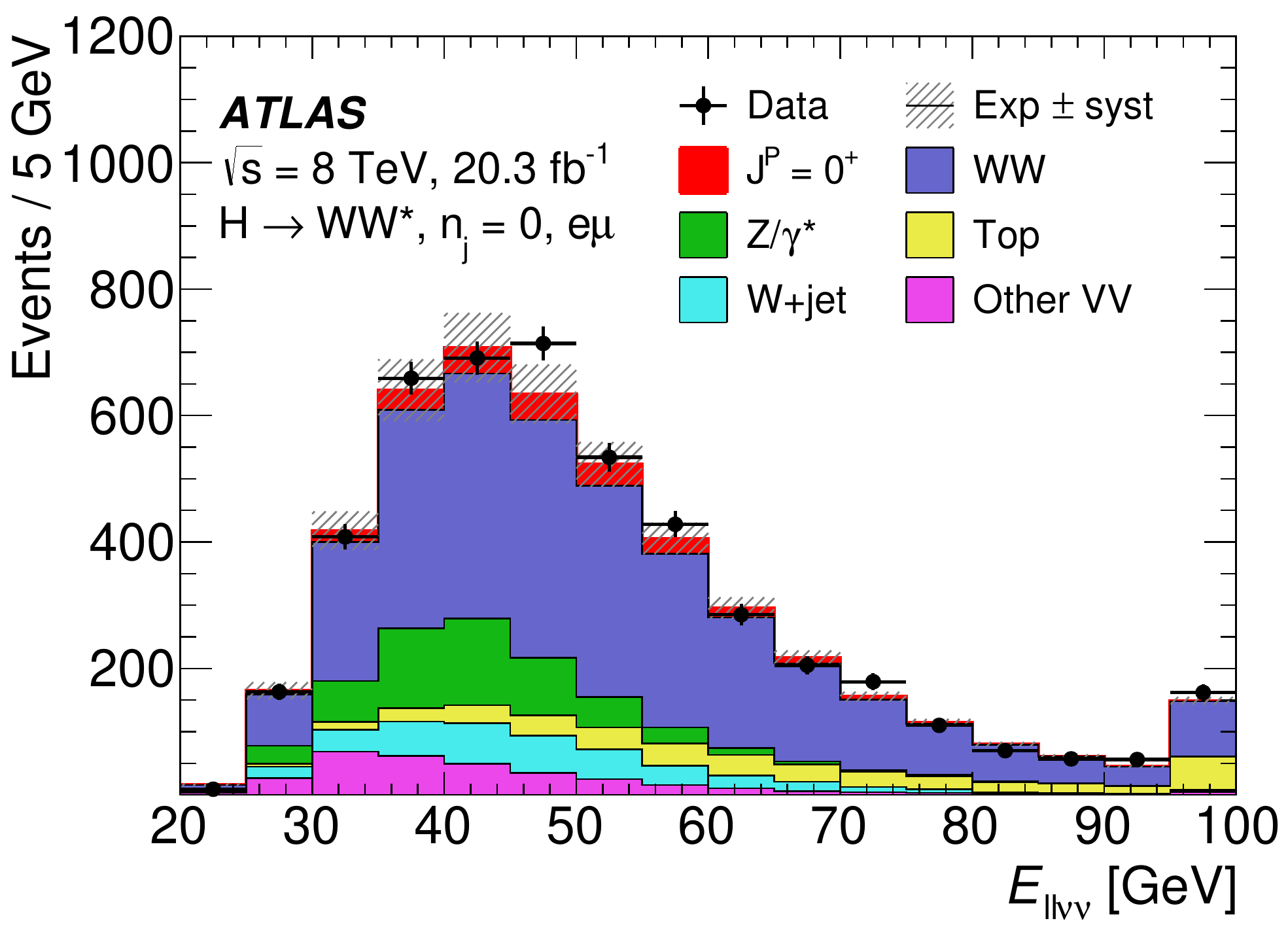}} 
    \caption{Expected and observed distributions of \ptll, \mll, \dphill, \mT, \ptmiss, \dpt\ and \Efun\ for the 0-jet category. The shaded band represents the systematic uncertainties described in Sects.~\ref{sec:backgrounds} and~\ref{systematics}. The signal is shown assuming an SM Higgs boson with mass $m_H=125$~\GeV. The backgrounds are normalised using control regions defined in Sect.~\ref{sec:backgrounds}. The last bin in each plot includes the overflow.} 
  \label{fig:vars_unblind_sr0jet}
\end{figure}

\begin{figure}
  \centering
\subfloat{\includegraphics[width=0.49\textwidth]{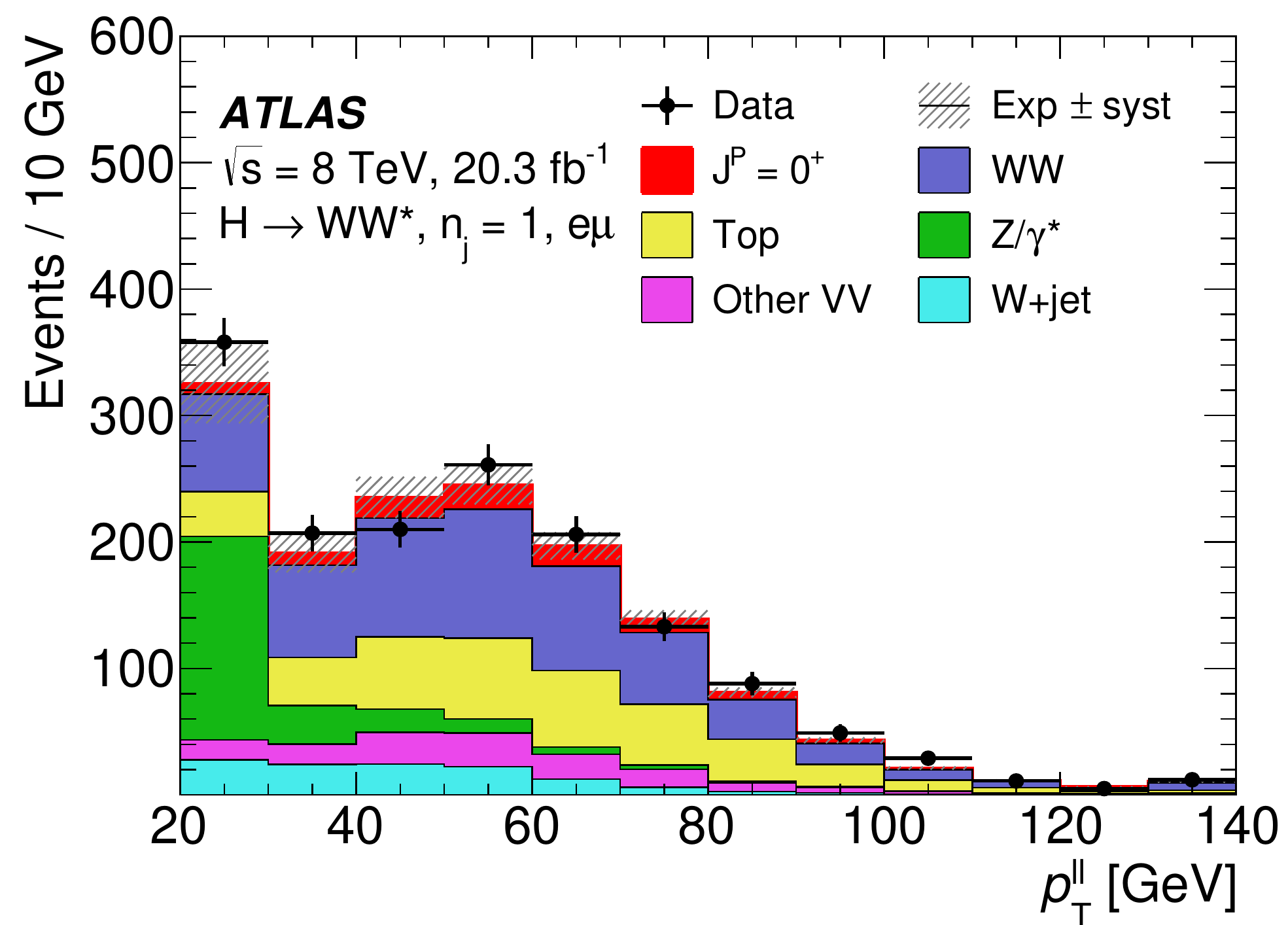}}
\subfloat{\includegraphics[width=0.49\textwidth]{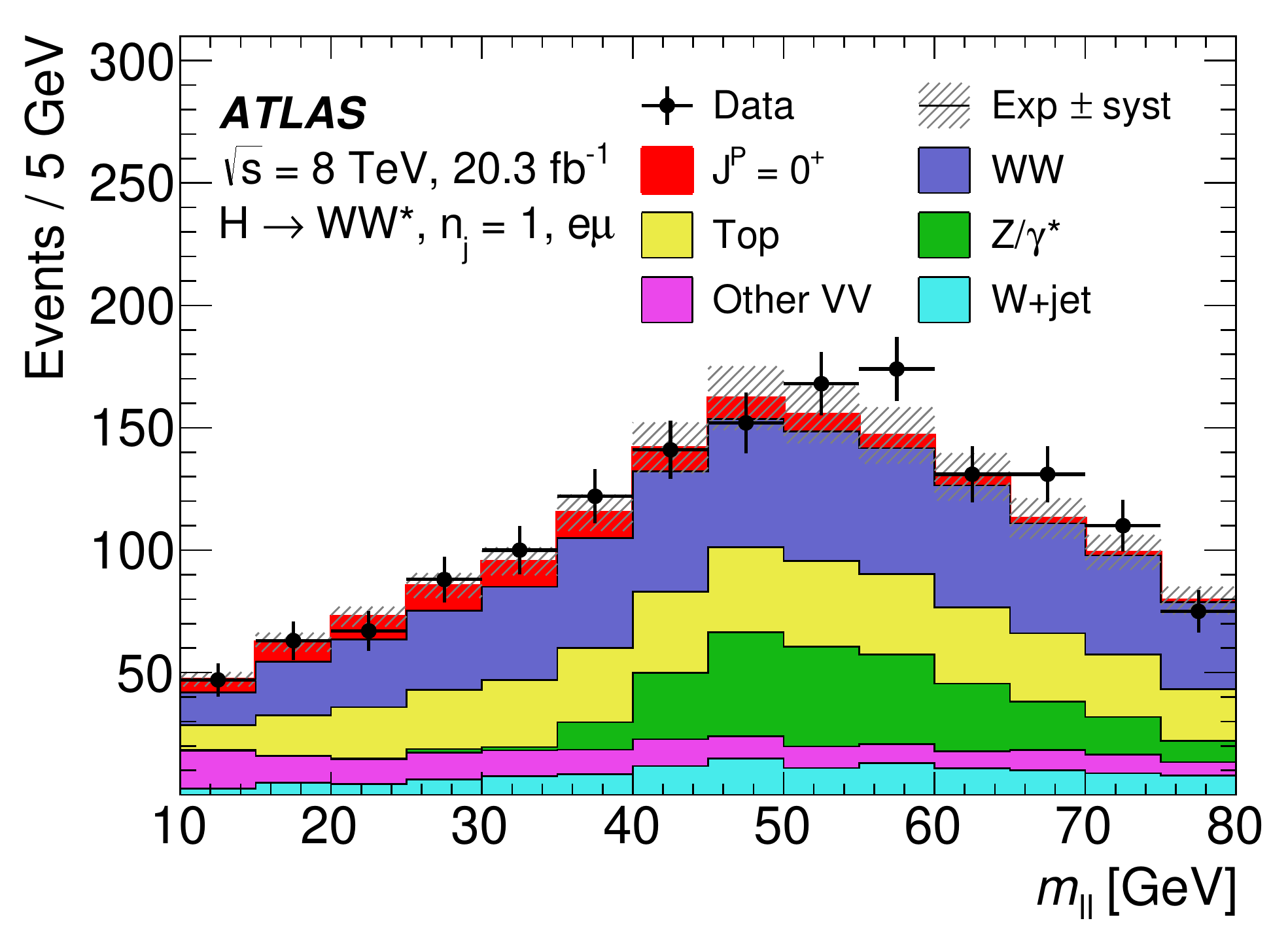}}\\
\subfloat{\includegraphics[width=0.49\textwidth]{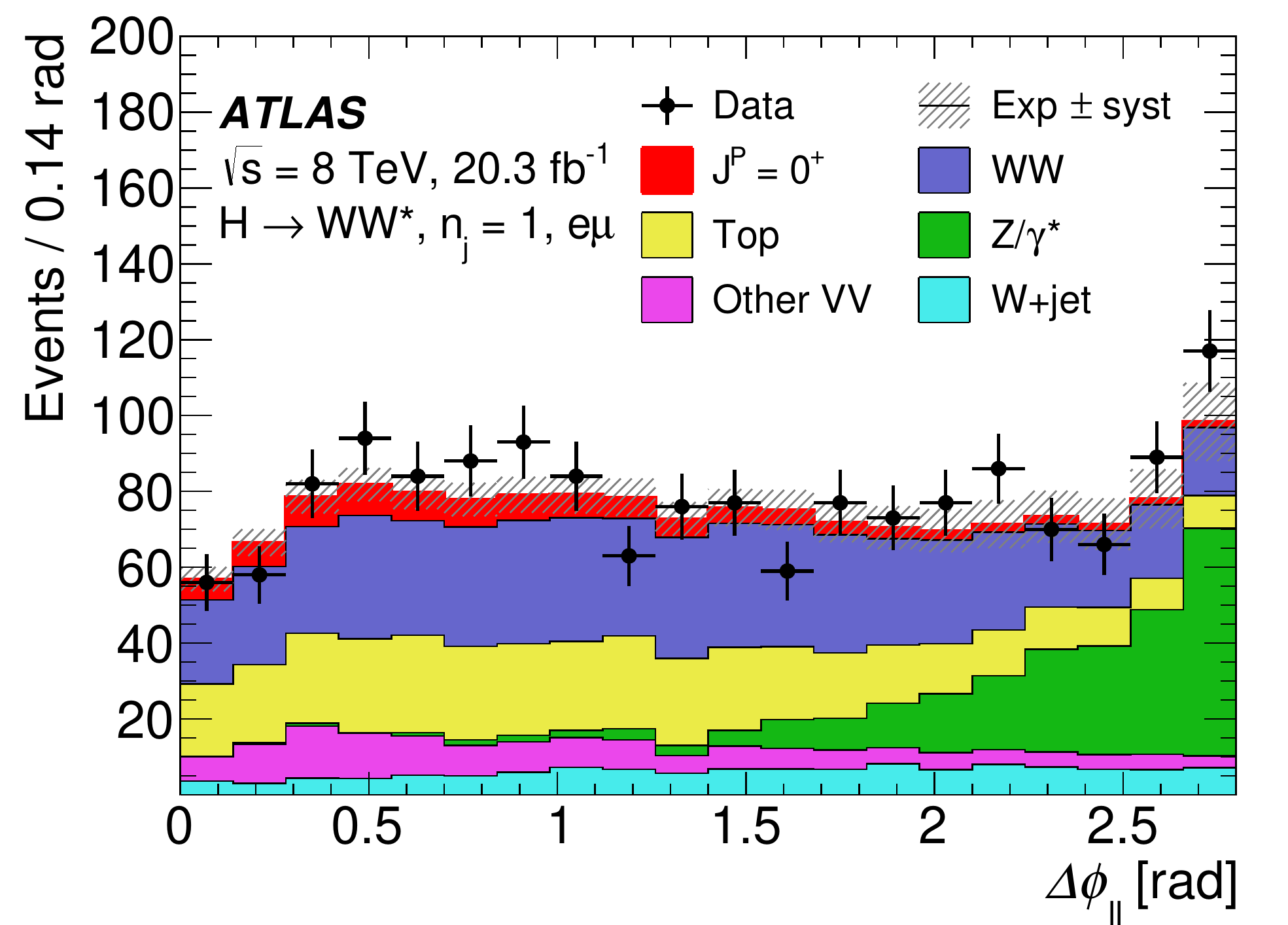}}
\subfloat{\includegraphics[width=0.49\textwidth]{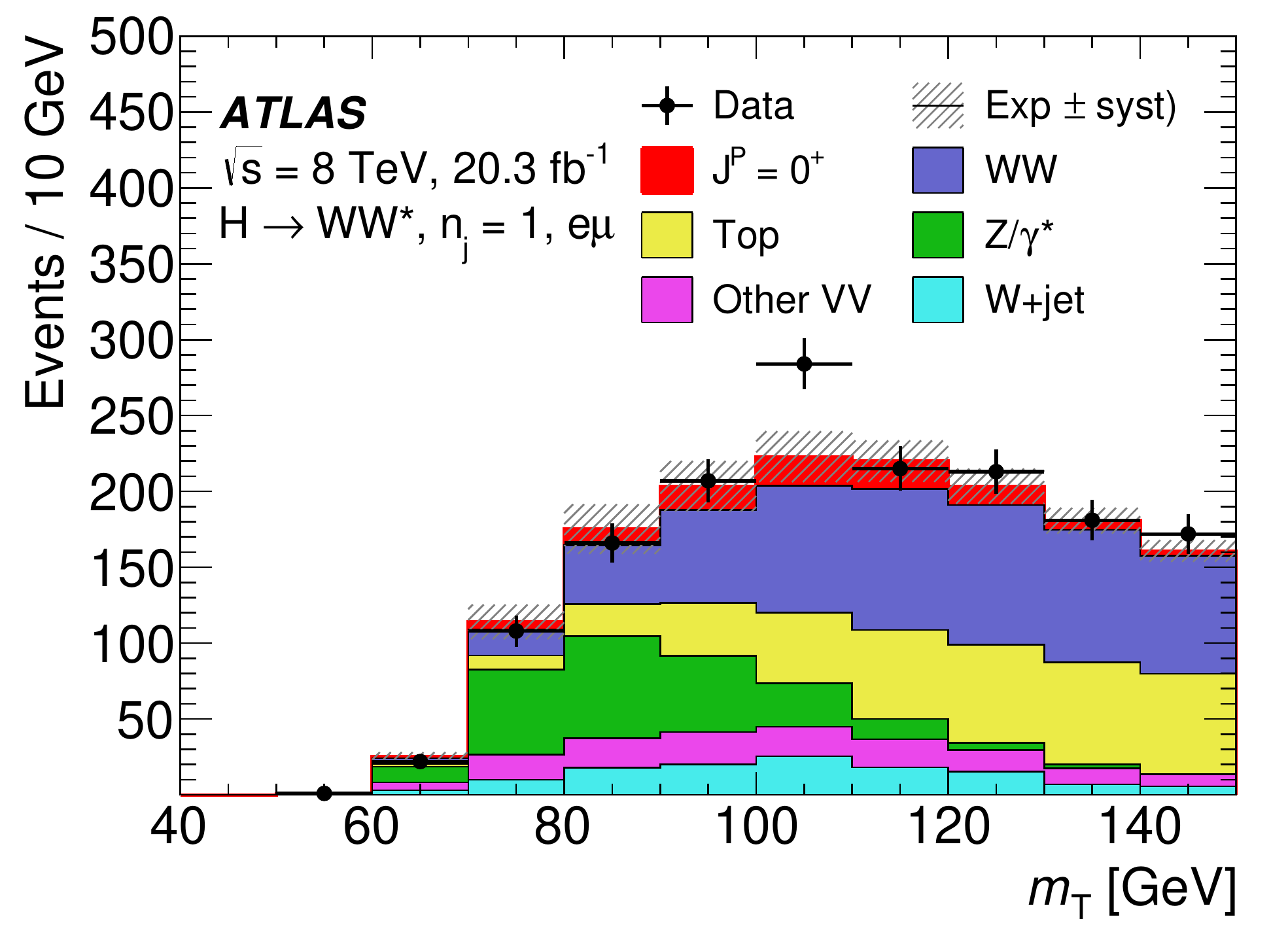}}
    \caption{Expected and observed distributions of  \ptll, \mll, \dphill\ and \mT\ for the 1-jet category. The shaded band represents the systematic uncertainties described in Sects.~\ref{sec:backgrounds} and~\ref{systematics}. The signal is shown assuming an SM Higgs boson with mass $m_H=125$~\GeV. The backgrounds are normalised using control regions defined in Sect.~\ref{sec:backgrounds}. The last bin in each plot includes the overflow.} 
  \label{fig:vars_unblind_sr1jet}
\end{figure}

\section{Backgrounds} 
\label{sec:backgrounds}

The background contamination in the signal region (SR) is briefly discussed in the previous section. This section is dedicated to a more detailed 
description of backgrounds and their determination. The following physics processes relevant for this analysis are discussed:
%
\begin{itemize}
  \item $\WW$: non-resonant $W$-boson pair production;
  \item top quarks (labelled as Top): top-quark pair production ($\ttbar$) and single-top-quark production ($t$);
  \item misidentified leptons (labelled as \Wjets): $W$-boson production, in association with
        a jet that is misidentified as a lepton, and dijet or multi-jet
        production with two misidentifications;
  \item $Z/\gamma^*$ decay to $\tau\tau$ final states.
\end{itemize}

Other smaller backgrounds, such as non-$WW$ dibosons ($\Wg$, $\Wgs$, $\WZ$ and $\ZZ$) labelled as $VV$ in the following, as well as the very small $Z/\gamma^* \rightarrow ee$ or $\mu\mu$ contribution, are estimated directly from simulation with the appropriate theoretical input as discussed in Sect.~\ref{sec:detector_samples}.

The dominant background sources are normalised either using only data, as in the case of the \Wjets\ background, or using data yields in an 
appropriate  control region (CR) to normalise the MC predictions, as for $WW$, \Ztt\ and top-quark backgrounds. The event selection in control 
regions is orthogonal to the signal region selection but as close as possible to reduce the extrapolation uncertainties from the CRs to the SR. The 
requirements that define these regions are listed in Table~\ref{tab:crcuts}. 

\begin{table}
\caption{\label{tab:crcuts}List of selection criteria used to define the orthogonal control regions for $WW$, top-quark and \Ztt\ backgrounds.}
\begin{center}
\begin{tabular}{c|c}
\dbline
  Control region & Selection \\
  \sgline
 $WW$ CR 0-jet & Preselection, $\ptll > 20$~\GeV , $80<\mll<150$~\GeV \\
 \multirow{2}*{$WW$ CR-1 jet} & Preselection, $b$-veto, $m_{\tau\tau} < m_Z-25$~\GeV \\
  & $m_{\rm T}^{\ell} > 50$~\GeV, $\mll > 80$~\GeV  \\
  \sgline
  Top CR 0-jet & Preselection, $\dphill < 2.8$, all jets inclusive \\
    Top CR 1-jet & At least one $b$-jet, $m_{\tau\tau} < m_Z-25$~\GeV \\
\sgline
  $\Ztt$ CR 0-jet &  Preselection, $\mll < 80$~\GeV, $\dphill > 2.8$ \\
    $\Ztt$ CR 1-jet &  Preselection, $b$-veto, $m_{\rm T}^{\ell} > 50$~\GeV, $\mll < 80$~\GeV, $|m_{\tau\tau}-m_Z|<25$~\GeV\\
    \dbline
\end{tabular}
\end{center}
\end{table}

The control regions, for example the $WW$ CR, are used to determine a normalisation factor, $\beta$, defined by the ratio of the observed 
to expected yields of $WW$ candidates in the CR, where the observed yield is obtained by subtracting the non-$\WW$ contributions 
from the data. The estimate  $B_\mathrm{SR}^{\mathrm{est}}$ for the background under consideration, in the SR, can be written as:

\begin{equation}
 \nq
 B_\mathrm{SR}^{\mathrm{est}}
  = B_\mathrm{SR}\,\CDOT\np\underbrace{N_\mathrm{CR}/B_\mathrm{CR}}_{\mbox{\scriptsize Normalisation\,$\beta$}}\no
  = N_\mathrm{CR}\,\CDOT\np\underbrace{B_\mathrm{SR}/B_\mathrm{CR}}_{\no\mbox{\scriptsize Extrapolation\,$\alpha$}}\nq \quad,
\label{eqn:est_ww}
\end{equation}

where $N_\mathrm{CR}$ and $B_\mathrm{CR}$ are the observed yield and the MC estimate in the CR, respectively, and $B_\mathrm{SR}$ is 
the MC estimate in the signal region. The parameter $\beta$ defines the data-to-MC normalisation factor in the CR, while the parameter $\alpha$
defines the extrapolation factor from the CR to the SR predicted by the MC simulation. With enough events in the CR, the large theoretical uncertainties 
associated with estimating the background only from simulation are replaced by the combination of two significantly smaller uncertainties: the statistical 
uncertainty on $N_\mathrm{CR}$ and the systematic uncertainty on $\alpha$.

The extrapolation factor $\alpha$ has uncertainties which are common to all MC-simulation derived backgrounds:
\begin{itemize}
\item uncertainty due to higher perturbative orders in QCD not included in the MC simulation, evaluated by varying the renormalisation and
factorisation scales by factors one-half and two;
\item uncertainty due to the PDF choice, estimated by 
taking the largest difference between the nominal PDF set (e.g. CT10) and two alternative PDF sets (e.g. MSTW2008~\cite{Martin:2009iq} and NNPDF2.3~\cite{Ball:2012cx}), with the  uncertainty determined from the error eigenvectors of the nominal PDF set added in quadrature;
\item uncertainty due to modelling of the underlying event, hadronisation and parton shower (UE/PS), evaluated by comparing the predictions from the nominal and alternative parton shower models, e.g. \PYTHIA\  and \HERWIG.
\end{itemize}

The section is organised as follows. Section~\ref{sec:wwcr} describes the $\WW$ background -- the dominant background in both the 0- and 1-jet categories. 
Section~\ref{sec:topcr} describes the background from the top-quark production, the second largest background in the 1-jet category. The $Z/\gamma^* \rightarrow \tau^+\tau^-$ background 
is described in Sect.~\ref{sec:zttcr}, while the data-derived estimate of the \Wjets\ background is briefly described in Sect.~\ref{sec:wjets}. The extrapolation 
factor uncertainties are summarised in Table~\ref{tab:alpha_uncert}. More details can be found in Ref.~\cite{ATLAS-CONF-2014-060}.

\begin{table}[htbp]
  \caption{\label{tab:alpha_uncert} Theoretical uncertainties (in \%) on the extrapolation factor $\alpha$ for $WW$, top-quark and \Ztt\ backgrounds. "Total" refers to the sum in quadrature of 
  all uncertainties. The negative sign indicates anti-correlation with respect to the unsigned uncertainties for categories in the same column.
  The uncertainties on the top-quark background extrapolation factor in the 0-jet category are discussed in Sect.~\ref{sec:topcr}.}
  \begin{center}
\begin{tabular}{c|cccccc|c}
\dbline
Category & Scale & PDF& Gen &  EW & UE/PS & $p_{\rm T}^Z$ & Total \\
  \sgline
  \multicolumn{8}{c}{$WW$ background } \\
   \sgline
  SR 0-jet & 0.9   & 3.8 & 6.9 & --0.8 & --4.1 & -- & 8.2 \\
  SR 1-jet & 1.2   & 1.9 & 3.3 & --2.1 & --3.2 & -- &5.3 \\
  \sgline
   \multicolumn{8}{c}{Top-quark background} \\ 
     \sgline
    SR 1-jet        & --0.8  & --1.4  & 1.9   & -- & 2.4 & -- & 3.5 \\
  $WW$ CR 1-jet &  0.6   & 0.3  &  --2.4 & -- & 2.0  & -- &  3.2 \\
    \sgline
   \multicolumn{8}{c}{\Ztt\ background } \\ 
     \sgline
  SR 0-jet           & --7.1   &  1.3 & -- & -- & --6.5   & 19 & 21.3\\
  SR 1-jet           & 6.6    & 0.66  & --  & -- & --4.2 & -- & 7.9\\

  $WW$ CR 0-jet  & --11.4 & 1.7 &  -- & --& --8.3    & 16& 21.4 \\
  $WW$ CR 1-jet  & --5.6   & 2.2  &  -- & -- & --4.8    & -- & 7.7\\
\dbline  
  
\end{tabular}\end{center}

\end{table}

\subsection{Non-resonant $W$-boson pairs }
\label{sec:wwcr}

Non-resonant $W$-boson pair production is the dominant (irreducible) background in this
analysis. Only some of the kinematic properties allow resonant and non-resonant production to be distinguished. The 
$WW$ background is normalised using a control region which differs from the signal region in having a different range of dilepton 
invariant mass, \mll. The leptons from non-resonant $WW$ production tend to have a larger opening
angle than the resonant $WW$ production. Furthermore, the Higgs-boson mass is lower than the mass of the system formed 
by the two $W$ bosons. Thus, the non-resonant $WW$ background is dominant at high \mll\ values.

The 0-jet $WW$ control region is defined after applying the \ptll\ criterion by changing the \mll\ requirement to $80<\mll<150$~\GeV.  The 1-jet $WW$ control 
region is defined after the $m_{\rm T}^{\ell}$ criterion by requiring $\mll > 80$~\GeV. The purity of the $WW$ 
control region is expected to be 69\% in the 0-jet category and 43\% in the 1-jet category. Thus, the data-derived normalisation of the main non-$WW$ backgrounds, the top-quark and Drell--Yan backgrounds, is applied in the $WW$ CR as described in the following two subsections. Other small backgrounds are normalised 
using MC simulation. The CR normalisation is applied to the combined  $WW$ estimate independently of the production ($qq, qg$ or $gg$) process. The 
\dphill\ and \mll\ distributions in the $WW$ control region are shown in Fig.~\ref{fig:vars_wwcrjet} for the 0-jet and 1-jet final states. 

Apart from the sources discussed in the previous section, the extrapolation factor $\alpha$ has uncertainties due to the generator 
choice, estimated by comparing the \POWHEG+\HERWIG\ and \amcatnlo+\HERWIG\ generators, and due to higher-order electroweak corrections
determined by reweighting the MC simulation to the NLO electroweak calculation. All uncertainties are summarised in Table~\ref{tab:alpha_uncert}.

\begin{figure}
  \centering
\subfloat{\includegraphics[width=0.48\textwidth]{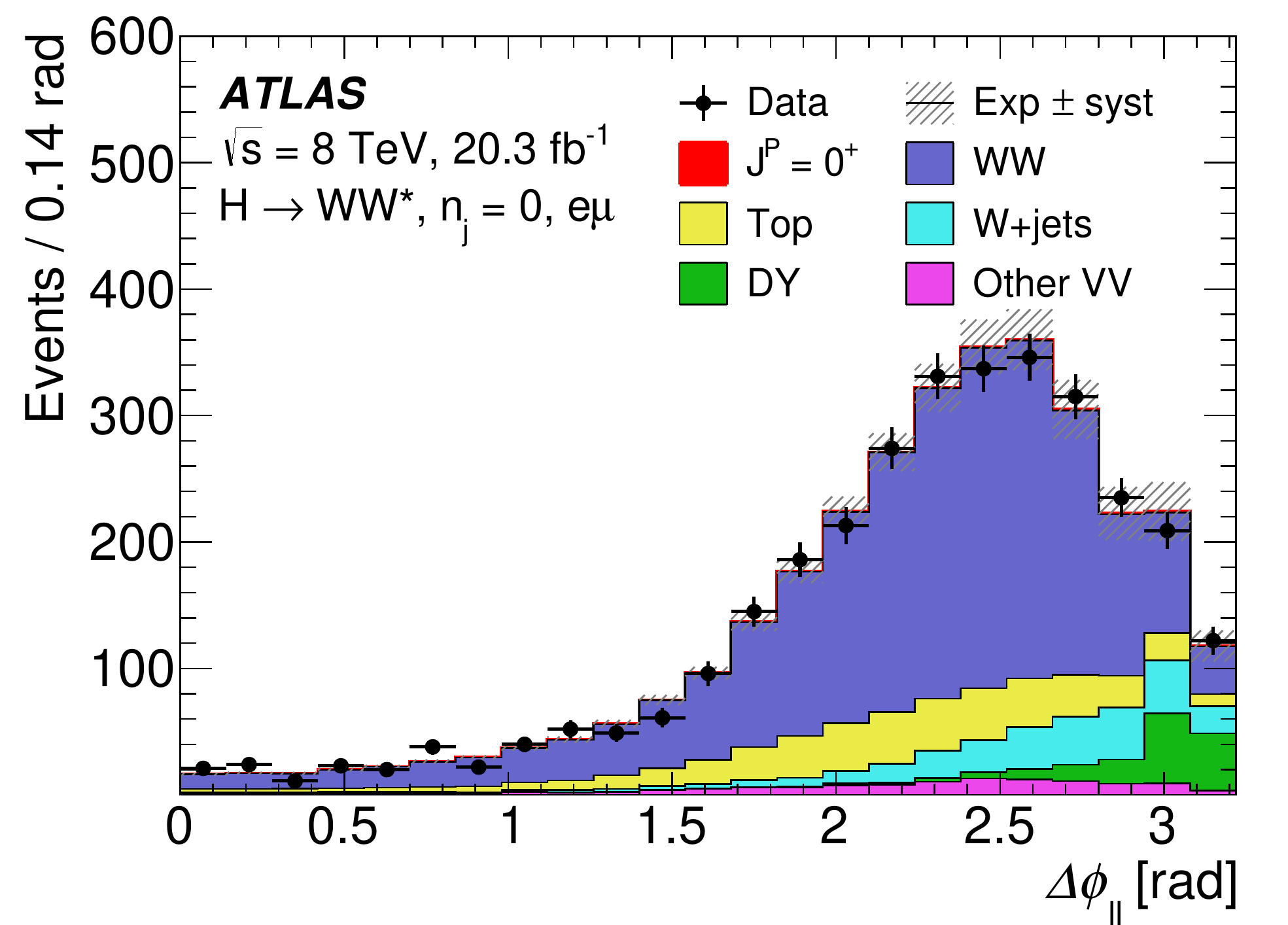}}
\subfloat{\includegraphics[width=0.48\textwidth]{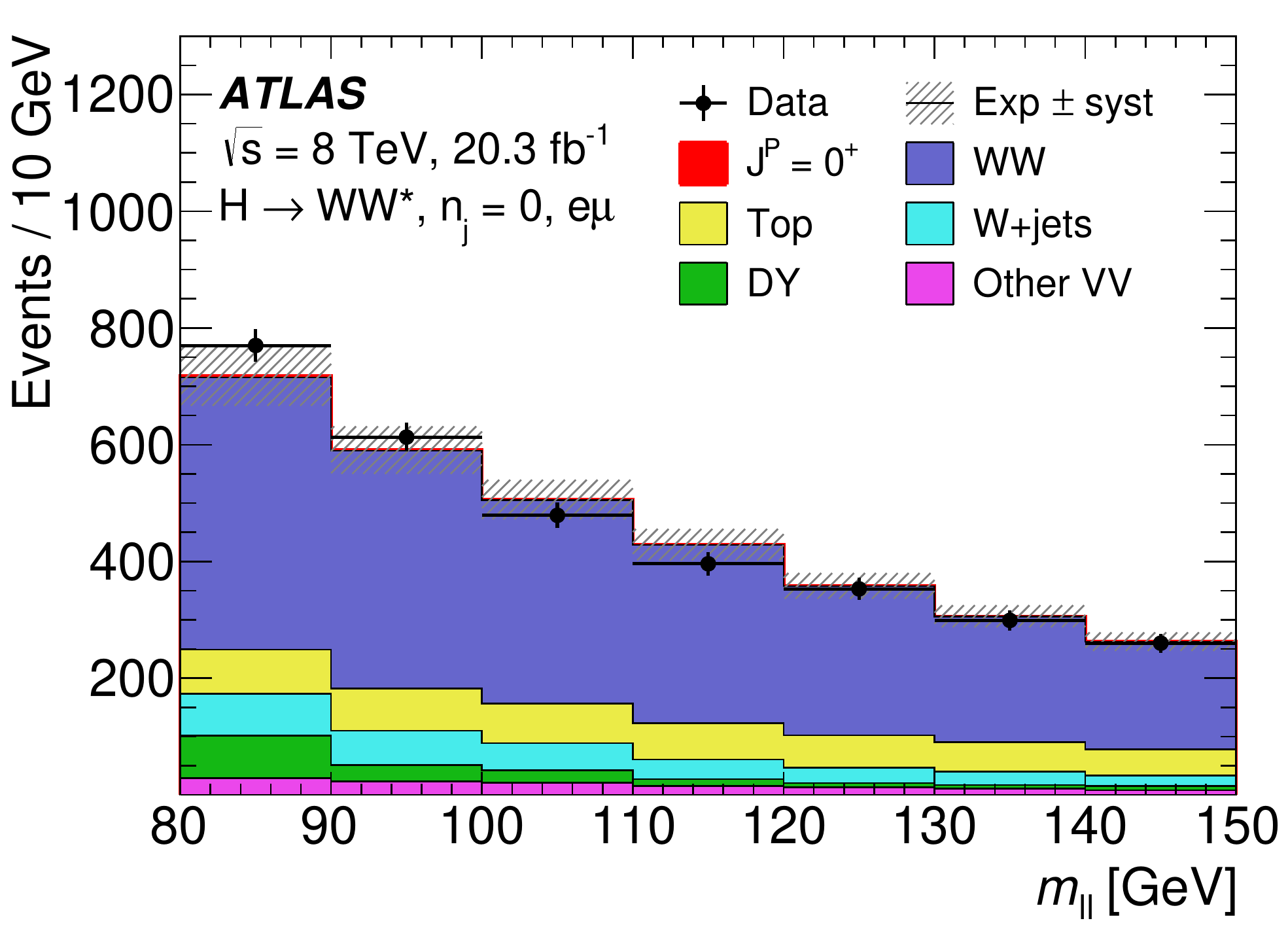}}\\
\subfloat{\includegraphics[width=0.48\textwidth]{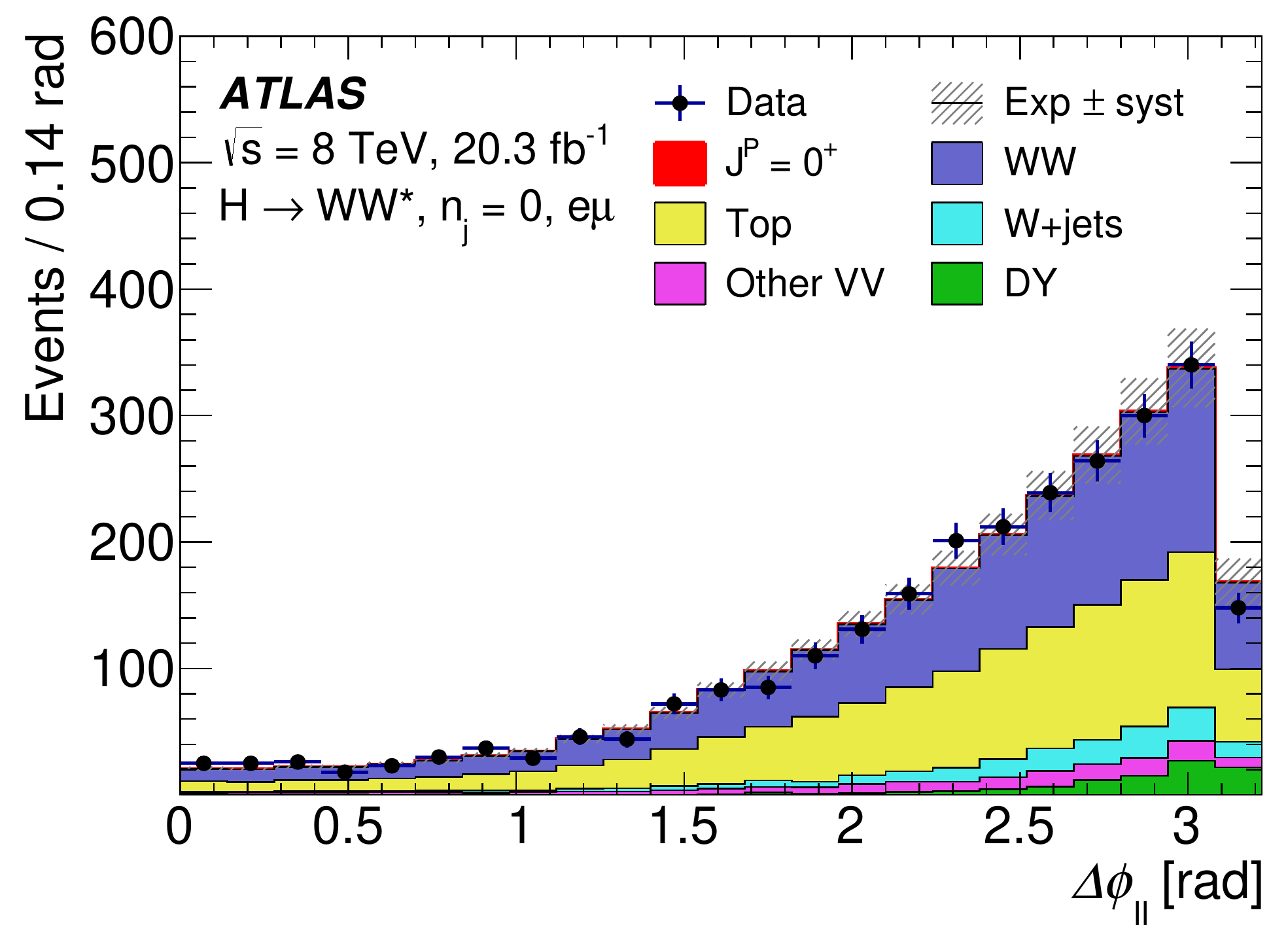}}
\subfloat{\includegraphics[width=0.48\textwidth]{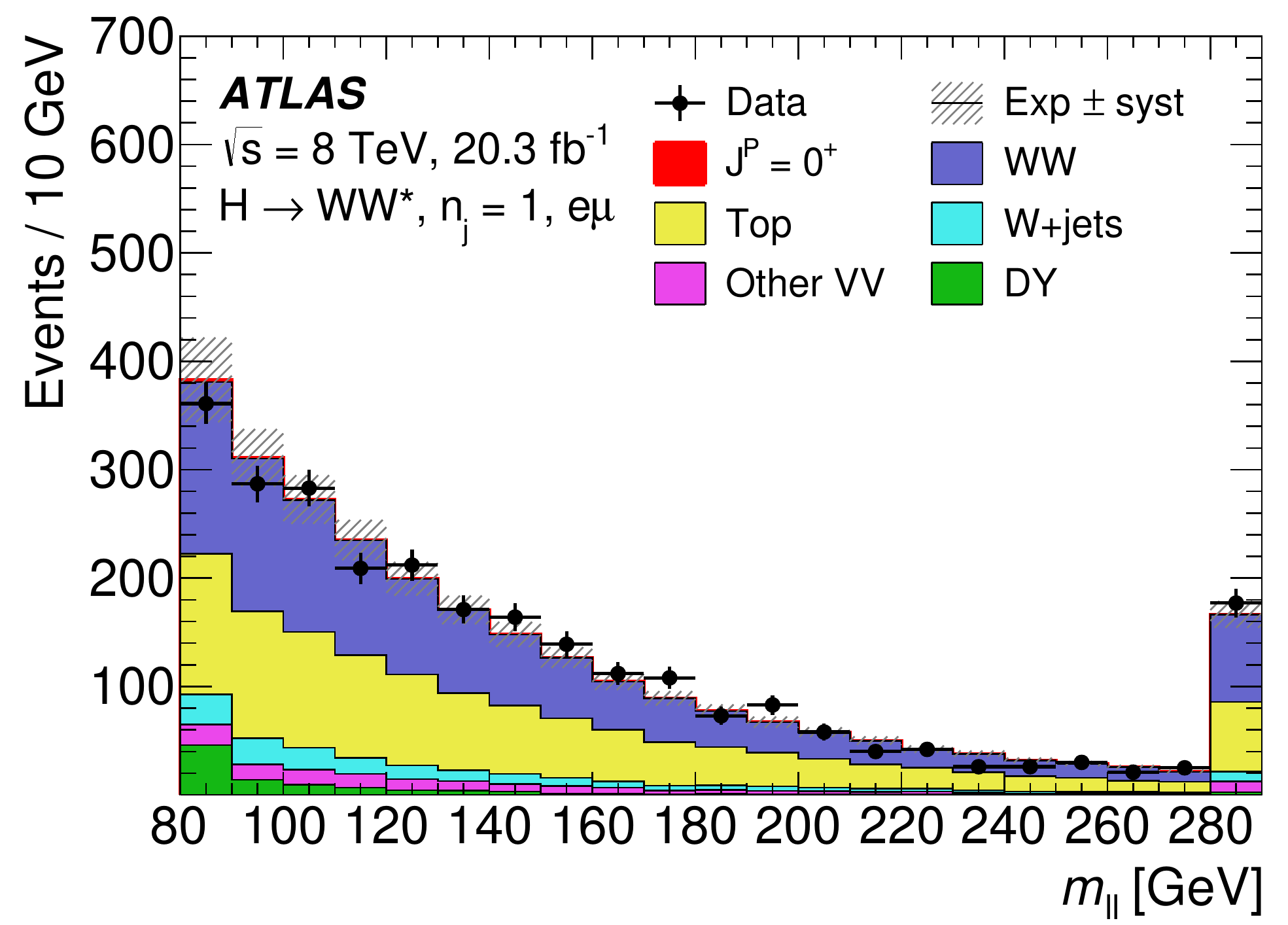}}
    \caption{The \dphill\ and \mll\ distributions in the $WW$ control region, for the 0-jet (top) and 1-jet (bottom) categories. The signal is shown  
    assuming an SM Higgs boson with mass $m_H=125$~\GeV. The signal contamination is negligible for the SM as well as for the alternative hypotheses.
The normalisation factors from the control regions described in Sect.~\ref{sec:backgrounds} are applied. The last bin in each plot includes the overflow.} 
  \label{fig:vars_wwcrjet}
\end{figure}

\subsection{Top quarks}
\label{sec:topcr}

The top-quark background is one of the largest backgrounds in this analysis. Top quarks can be produced in pairs (\ttbar) or individually in single-top processes in association with a $W$ boson ($Wt$) or lighter quark(s) (single-$t$).
The  top-quark background normalisation from data is derived independently of the production process.

For the 0-jet category, the control region is defined by applying the preselection cuts including the missing transverse momentum threshold, with an additional 
requirement of $\dphill < 2.8$ to reduce the $\Ztt$ background. The top-quark background 0-jet CR is inclusive in the number of jets and has a purity of 74\%. The 
extrapolation parameter $\alpha$ is determined as described in Eq.~(\ref{eqn:est_ww}). The value of $\alpha$ is corrected using data in a sample containing 
at least one $b$-tagged jet~\cite{ATLAS-CONF-2014-060}. 

The resulting normalisation factor is $1.08\pm0.02$ (stat.). The total uncertainty on the normalisation factor is 8.1\%. The total uncertainty includes variations 
of the renormalisation and factorisation scales, PDF choice and parton shower model. Also the uncertainty on the \ttbar\ and $Wt$ production cross-sections and on the 
interference of these processes is included. An additional theoretical uncertainty is evaluated on the efficiency of the additional selection after the jet-veto requirement.
Experimental uncertainties on the simulation-derived components are evaluated as well.

In the 1-jet category, the top-quark background is the second leading background, not only in the signal region but also in the $WW$ control region, where the 
contamination by this background is about 40\%. Thus two extrapolation parameters are defined: $\alpha_{\mathrm{SR}}$ for the extrapolation to the 
signal region and $\alpha_{WW}$ for the extrapolation to the $WW$ control region. The 1-jet top-quark background control region is defined after the preselection 
and requires the presence of exactly one jet, which must be $b$-tagged.  Events with additional $b$-tagged jets with $20<\pt< 25$~\GeV\ are vetoed, 
following the SR requirement. Selection criteria on $m^{\ell}_{\rm T}$ and $m_{\tau\tau}$ veto are applied as well. The \dphill\ and \mll\ distributions 
in the 1-jet CR are shown in Fig.~\ref{fig:vars_topcrjet}.

\begin{figure}
  \centering
\subfloat{\includegraphics[width=0.48\textwidth]{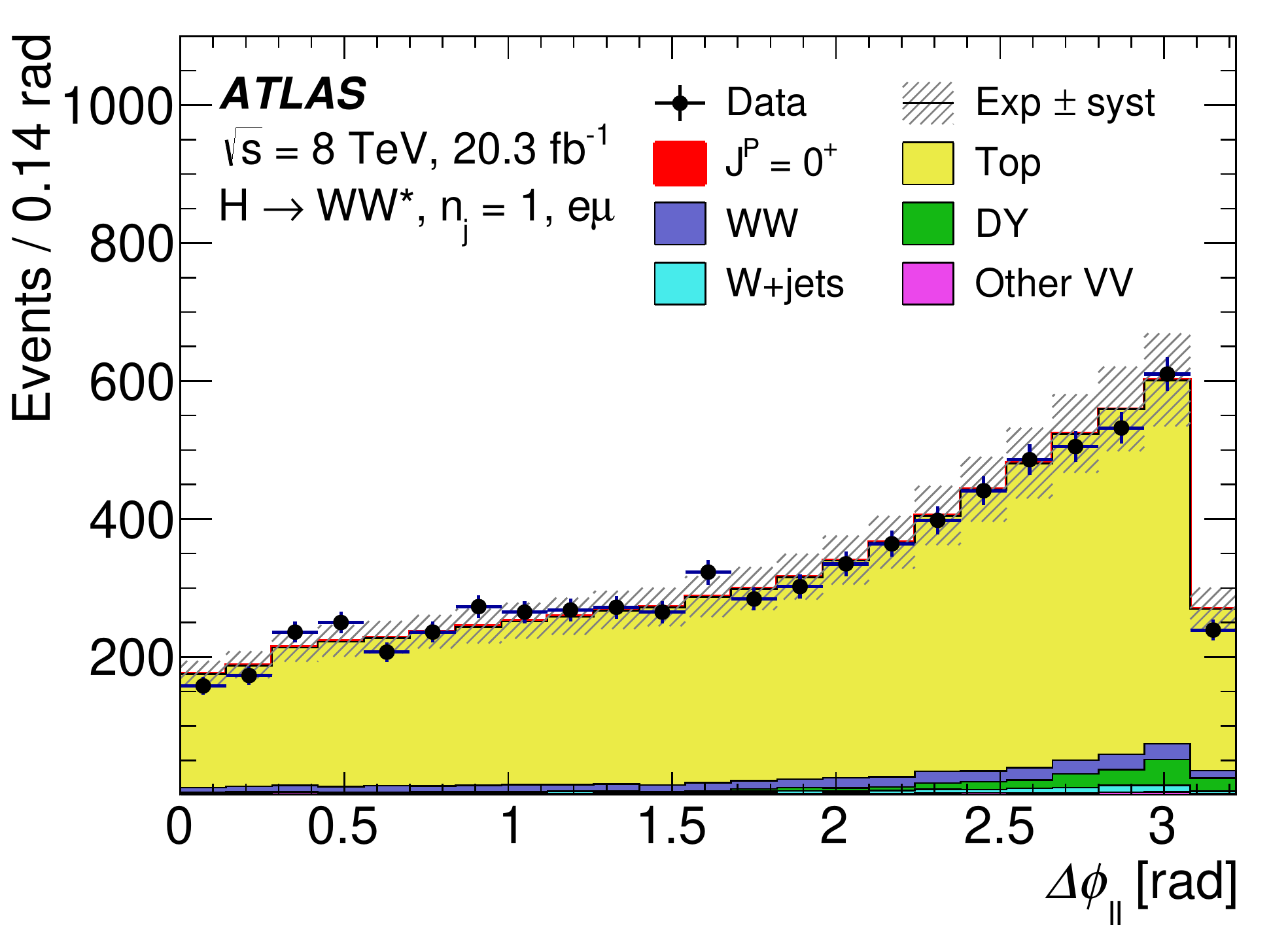}}
\subfloat{\includegraphics[width=0.48\textwidth]{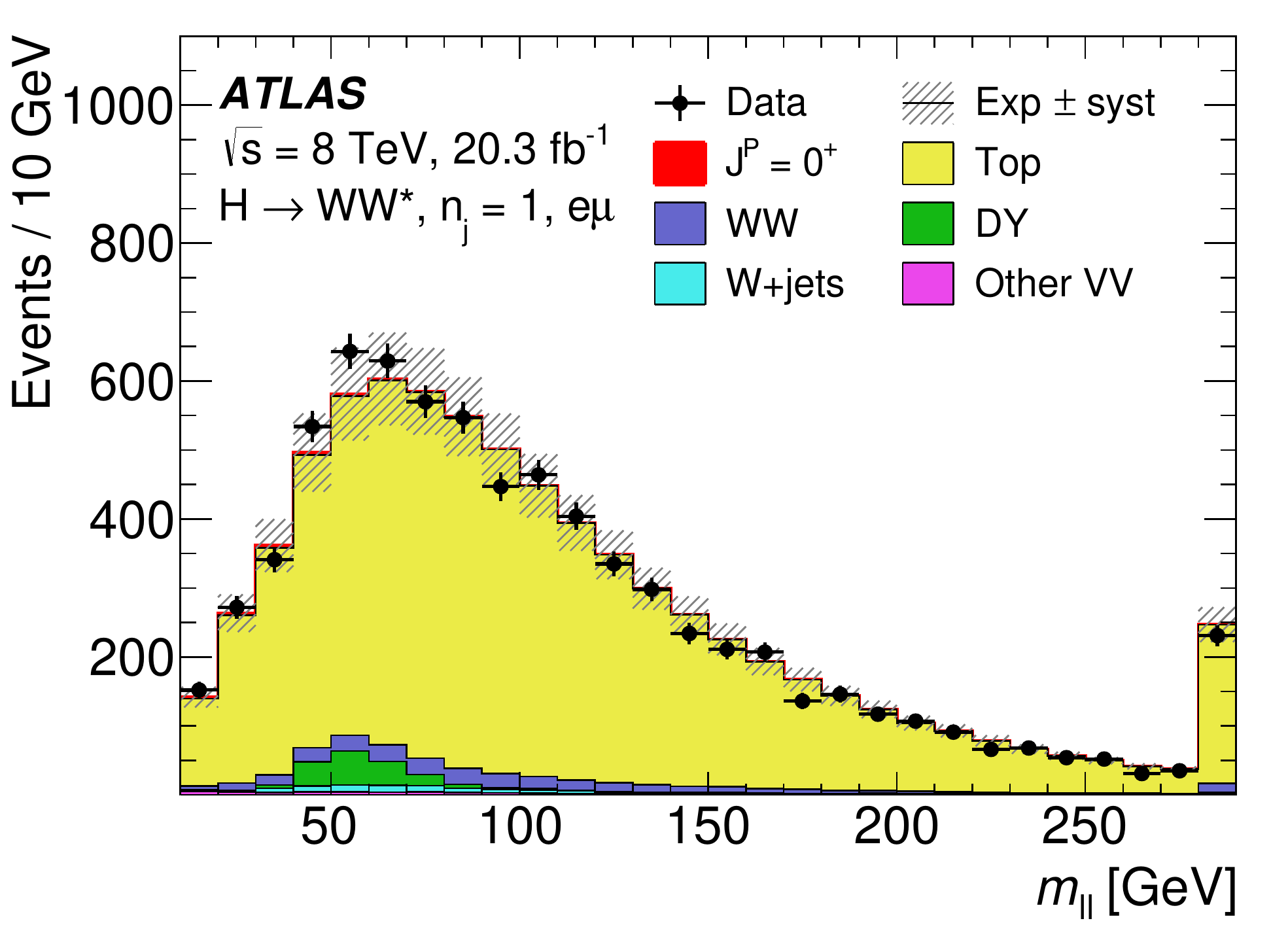}}
    \caption{The \dphill\ and \mll\ distributions in the top-quark background control region for 
    the 1-jet category. The signal is shown  assuming an SM Higgs boson with mass $m_H=125$~\GeV. The signal contamination is negligible for the SM as well as for the alternative hypotheses. The normalisation factors from the control regions described in Sect.~\ref{sec:backgrounds} are applied. The last bin in each plot includes the overflow.} 
  \label{fig:vars_topcrjet}
\end{figure}

The extrapolation uncertainty is estimated using the above mentioned sources of theoretical uncertainties and the additional uncertainties specific to the top-quark background: \ttbar\ and single-top cross-sections and the interference between single and pair production of top quarks. A summary of 
the uncertainties is given in Table~\ref{tab:alpha_uncert}.

\subsection{Drell--Yan}
\label{sec:zttcr}

The Drell--Yan  background is dominated by  \Ztt\ events with $\tau$-leptons decaying leptonically. The \Ztt\ 0-jet control region is defined by 
applying the preselection requirements, adding $\mll<80$~\GeV\ and reversing the \dphill\ criterion, 
$\dphill > 2.8$. The purity of this control region is expected to be 90\%. The \Ztt\ 1-jet control region is defined by applying the preselection requirements, 
$b$-veto, $m_{\rm T}^{\ell} > 50$~\GeV\ as in the signal region but requiring $|m_{\tau\tau}-m_Z|<25\GeV$. The purity of the 1-jet
control region is about 80\%.

The \Ztt\ predictions in the 0- and 1-jet categories are estimated using the extrapolation from the control region to the signal region and to the $WW$ control region, as there is a 4 --5\% contamination of \Ztt\ events in the $WW$ control region. The \dphill\ and \mll\ distributions in 
the \Ztt\ control region are shown in Fig.~\ref{fig:vars_ZZcrjet} for the 0-jet and 1-jet final states.

\begin{figure}
  \centering
\subfloat{\includegraphics[width=0.48\textwidth]{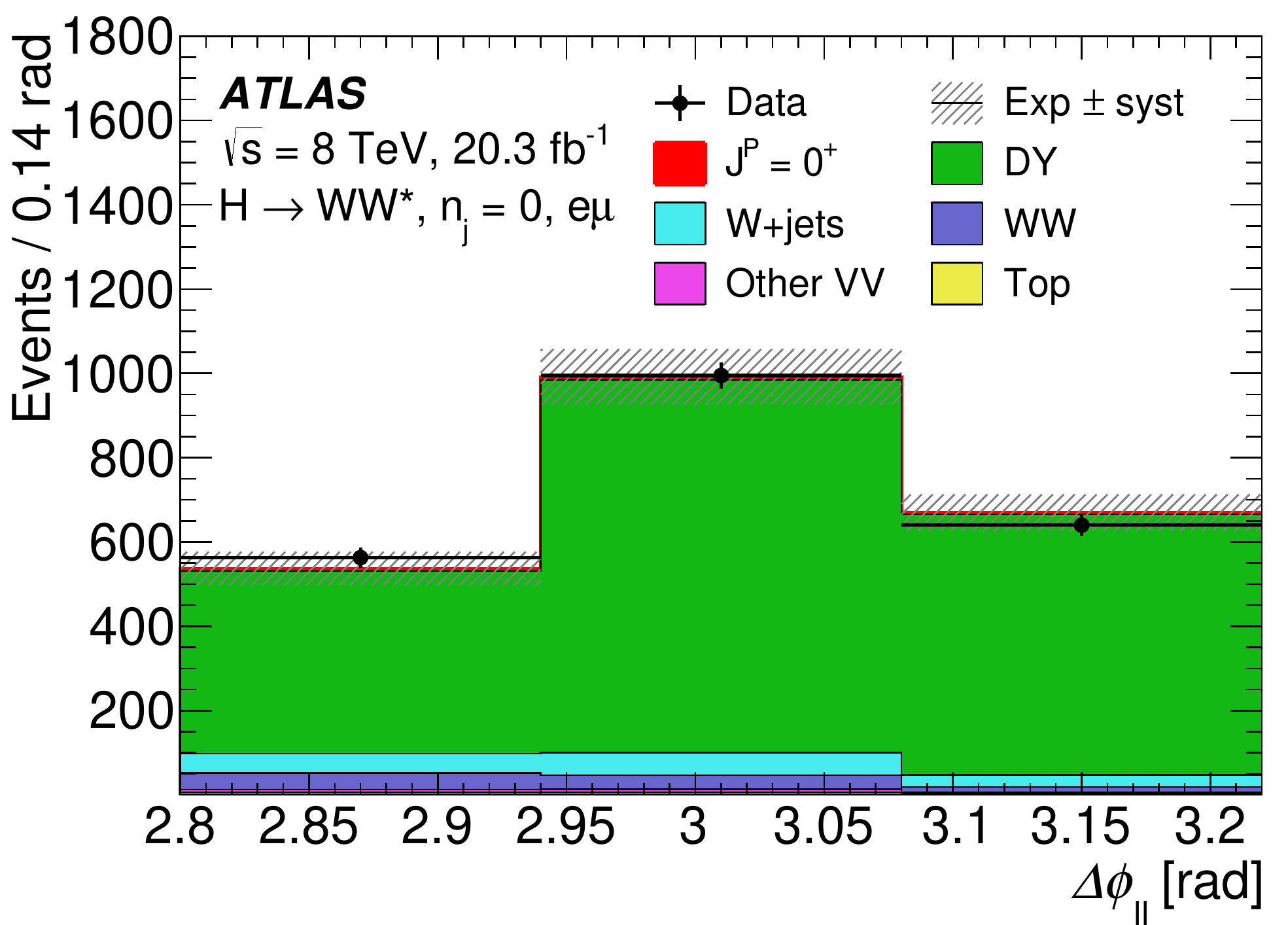}}
\subfloat{\includegraphics[width=0.48\textwidth]{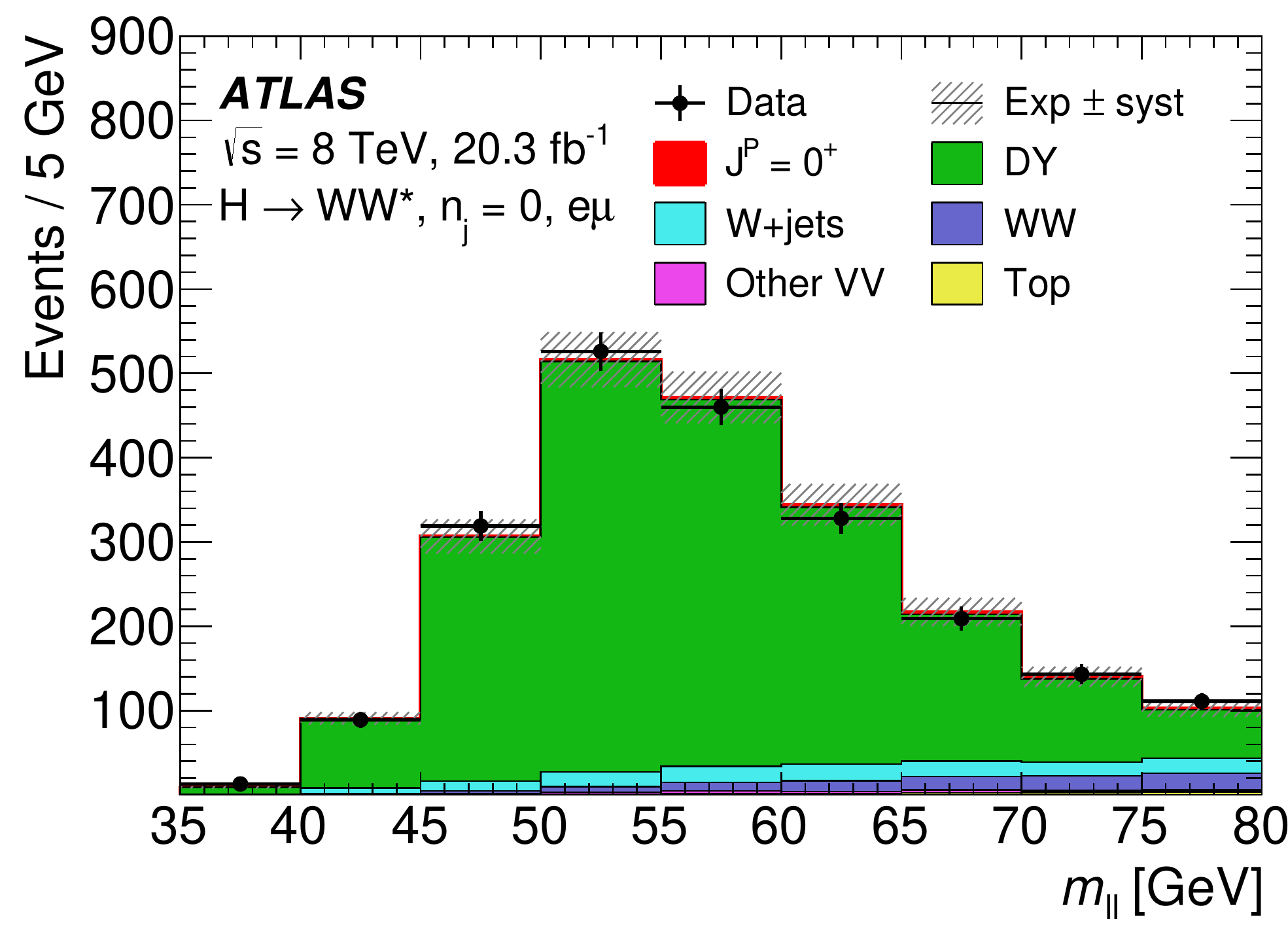}}\\
\subfloat{\includegraphics[width=0.48\textwidth]{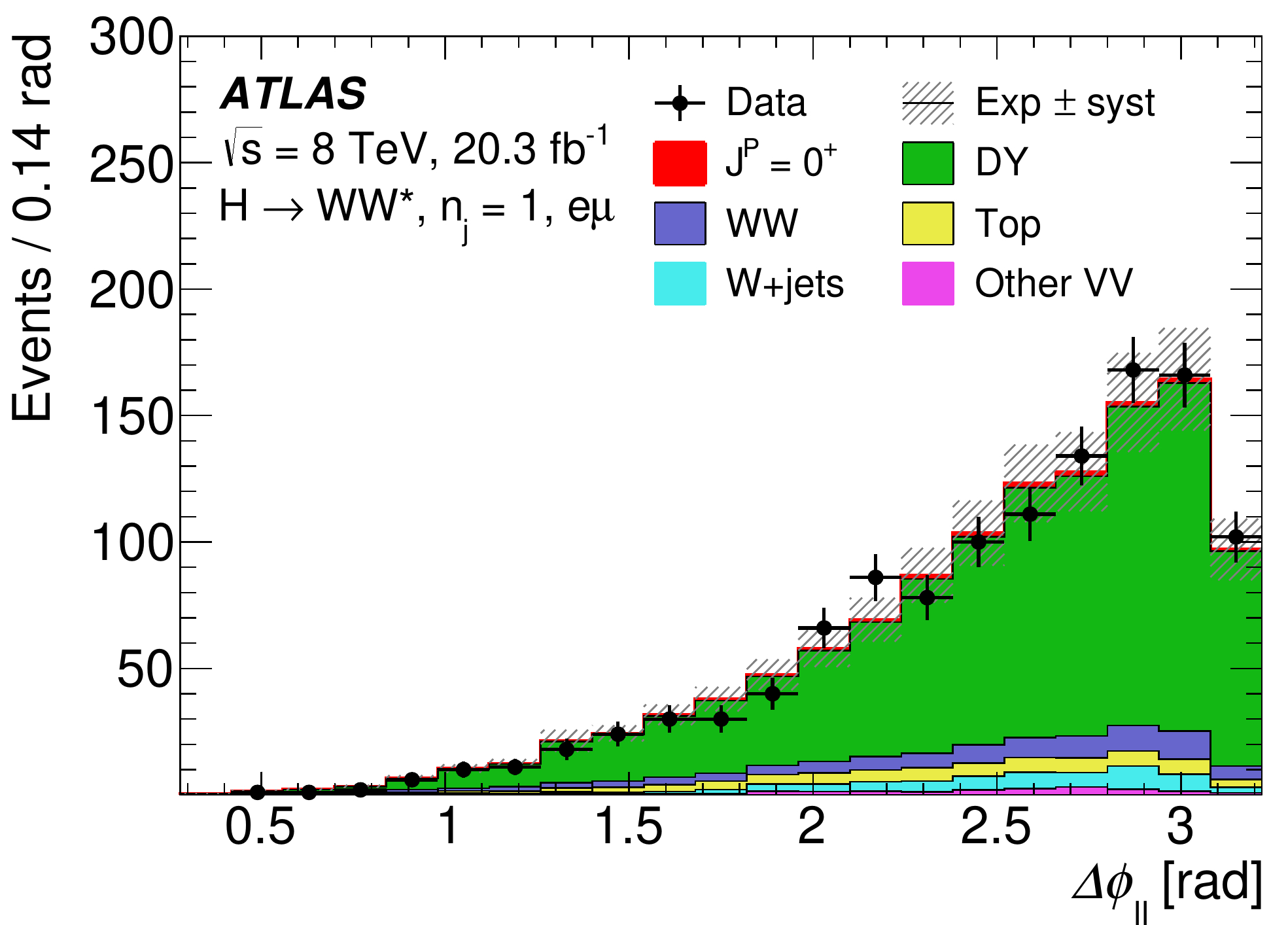}}
\subfloat{\includegraphics[width=0.48\textwidth]{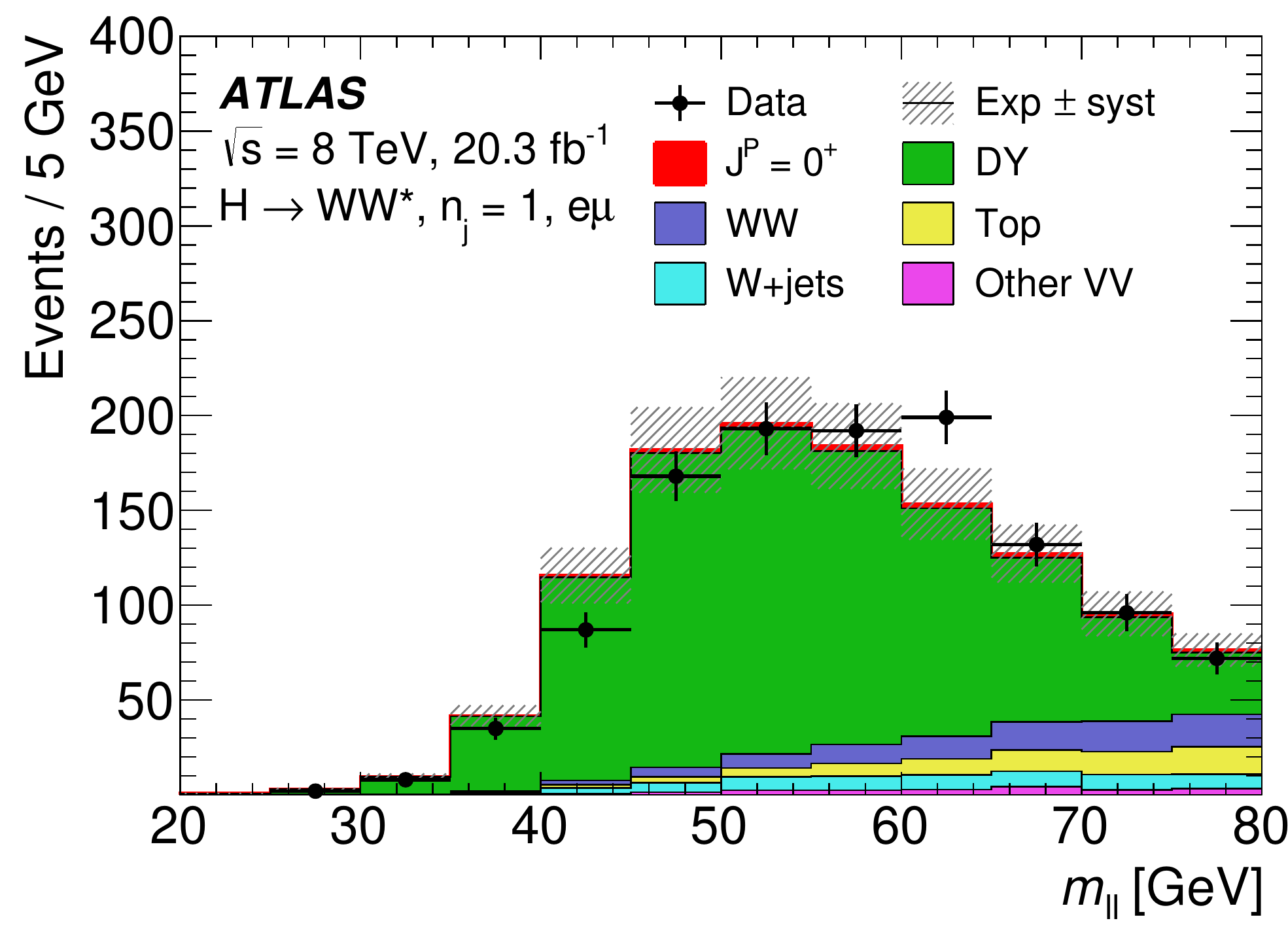}}
    \caption{The \dphill\ and \mll\ distributions in the \Ztt\ control region, for the 0-jet (top) and 1-jet (bottom) categories. The signal 
    is shown assuming an SM Higgs boson with mass $m_H=125$~\GeV. The signal contamination is negligible for the SM as well as for the alternative hypotheses. The normalisation factors from the control regions described in Sect.~\ref{sec:backgrounds} are applied. } 
  \label{fig:vars_ZZcrjet}
\end{figure}

A mismodelling of the transverse momentum of the $Z$ boson $p_{\rm T}^Z$, reconstructed as \ptll, is observed in the DY-enriched region. The mismodelling is more
pronounced in the 0-jet category. The \ALPGEN+\HERWIG\ MC generator does not adequately model the parton shower of the soft jets which balance \ptll\ in 
events with no selected jets. A correction, based on weights derived from a data-to-MC comparison in the $Z$ mass peak, is therefore 
applied to MC events in bins of $p_{\rm T}^{\ell \ell}$ in the 0-jet category. The weights are applied to $p_{\rm T}^Z$ at generator-level for all lepton flavour decays. 

Apart from the above mentioned sources of theoretical uncertainties, one additional uncertainty on the $p_{\rm T}^Z$-reweighting in the 0-jet category is estimated by comparing the difference between the nominal (derived in the $Z$ mass peak) and the alternative (derived in the $Z$ mass peak but after the $\ptmiss > 20$~\GeV\ criterion) set of weights. All uncertainties are summarised in Table~\ref{tab:alpha_uncert}.

\subsection{Misidentified leptons}
\label{sec:wjets}

The \Wjets\ background is estimated in the same way as in Ref.~\cite{ATLAS-CONF-2014-060}, where a detailed description of the method 
can be found. The \Wjets\ control sample contains events where one of the two lepton candidates satisfies the identification
and isolation criteria for the signal sample, and the other lepton fails to meet these criteria but satisfies less restrictive criteria
(these lepton candidates are called "anti-identified''). Events in this sample are otherwise required to satisfy all of the
signal selection requirements. The dominant component of this sample (85\% to 90\%) is due to $W$+jets events in which
a jet produces an object reconstructed as a lepton. This object may be either a non-prompt lepton from the decay of a
hadron containing a heavy quark, or a particle (or particles) originating from a jet and reconstructed as a lepton candidate.

The $W$+jets contamination in the signal region is obtained by scaling the number of events in the data control sample by an extrapolation factor. 
This extrapolation factor is measured in a data sample of jets produced in association with $Z$~bosons reconstructed in either the $ee$ 
or $\mu\mu$ final state (referred to as the $Z$+jets control sample below). The factor is the ratio of the number of identified lepton candidates 
satisfying all lepton selection criteria to the number of anti-identified leptons measured in bins of anti-identified lepton \pT\ and \myeta.
Each number is corrected for the presence of processes other than $Z$+jets. 

The composition of the associated jets  -- namely the fractions of jets due to the production of heavy-flavour quarks, light-flavour quarks and
gluons -- in the $Z$+jets sample and the $W$+jets sample are different. Monte Carlo simulation is used to correct the extrapolation factors
and to determine the associated uncertainty. Other important uncertainties on the \Zjets\ extrapolation factor are due to the limited number of jets that 
meet the lepton selection criteria in the $Z$+jets control sample and the uncertainties on the contributions from other physics processes. 

The total systematic uncertainty on the corrected extrapolation factors varies as a 
function of the \pT\ of the anti-identified lepton; this variation is from 29\% to 61\% for anti-identified electrons and 25\% to 46\% for anti-identified muons. The 
systematic uncertainty on the corrected extrapolation factor dominates the systematic uncertainty on the \Wjets\ background.

\section{BDT analysis} 
\label{sec:bdt_analysis}
Both the spin and the CP analysis employ a BDT algorithm\footnote{A decision tree is a collection of cuts used to classify events 
as signal or background. The classification is based on a set of discriminating variables  (BDT input variables) 
on which the algorithm is trained. The input events are repeatedly split using this information. 
At each split, the algorithm finds the variable and the optimal selection cut on this variable, 
that give the best separation between signal and background. Finally, an overall output weight 
(BDT output) is assigned to each event: the larger the weight, the more signal-like the event is 
classified to be. More details can be found in Ref.~\cite{TMVA}.} to distinguish between different signal hypotheses. 
In all cases, two discriminants are trained to separate the signals from each other, or from the various background
components, using the discriminating variables described in Sect.~\ref{sec:variables}. The resulting two-dimensional BDT output is then used to construct a binned likelihood,
which is fitted to the data to test its compatibility with the SM or BSM Higgs hypotheses, using the fit procedure presented in Sect.~\ref{sec:fit_uncertainties}.

Before the training, the same preselection and some of the selection cuts listed in Table~\ref{tab:comparecuts} are applied to data and on all MC predictions for background and signal. The additional selection requirements adopted for both the 0- and 1-jet categories are $\mll<100$~\GeV\ and on \pTH\ for the spin-2 non-universal coupling models. The loosening of the \mll\ requirement with respect to the one applied in the full event selection is meant to increase the number of MC events for training. 
In the 0-jet category a requirement $\ptll>20$~\GeV\ is applied while the~\dphill\ cut is omitted, whereas the latter is needed in the 1-jet category due to the large DY background.
All background samples are used in the training and each one is weighted by the corresponding production cross-section. 

\subsection{Spin analysis}
The spin analysis presented here follows closely the strategy of Ref.~\cite{HiggsSpin2013} for the 0-jet category, while the 1-jet category has been added and is treated likewise. For each category, one BDT discriminant (called BDT$_0$ in the following) is trained to discriminate between the 
SM hypothesis and the background, and a second one (BDT$_2$) to discriminate between the alternative spin-2 hypotheses and the background. This results 
in five BDT$_2$ trainings for the alternative spin-2 models defined in Sect.~\ref{sec:choicespin2} and one BDT$_0$ training for the SM Higgs boson. 

The distributions of the input variables used for BDT$_0$ and BDT$_2$ in the 0-jet and 1-jet categories, respectively, are shown in
Figs.~\ref{fig:signal_cutvars_0jet} and ~\ref{fig:signal_cutvars_1jet} (see Sect.~\ref{sec:variables}).

The BDT discriminant distributions (also referred to as BDT output distributions) for the 0-jet and 1-jet signal region are shown in Figs.~\ref{figure:spin_sr_bdtoutput_2pmin_1a}
and~\ref{fig:spin_sr_bdtoutput_2pnonuniversal_125_0} for the case of  universal couplings
and of non-universal ones with  $\pTH< 125$~\GeV, respectively.
The plots for non-universal couplings and $\pTH<300$~\GeV\ are very similar to the ones obtained  using the requirement $\pTH<125$~\GeV\ except for the BSM signal 
distribution.  The SM Higgs signal is normalised using 
the SM Higgs-boson production cross-section. Good agreement between data and MC simulation is observed in those distributions once the SM signal is included.

\begin{figure}[htb]
\begin{center}
\subfloat{\includegraphics[width=0.49\textwidth]{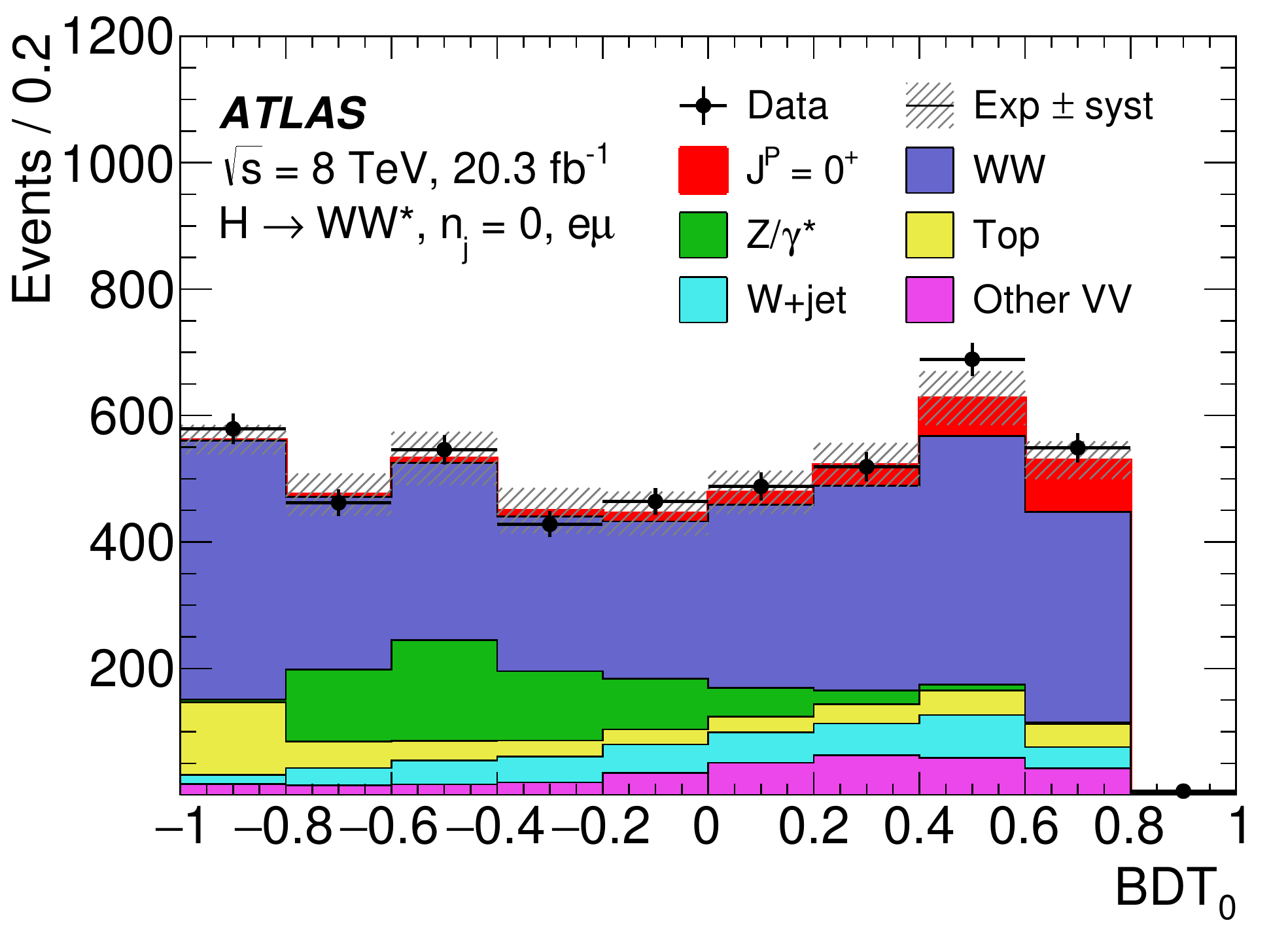}}
\subfloat{\includegraphics[width=0.49\textwidth]{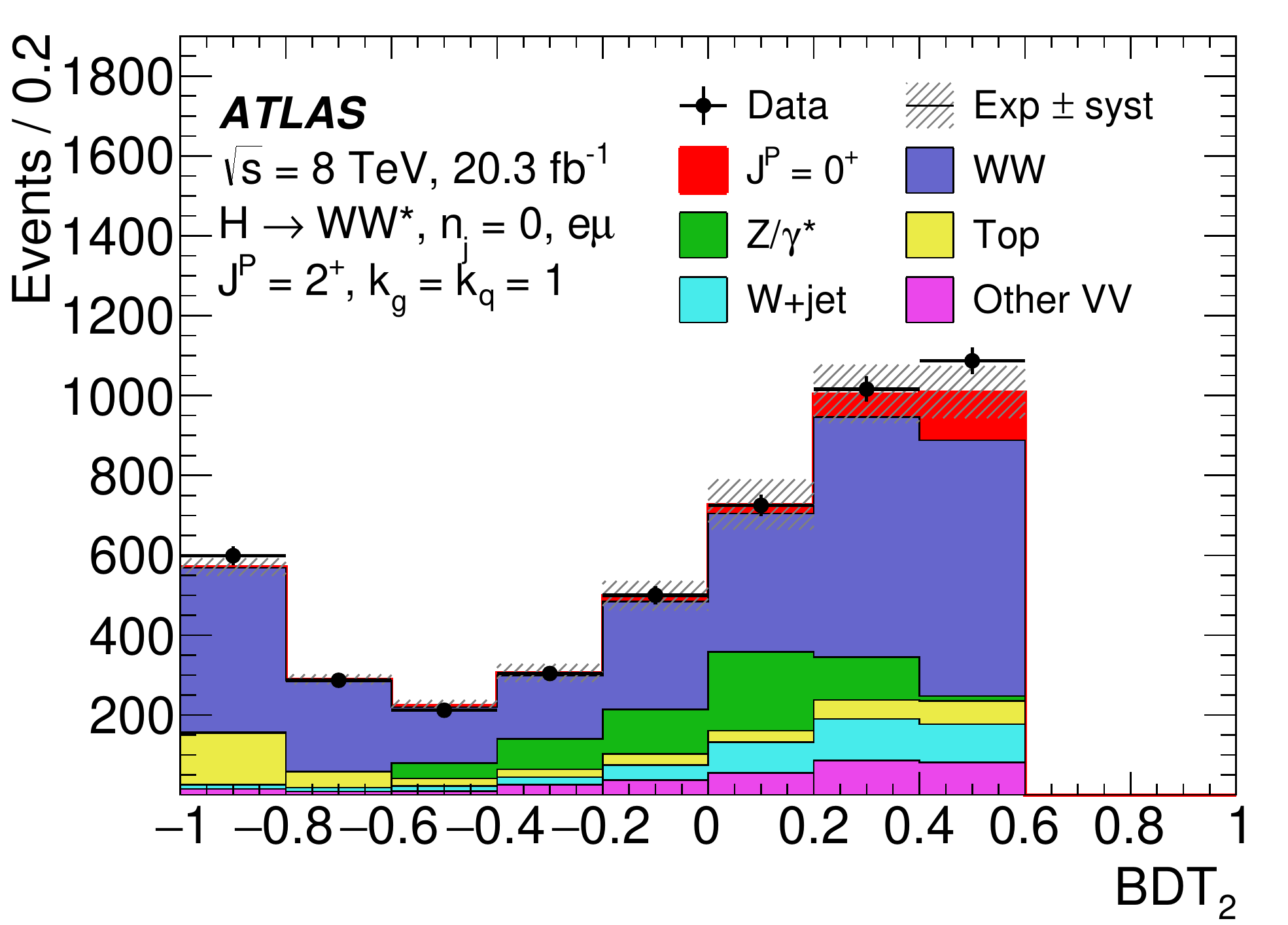}}\\
\subfloat{\includegraphics[width=0.49\textwidth]{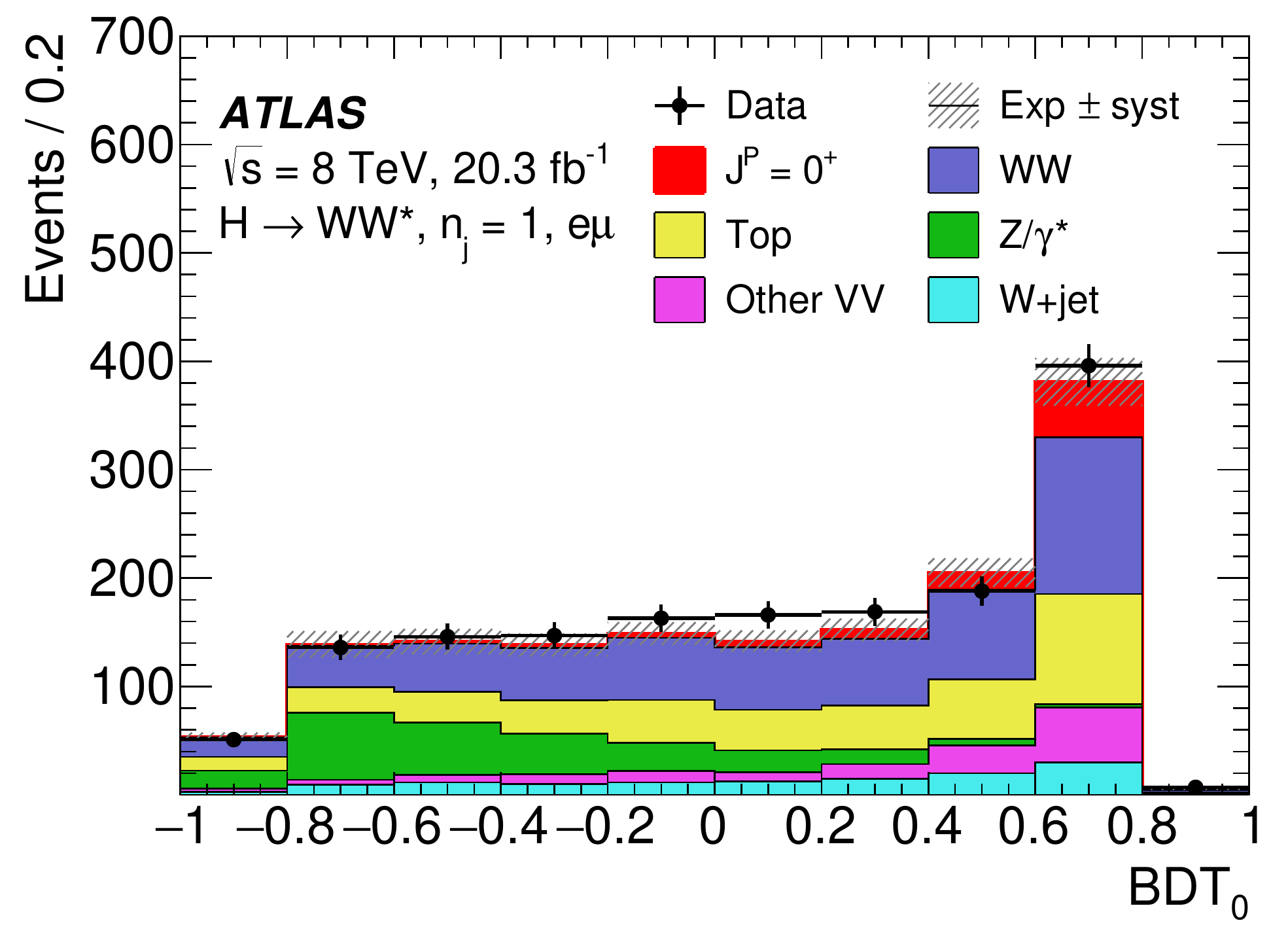}}
\subfloat{\includegraphics[width=0.49\textwidth]{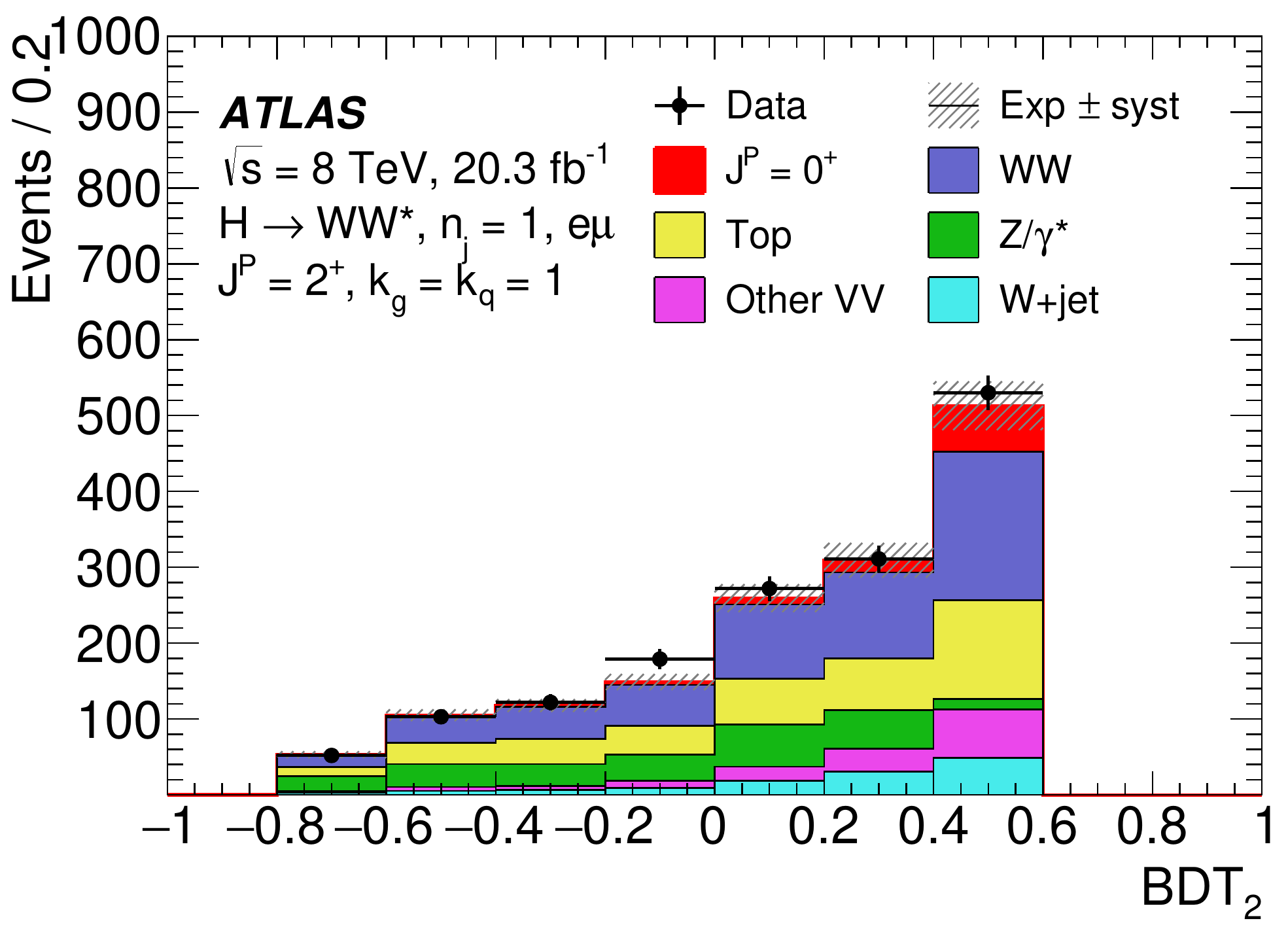}}
\caption{The distributions of the output of BDT$_0$, discriminating between the SM hypothesis and the background, and BDT$_2$, discriminating 
between the alternative spin-2 hypothesis and the background, in the signal region for the spin-2 model with universal couplings. The signal is shown for the SM Higgs-boson hypothesis with $m_H= 125$~\GeV. The background yields are corrected with the normalisation factors determined in the control regions.}
\label{figure:spin_sr_bdtoutput_2pmin_1a}

\end{center}
\end{figure}

\begin{figure}
\centering
\subfloat{\includegraphics[width=0.44\textwidth]{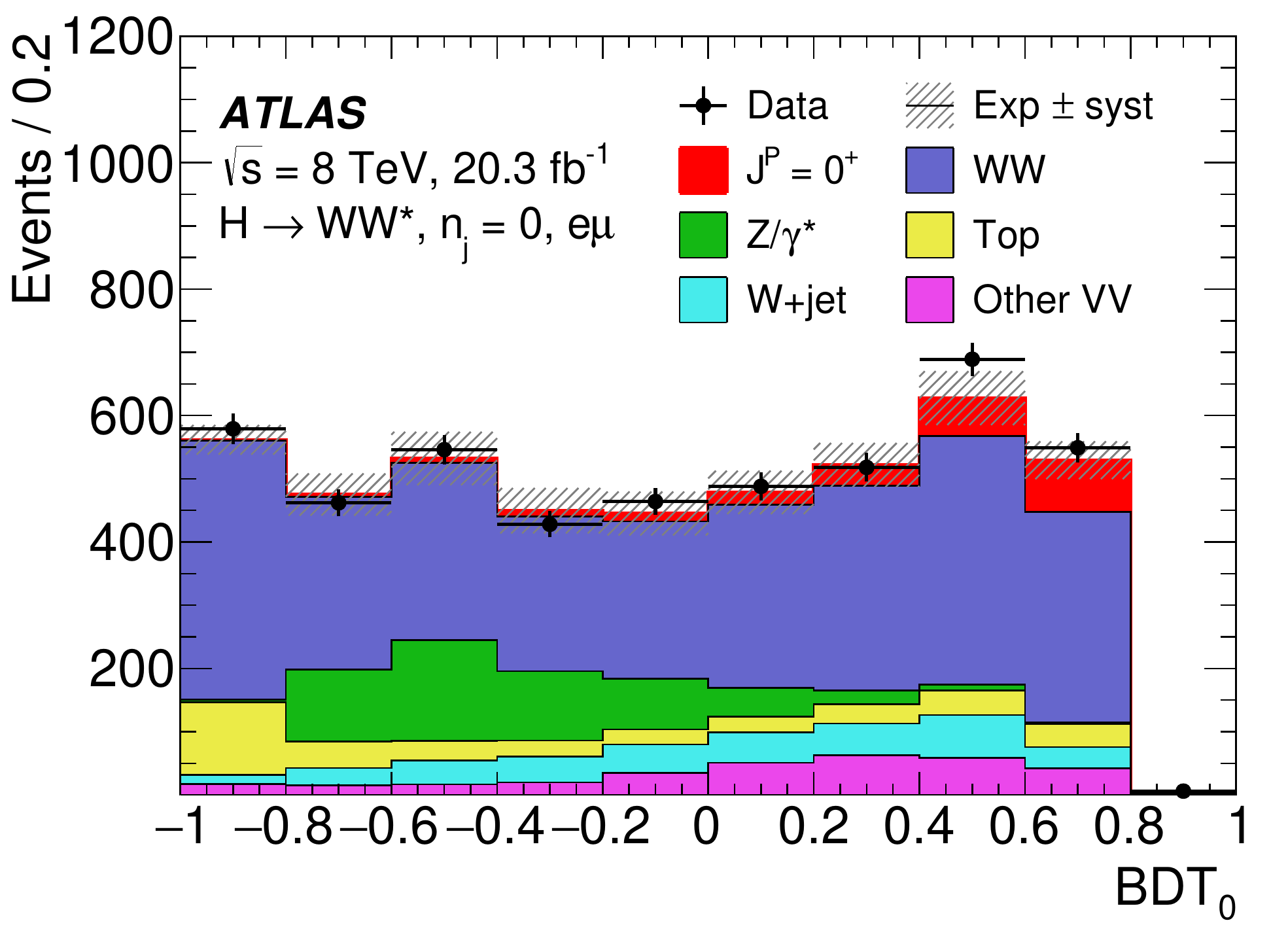}}
\subfloat{\includegraphics[width=0.44\textwidth]{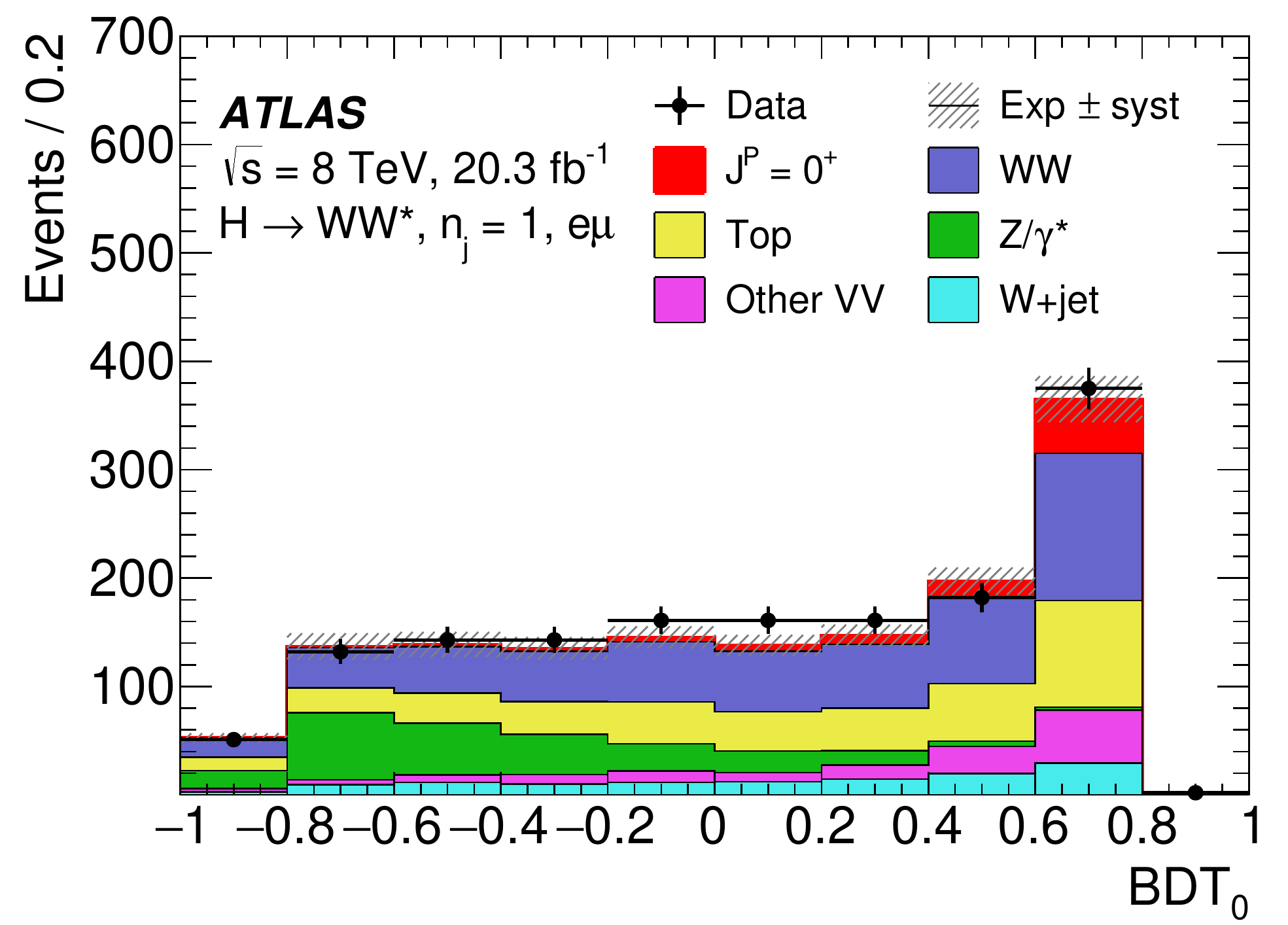}} \\
\subfloat{\includegraphics[width=0.44\textwidth]{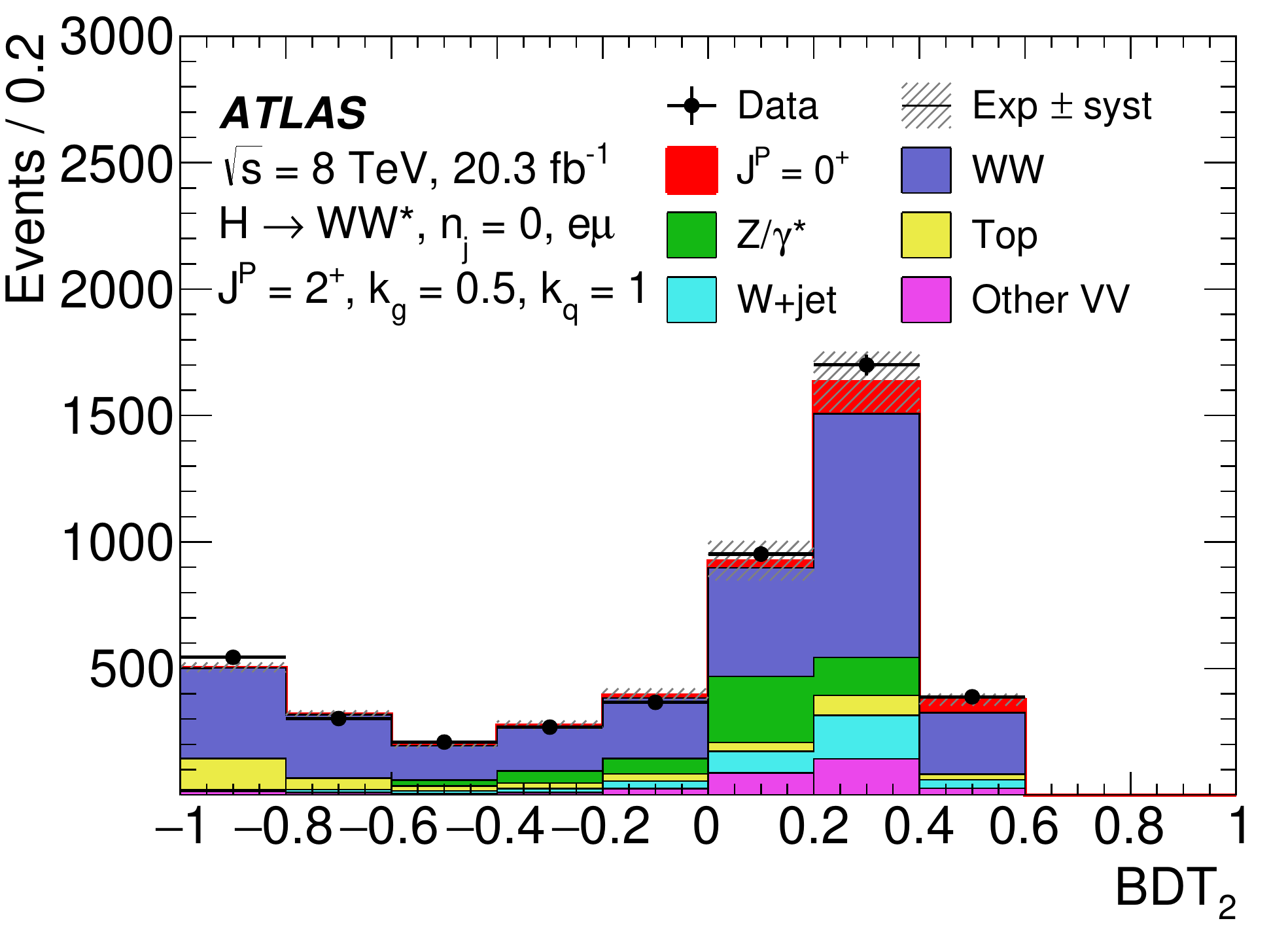}} 
\subfloat{\includegraphics[width=0.44\textwidth]{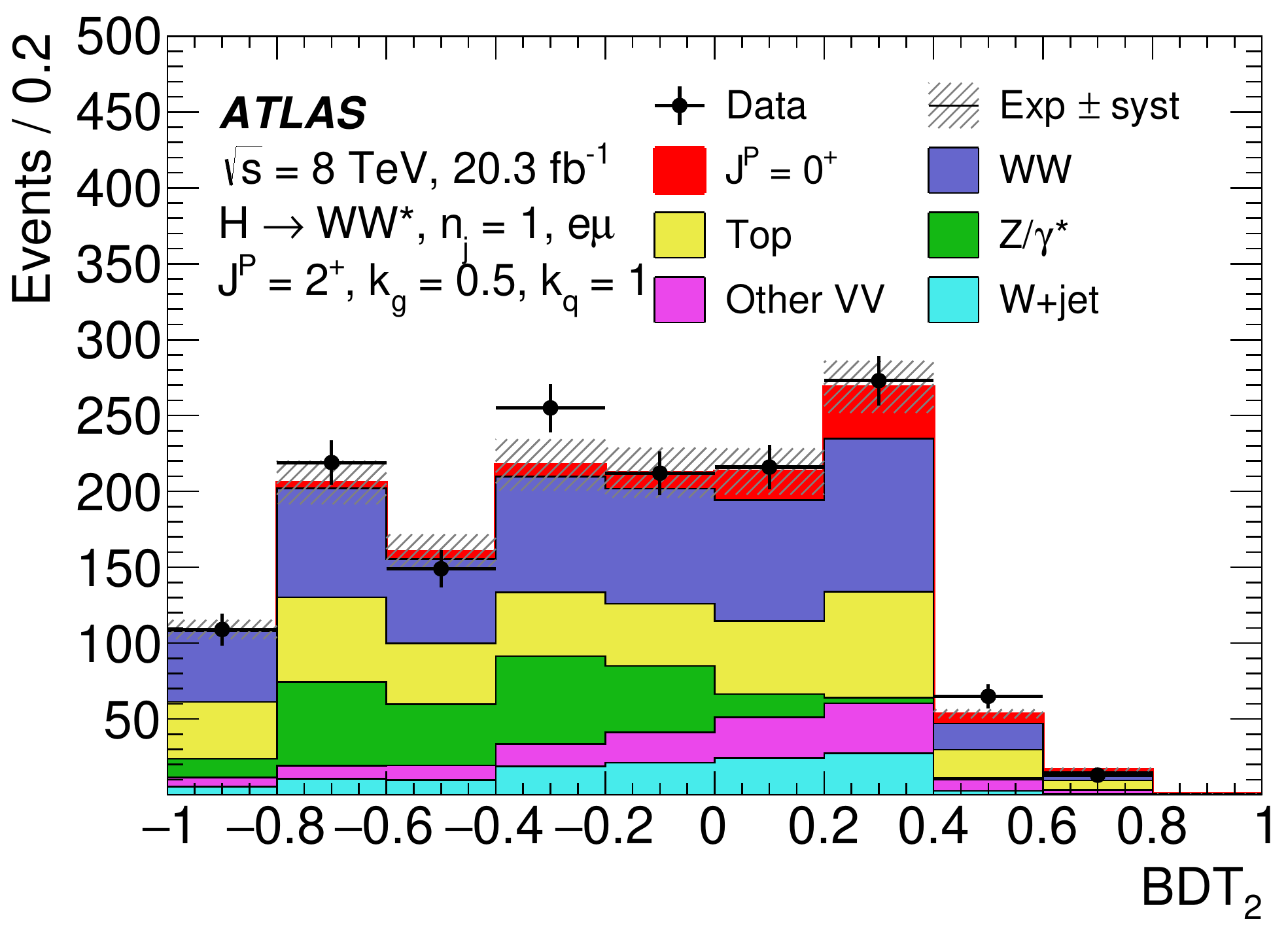}} \\
\subfloat{\includegraphics[width=0.44\textwidth]{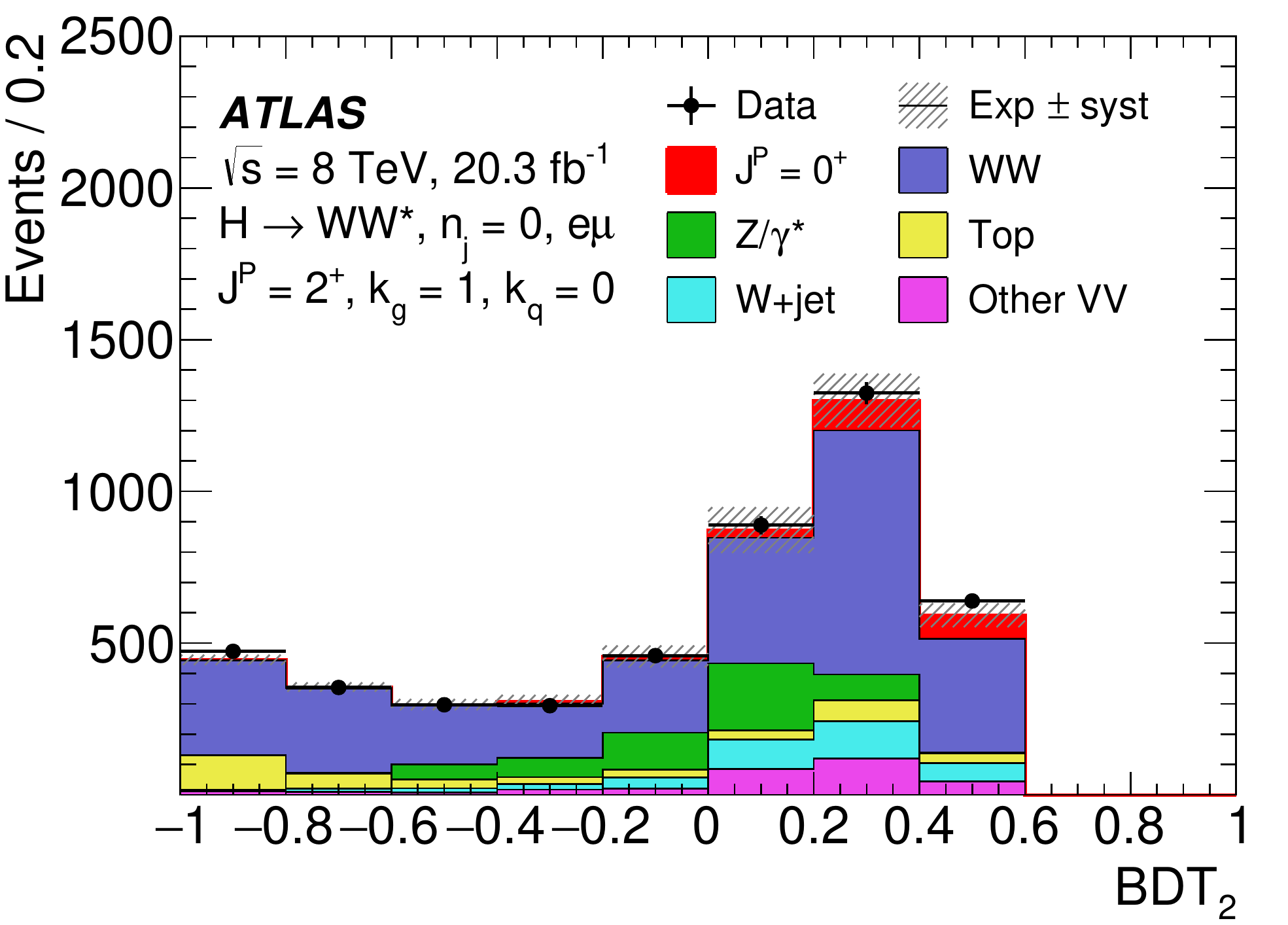}} 
\subfloat{\includegraphics[width=0.44\textwidth]{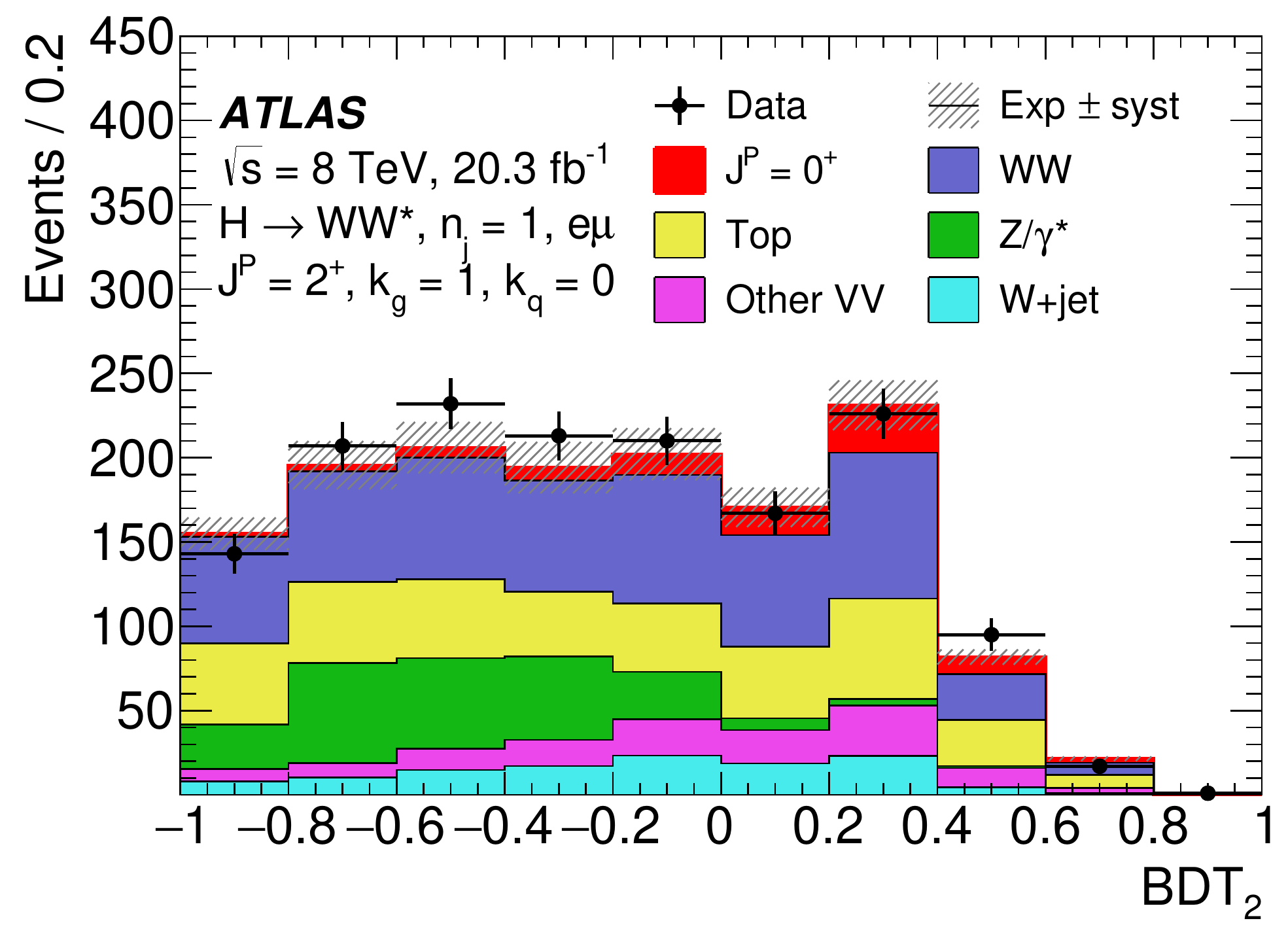}}
\caption{BDT$_0$ and BDT$_2$ output distributions in the signal region for spin-2 models with non-universal couplings. The signal is shown for the SM Higgs-boson hypothesis with $m_H = 125$~\GeV. The $\pTH < 125$~\GeV\ selection requirement is applied to all signal and background
processes, corrected with the normalisation factors determined in the control regions. }
\label{fig:spin_sr_bdtoutput_2pnonuniversal_125_0}
\end{figure}

\subsection{CP analysis}

The CP analysis -- which includes both the fixed-hypothesis test and the CP-mixing scan -- uses only the 0-jet category. In this case as well, two BDT discriminants are trained: the first, BDT$_0$, is identical to the one described above for the spin analysis (SM Higgs-boson signal versus background, using \mll, \ptll, \dphill\ and \mT\ as input variables, as shown in Fig.~\ref{fig:signal_cutvars_0jet}). The second BDT, however, called \bdtcp\ in the following, is trained to discriminate between the SM signal and signal for the alternative hypothesis without any background component.
The training obtained using the two pure  CP-even or CP-odd hypotheses is then applied to all the CP-mixing scenarios. 
As described in Sect.~\ref{sec:variables},
the \bdtcp\ training uses different input variables: \mll, \dphill, \ptll\ and \ptmiss\ for the CP-even scenario, as shown in Fig.~\ref{fig:inputvars_shapes_training_CP_even},
and \mll, \dphill, \Efun\ and \dpt\ for the CP-odd scenario, as shown in Fig.~\ref{fig:inputvars_shapes_training_CP_odd}.

The different training strategy adopted for \bdtcp\ and BDT$_2$ is motivated
by the intrinsic difference between the spin and CP analyses: while, in the former case, 
the spin-2 signal is more background-like (its shape is similar to that of the dominant $WW$ background), in the latter case, the different signal hypotheses 
result in shapes of the input variable distributions which are quite similar to each other, while they remain different from the background shape. Therefore, for the CP analysis, the best separation power is obtained by training \bdtcp\ to discriminate
between the SM and BSM hypotheses. 

The \bdtcp\ output distributions for the SM versus BSM CP-odd and CP-even hypotheses are shown in Fig.~\ref{fig:CPmixBDToutput}. Good agreement between data and MC simulation is also found in this case, once the SM Higgs-boson signal is included. 

\begin{figure}[htb]
\centering
\subfloat{\includegraphics[width=0.49\textwidth]{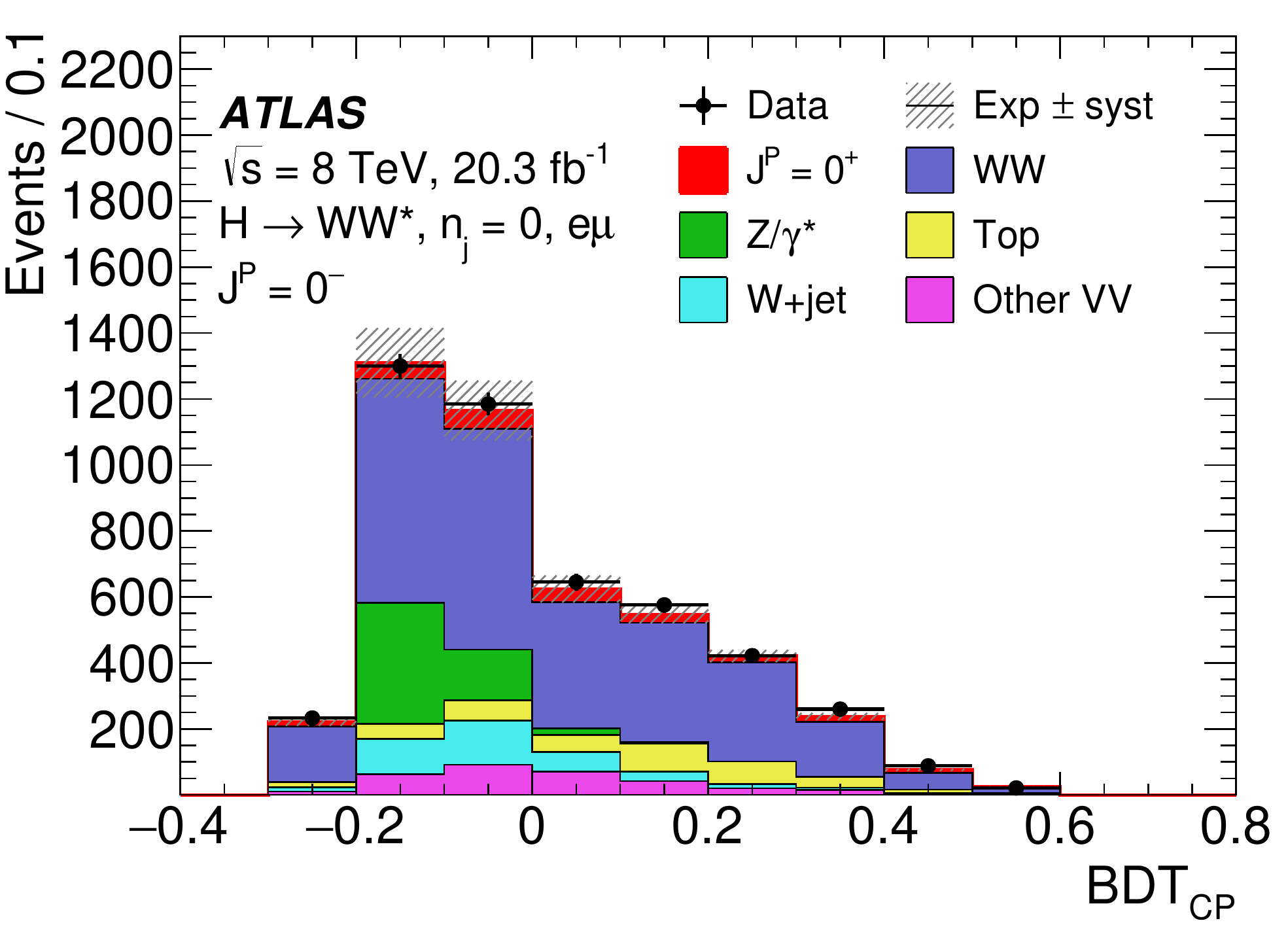}}
\subfloat{\includegraphics[width=0.49\textwidth]{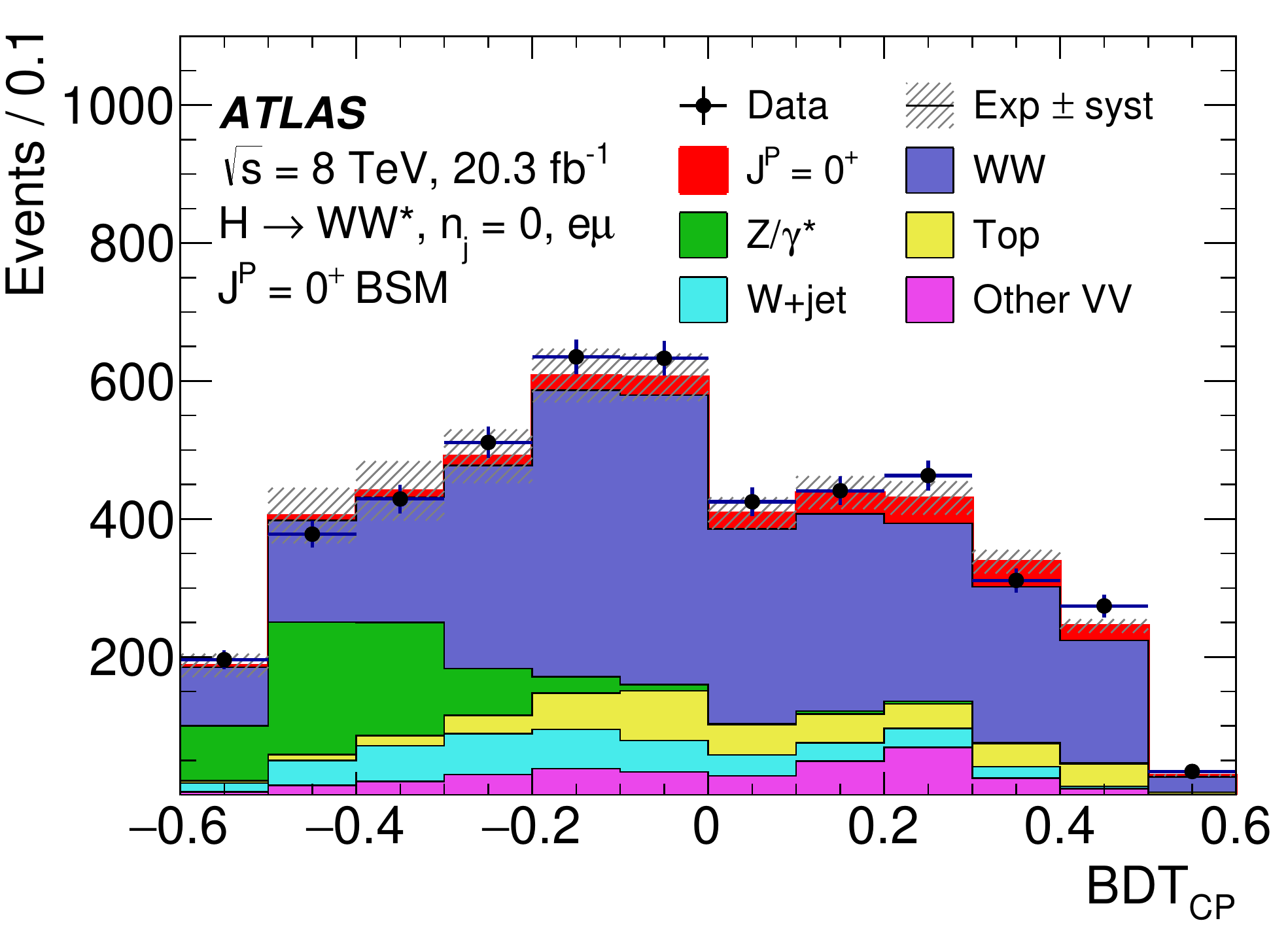}}

\caption{Distributions of the output of \bdtcp, discriminating 
between the SM signal and the signal for the alternative hypothesis, in the signal region for the SM versus BSM CP-odd (left) and SM versus BSM CP-even (right) hypotheses. The signal is shown for the SM Higgs-boson hypothesis with $m_H = 125$~\GeV. The background yields are corrected with the normalisation factors determined in the control regions. }
\label{fig:CPmixBDToutput}
\end{figure}

\section{Fit procedure} 
\label{sec:fit_uncertainties}
This section discusses the statistical approach adopted in this paper. First, the rebinning of the two-dimensional BDT output distribution is discussed. 
The rebinning is applied for both analyses: the fixed-hypothesis tests and the CP-mixing analysis. Afterwards the statistical procedure for the individual analyses is presented.

The two-dimensional BDT$_0$ $\times$ BDT$_2$ output (or BDT$_0$ $\times$ \bdtcp\ for the CP analysis) distribution is unrolled row by row to a one-dimensional distribution. After the unrolling, bins with less than one background event are merged. The latter threshold is applied to the sum of
weighted background events, i.e. after the normalisation to the corresponding cross-section and luminosity and the application of the post-fit scale factors 
to the background processes. This is done independently in the 0-jet and 1-jet categories and for all benchmarks and scans where a retraining of the BDT 
has occurred. Such a procedure is not intended to improve the expected sensitivity per se, rather to stabilise the fit in the presence of a large number of free parameters.

\subsection{Procedure for the fixed-hypothesis test}
\label{sec:stat_spin}

The statistical analysis of the data employs a binned likelihood ${\cal L}(\varepsilon,\mu,{\boldsymbol{\theta}})$ constructed with one parameter of interest, $\varepsilon$, which 
represents the fraction of SM Higgs-boson events with respect to the expected signal yields, and can assume only discrete values $\varepsilon=0$ (for the alternative ALT hypothesis) 
and $\varepsilon=1$ (for the SM hypothesis).  

Template histograms representing the nominal signal and background rates are used to construct ${\cal L}(\varepsilon,\mu,{\boldsymbol{\theta}})$, summing over the bins ($N_{\rm bins}$) of the unrolled BDT output distributions, per jet category in the spin-2 analysis case. 
$S_{\mathrm{SM},i}$ and $S_{\mathrm{ALT},i}$ are the signal yields for the SM and alternative hypothesis, respectively, while $B_i$ 
refers to the total background. Systematic uncertainties are represented through the $N_{\rm sys}$ nuisance parameters ${\boldsymbol{\theta}}$, constrained by the auxiliary measurements 
$\mathcal{A}(\tilde{ \boldsymbol{{\theta}}}|{\boldsymbol{\theta}} )$, where $\tilde{\boldsymbol\theta}$ is the central value of the measurement. The full likelihood can then be written as:

\begin{equation}
\mathcal{L}(\varepsilon, \mu, \vec{\theta}) = {\displaystyle\prod_{i}^{N_{\rm bins}}P(N_{i}|\, \mu(\varepsilon \, S_{\mathrm{SM},i}(\vec{\theta}) + (1-\varepsilon) \, S_{\mathrm{ALT},i}(\vec{\theta})) + B_{i}(\vec{\theta}))}\times\displaystyle\prod_{i}^{N_{\rm sys}}\mathcal{A}({\tilde{\theta}}_{i}|{\theta}_{i})\;.
\end{equation}

The analysis is designed to rely on shape information to distinguish 
between different signal hypotheses. The overall signal normalisation $\mu$ is obtained from the fit and, in the case of the spin analysis, 
as a combination over both jet categories.
Further details of the various likelihood terms can be found in Ref.~\cite{ATLAS-CONF-2014-060}.

The compatibility of the data and two signal hypotheses is then estimated using a test statistic defined as:

\begin{equation}
q = \ln\frac{\mathcal{L}(\varepsilon=1,\hat{\hat{\mu}}_{\varepsilon=1}, \hat{\hat{\boldsymbol\theta}}_{\varepsilon=1})}
                  {\mathcal{L}(\varepsilon=0, \hat{\hat{\mu}}_{\varepsilon=0},\hat{\hat{\boldsymbol\theta}}_{\varepsilon=0})}\;. 
\end{equation}
For both the numerator and denominator, the likelihood is maximised independently over all nuisance parameters to obtain the maximum likelihood estimators
$\hat{\hat{\mu}}$ and $\hat{\hat{\boldsymbol\theta}}$. 
Pseudo-experiments for the two hypotheses ($\varepsilon=0,1$) are used to obtain the corresponding 
distributions of the test statistic $q$ and subsequently to evaluate the $p$-values, which define the expected and observed sensitivities for various hypotheses. The expected
$p$-values are calculated using the fitted signal strength in data, \pexpsm\ for the SM hypothesis, and  \pexpalt\ for the alternative hypothesis. In addition, for the SM hypothesis the expected $p$-value fixing the signal normalisation to the SM prediction, \pexpmusm, is given. The observed $p$-values, \pobssm\ and \pobsalt, are defined as the probability of obtaining a $q$-value smaller (larger) than the observed value under the SM (alternative) signal hypothesis.
Pseudo-experiments are needed because the asymptotic approximation~\cite{asymptotics} does not hold when the parameter of interest, $\varepsilon$ in this case,
takes only discrete values (0 or 1), and in particular $-2 \ln({\cal L})$ does not follow a $\chi^2$ distribution. 

The confidence level (CL) for excluding an alternative BSM hypothesis in favour of the SM is evaluated by means of a CL estimator \cite{cls}:

\begin{equation}
 \mbox{CL}_\text{s}=\frac{p^{\rm ALT}_{\rm obs}}{1-p^{\rm SM}_{\rm obs}}\, ,
\end{equation}
which normalises the rejection power of the alternative hypothesis, $p^{\mbox{\tiny ALT}}$, to the compatibility of the data with the SM case, $1-p^{\mbox{\tiny SM}}$.

\subsection{Procedure for CP-mixing analysis}
\label{sec:stat_cp}

The likelihood definition for the CP-mixing analysis is the same as for the spin analysis, with $\varepsilon=1$ corresponding to the SM
signal hypothesis and $\varepsilon=0$ corresponding to the alternative CP hypothesis.  Whereas for the fixed-hypothesis test, the sensitivities are 
estimated by means of pseudo-experiments and follow the procedure explained above, for the CP-mixing analysis, the simpler asymptotic approximation 
is used, since the fraction of BSM signal events is now considered a continuous parameter. Results using the asymptotic approximation are cross-checked with pseudo-data for a few values of the scan parameter.

The fits to data and to the MC expectation under the SM hypothesis are performed for each value of the scan parameter. Two fits to the SM expectation are 
evaluated: fixing the signal normalisation to the SM expectation and to the observed SM signal normalisation. From the fit, the value of the log-likelihood (LL) 
is extracted, as a function of the CP-mixing fraction. The maximum of the LL curve is determined and its difference from all other values is 
computed, $-2\Delta\mbox{LL}$.  The $1\sigma$ and $2\sigma$ confidence levels are then found at $-2\Delta\mbox{LL}=1$ and $-2\Delta\mbox{LL}=3.84$, respectively.

\section{ Systematic uncertainties}
\label{systematics}

This section describes the systematic uncertainties considered in this analysis, which are divided into two categories: 
experimental uncertainties and theoretical ones which affect the shape of the BDT output distribution. 
The systematic uncertainties specific to the normalisation of individual backgrounds are described in Sect.~\ref{sec:backgrounds}.

\subsection{Experimental uncertainties}

The jet-energy scale and resolution and the $b$-tagging
efficiency are the dominant sources of experimental uncertainty in this category, followed by the lepton resolution, identification and trigger efficiencies and the missing  transverse momentum measurement. The latter is
calculated as the negative vector sum of the momentum
of objects selected according to the ATLAS identification algorithms, such as leptons, photons, and jets, and of the remaining soft objects (referred to as soft terms in the following) that typically have low values of \pT~\cite{ATLAS-CONF-2014-060}.
The various systematic contributions taken into account in the analysis are listed in Table~\ref{tab:systdef}. More information on the experimental systematic uncertainties can be found in Ref.~\cite{ATLAS-CONF-2014-060}.

\begin{table}[h!]
        \caption{\label{tab:systdef} Sources of experimental systematic uncertainty considered in the analysis. The source and magnitude of the uncertainties and their impact on the reconstructed objects is indicated.}
  \begin{center}
    \small
    \begin{tabular}{l|l}
      \dbline
      Source of uncertainty         & Treatment in the analysis and its magnitude \\
      \sgline
    Jet energy scale                 & 1 --7\% in total as a function of jet $\eta$ and \pT \\
				                 & \\
                                                      
    Jet energy resolution   & 5 --20\% as a function of jet $\eta$ and \pT \\
	                                                 & Relative uncertainty on the resolution is 2 --40\% \\
				                 & \\

    $b$-tagging                             & $b$-jet identification: 1 --8\% decomposed in \pT\ bins \\
                                                       & Light-quark jet misidentification: 9 --19\% as a function of $\eta$ and \pT \\
                                                       & $c$-quark jet misidentification: 6 --14\% as a function of \pT \\                                                                                                              
    					        & \\

    Leptons                                       & Reconstruction, identification, isolation, trigger efficiency: below 1\% \\
    						  & except for electron identification: 0.2 --2.7\% depending on $\eta$ and \pT \\
    						   & Momentum scale and resolution: $< 1$\% \\ 
    						   & \\
						   
    Missing transverse momentum   & Propagated jet-energy and lepton-momentum scale uncertainties \\
				                     & Resolution (1.5 --3.3 \GeV) and scale variation (0.3 --1.4 \GeV) \\
				                    & \\
    Pile-up                                        & The number of pile-up events is varied by 10\% \\
    	                                                & \\

    Luminosity                                 & 2.8\% \cite{lumi} \\
    \dbline
    \end{tabular}

  \end{center}
\end{table}

In the likelihood fit, the experimental uncertainties are varied in a correlated way across all backgrounds and across signal and 
control regions, so that the uncertainties on the extrapolation factors $\alpha$ described in Sect.~\ref{sec:backgrounds} are correctly 
propagated. All sources in Table~\ref{tab:systdef} are analysed to evaluate their impact on both the yield normalisation and on the 
shape of the BDT discriminant distributions. Shape uncertainties are ignored if they are smaller than 5\% (smaller than the statistical 
uncertainty) in each bin of the distributions under study. Normalisation uncertainties are ignored as well if they are below 0.1\%.

 
\subsection{Modelling uncertainties}

The dominant background is SM $WW$ production, and therefore uncertainties on the shape and yield in 
the signal region for this background require special attention. The uncertainties on the $WW$ normalisation are 
discussed in Sect.~\ref{sec:wwcr}; the shape uncertainties are addressed in this section. 

An important uncertainty arises from the modelling of the shape of the $WW$ background in the signal region, which is obtained using the  same procedure adopted in the evaluation of the theoretical uncertainty on  the $WW$ extrapolation parameter. The scale uncertainty on the MC prediction of the BDT discriminants was studied by varying the factorisation 
and renormalisation scales up and down by a factor of two.  The parton shower and generator uncertainties are estimated by 
comparing the \HERWIG\ and \PYTHIA\ parton shower programs and by  comparing \POWHEG+\HERWIG\ and \amcatnlo+\HERWIG, respectively. Finally, the PDF uncertainty is estimated 
by combining the CT10 PDF error set  with the difference between the central values of NNPDF2.3 and CT10. 
The procedure is repeated for each of the final BDT output distributions and for each benchmark of the spin and parity analyses. 

Modifications to the shape of the final BDT distribution from PDF and scale variations are found to be negligible, and well within the statistical 
uncertainty of the Monte Carlo predictions. Therefore they are included in the fit model only as overall normalisation effects.
The parton shower and generator uncertainties were found to be statistically significant; therefore, a bin-by-bin shape uncertainty is applied. 

The interference between the $gg \rightarrow WW$ and the $gg \rightarrow H$ processes is not taken into account in this study because of its negligible effect. In fact it results in a 4\% decrease in the total yield of events after the selection criteria and is of the same order as in Ref.~\cite{ATLAS-CONF-2014-060}. These results confirm the expectations in Ref.~\cite{Kauer:2012hd}.

The signal final-state observables are affected by the underlying Higgs-boson \pT\ distribution. The Higgs-boson \pT\ distribution for 
a spin-0 particle is given by the \pTH-reweighted \POWHEG+\PYTHIA\ generator prediction as mentioned in Sect.~\ref{sec:detector_samples}.
All spin-0 samples are reweighted to the same \pTH\ distribution to avoid any impact of the difference in the Higgs-boson \pT\ predictions between \mgaMC\ and \POWHEG\ on the CP-analysis results. No additional shape uncertainty is considered.
For the spin-2 benchmarks no theoretical uncertainties on the Higgs-boson \pTH\ are considered, because they are negligible compared to the effect of the 
choice of \pTH\ requirement in the non-universal couplings models. 

\subsection{Ranking of systematics}

The impact of each systematic variation on the $\mbox{CL}_{\text s}$ estimator gives the measure of the relevance of the systematic uncertainty
on the obtained result. The systematic uncertainties that are found to be most important in the various fixed-hypothesis tests are listed for the 
different cases in Table~\ref{tab:ranking}. 

\begin{table}
\caption{\label{tab:ranking} From top to bottom, systematic uncertainties (in \%) with the largest impact on the spin-2 universal couplings, BSM CP-odd and CP-even Higgs-boson fixed-hypothesis tests. This ranking is based on the impact of each systematic uncertainty on the $\mbox{CL}_{\text s}$ estimator (see Sect.~\ref{sec:fit_uncertainties}).
For the exact meaning of the different uncertainties related to the misidentified lepton rates (the \Wjets\ background estimate uncertainty), see Sect.~\ref{sec:wjets} and Ref.~\cite{ATLAS-CONF-2014-060}.}
\centering
 \small{ 
  \begin{tabular}{cc|cc|cc}

\dbline
  \multicolumn{2}{c|}{Spin-2} &  \multicolumn{2}{c|}{BSM CP-odd} & \multicolumn{2}{c}{BSM CP-even}\\
    \sgline
$WW$ generator:          &     2.6    & $WW$ generator:      & 0.73    & $WW$ UE/PS: & 21\\
$\pT^Z$ reweighting:    &   1.2     & $WW$ UE/PS:            & 0.66    &  Misid. rate (elec. stats): &9.2\\
Misid. rate (elec. stats):     &  1.1      & QCD scale $Wg^*$:  &  0.45   & Misid. rate (elec. flavour):&8.4\\
Misid. rate (elec. flavour):    &  1.0     & $\pT^Z$ reweighting: & 0.43    & Misid. rate (muon flavour): &7.4\\
 $WW$ UE/PS:               & 0.86    &  QCD scale $VV$ :      & 0.39   & Misid. rate (muon stats):  &7.3\\
Misid. rate (muon stats):    & 0.81    & QCD scale $Wg$:       & 0.38  & Misid. rate (elec. other): &7.3\\
 \Ztt\ generator:               &  0.76   & Misid. rate (elec. stats):    & 0.37   & $WW$ PDF $qq$-production: &6.9 \\
Misid. rate (muon flavour):  & 0.75     & Misid. rate (elec. other):   & 0.34  & $WW$ PDF $gg$-production: & 6.9\\
Misid. rate (elec. other):    &  0.67    & Misid. rate (elec. flavour):   & 0.33   & $WW$ generator: & 3.6\\
\dbline
 \end{tabular}
 }

 \end{table}

The $WW$ modelling uncertainty dominates in all three benchmarks, and another common large uncertainty is due to the $W$+jets 
background estimate.  The spin-2 and CP-odd analyses are affected by the  \Ztt\ modelling uncertainty. In addition, the CP-odd 
analysis is impacted by the modelling uncertainties on the non-$WW$ 
background. The impact of systematics on the $\mbox{CL}_\text{s}$ estimator is
larger for the CP-even case than for other benchmarks because of the lower sensitivity of the CP-even analysis.

\section{Results} 
\label{sec:results}

The results of the studies of the spin and parity quantum numbers are presented in this section.
The SM $J^{P}=0^{+}$ hypothesis is tested against several alternative spin/parity hypotheses,
and the mixture of the SM Higgs and a BSM CP-even or CP-odd Higgs bosons is studied by scanning all possible mixing combinations.

This section is organised as follows. The event yields and the BDT output distributions after the
fit to data are presented in Sect.~\ref{sec:postfit_yields}. The results of the fixed-hypotheses tests
for spin-2 benchmarks are discussed in Sect.~\ref{sec:spin2unbresults} and the results for 
spin-0 and CP-mixed tests are shown in Sect.~\ref{sec:CPunbresults}.

\subsection{Yields and distributions}
\label{sec:postfit_yields}

The post-fit yields for all signals and backgrounds are summarised in Table~\ref{tab:postfitNorm} for the spin and CP analyses. They account 
for changes in the normalisation factors and for pulls of 
the nuisance parameters. All the systematic uncertainties discussed in Table~\ref{tab:alpha_uncert} and Sect.~\ref{systematics}
are included in the fit.
The fitted signal yields vary significantly in the BSM scenarios because of the differences in the shapes of the input variable distributions between the benchmark
models. A striking example is given by the benchmark models with non-universal couplings: the fitted signal yield varies considerably between the $\pTH < 125\GeV$ 
and $\pTH < 300\GeV$ selections because of the presence of the tail at high \pTH\ values  discussed in Sect.~\ref{sec:spin2theory}. The yield fitted 
under the SM hypothesis, $270\pm 70$ events (see Table~\ref{tab:postfitNorm}), is in good agreement with the signal expectation of 238 events, 
corresponding to the ggF signal strength measured in Ref.~\cite{ATLAS-CONF-2014-060}.

\begin{table}[ht!]
\caption{Post-fit event yields for the 0- and 1-jet categories for various signal hypotheses.
The number of events observed in data, the signal and the total background yields, including their respective post-fit systematic
uncertainties, are shown in the top part of the table, assuming in each case the alternative signal hypothesis. 
The spin-2 \kg\ = \kq\ benchmark is used as an example in the bottom part of the table, to show in more detail the results under the SM Higgs-boson hypothesis.
For this fit, the individual backgrounds are listed for completeness (see Sect.~\ref{sec:backgrounds}).}

\begin{center}
\begin{tabular}{c|cc|cc}
\dbline
\multirow{2}*{Benchmark} & \multicolumn{2}{c|}{Signal} & \multicolumn{2}{c}{Total background}  \\
& 0-jet & 1-jet & 0-jet & 1-jet \\
\sgline

\kg\ = \kq &  $360 \pm 100$ & $126 \pm 34$ & $4370 \pm 240$ & $1430 \pm 60$ \\

\sgline
\kg\ = 0.5, \kq\ = 1, $\pTH < 125 \GeV$ & $300 \pm 100$ &  $103 \pm 33$ & $ 4430 \pm 240$ &  $1390 \pm 60$\\
\kg\ = 0.5, \kq\ = 1, $\pTH < 300 \GeV$ & $230 \pm  80$ &  $ 82 \pm 29$ & $ 4490 \pm 230$ &  $1460 \pm 70$ \\

\sgline

\kg\ = 1, \kq\ = 0, $\pTH < 125 \GeV$ & $320 \pm 90$& $111 \pm 32$ & $ 4410 \pm 240$ & $ 1390 \pm 60$ \\
\kg\ = 1, \kq\ = 0, $\pTH < 300 \GeV$ & $200 \pm 80$ & $71 \pm 28$ & $ 4520 \pm 240$ & $ 1480 \pm 70$ \\

\sgline 

BSM CP-odd  & $240 \pm 80$ & -- & $ 4490 \pm 260$  & -- \\
\sgline 

BSM CP-even & $180 \pm 60$ & -- & $ 4530 \pm 240$  & -- \\

\dbline
\end{tabular}
\end{center}

\begin{center}
\begin{tabular}{c|c|cc|ccccc}
\dbline

 &\multirow{2}*{Data} &\multirow{2}*{Signal} &  \multirow{2}*{Tot. bkg.}   & \multirow{2}*{$WW$} & \multirow{2}*{Top} & \multirow{2}*{DY} & \multirow{2}*{\Wjets} & \multirow{2}*{Other}\\
 &&&&&&\\
 \sgline
 SM 0-jet & 4730 & $270 \pm 70$ & $ 4460 \pm 240$ & 2904 & 376 & 464 & 370 & 345\\
 SM 1-jet & 1569 & $ 95 \pm 26$ & $ 1450 \pm 70$  &  607 & 355 & 233 & 124 & 133 \\
 \dbline
\end{tabular}
\end{center}

\label{tab:postfitNorm}

\end{table}

\subsection{Spin-2 results}
\label{sec:spin2unbresults}

The compatibility of the spin-2 signal model with the observed data is calculated following the prescription explained in 
Sect.~\ref{sec:stat_spin} for five different benchmarks discussed in Sect.~\ref{sec:choicespin2}. The 
expected distributions of the test statistic $q$, derived from pseudo-experiments, are shown for the universal couplings case
in Fig.~\ref{fig:spin2teststat_unblinded} for 0- and 1-jet combined. The $q$ distributions 
are symmetric and have no overflow or underflow bins.
The expected and observed significances and \CLs\ are summarised in Table~\ref{tab:results}. The expected significance \pexpsm\ using the observed SM normalisation is higher than \pexpmusm, because the observed SM yields in Table~\ref{tab:postfitNorm} are larger than the expected SM yields in Table~\ref{tab:Cutflow}. The SM hypothesis is favoured in all tests in data and the alternative model is 
disfavoured at 84.5\% CL for the model with universal couplings and excluded at 92.5\% to 99.4\% CL 
for the benchmark models with non-universal couplings. The exclusion limits for non-universal couplings are stronger for
a \pTH\ cut above 300~\GeV\ because of the enhanced sensitivity at high values of the Higgs-boson \pt.

The one-dimensional distribution of the unrolled post-fit BDT output distribution is presented in 
Fig.~\ref{fig:spin2BDTbkgsub125_0j} for the  $\kg =1$, $\kq =0$ and $\pTH<125$~\GeV\ scenario in the 0-jet case.  The distributions 
are shown for the SM and alternative signal hypotheses separately and compared with the data after the subtraction of  all backgrounds. Both the signal and background 
yields are normalised to the post-fit values. The distributions are ordered in terms of increasing signal yield and, for visualisation purposes,
only contain 
bins that have at least three signal events and a signal-to-background ratio of at least 0.02.

\begin{table}
\caption{Summary of expected and observed sensitivities for various alternative spin/CP benchmarks compared to the SM Higgs-boson 
hypothesis. The expected and observed $p$-values and the observed $1-\CLs$ value as defined in Sect.~\ref{sec:fit_uncertainties} are shown 
for various benchmarks. The results are computed taking into account systematic uncertainties, using the combined 0-jet and 1-jet categories for 
the spin analysis and only the 0-jet category for the CP analysis.}
\begin{center}
\begin{tabular}{l|ccccc|c}
\dbline
Channel & \pexpmusm & \pexpsm & \pexpalt &\pobssm &\pobsalt & $1-\CLs$ \\
\sgline
\multicolumn{7}{c}{Spin-2, $\kg\ = \kq$}\\
\sgline
0+1-jet & 0.131 & 0.039 & 0.033 & 0.246 & 0.117& 84.5\%  \\
\sgline
\multicolumn{7}{c}{Spin-2, $\kg\ = 0.5$, $\kq\ = 1$, $\pTH < 125 \GeV$}\\
\sgline
0+1-jet & 0.105 & 0.047 & 0.022& 0.685& 0.007 & 97.8\%  \\
\sgline
\multicolumn{7}{c}{Spin-2, $\kg\ = 0.5$, $\kq\ = 1$, $\pTH < 300 \GeV$}\\
\sgline
0+1-jet & 0.023 &0.014 & 0.004 & 0.524 & 0.003 & 99.3\% \\
\sgline
\multicolumn{7}{c}{Spin-2, $\kg\ = 1$, $\kq\ = 0$, $\pTH < 125 \GeV$}\\
\sgline
0+1-jet  & 0.109 &0.041 & 0.029 & 0.421 & 0.044 & 92.5\% \\
\sgline
\multicolumn{7}{c}{Spin-2, $\kg\ = 1$, $\kq\ = 0$, $\pTH < 300 \GeV$}\\
\sgline
0+1-jet & 0.015 &0.016 & 0.004 & 0.552 & 0.003 & 99.4\% \\
\sgline

\multicolumn{7}{c}{BSM CP-odd}\\
\sgline
0-jet & 0.078 &0.062& 0.032 & 0.652 & 0.012 & 96.5\%\\
\sgline
\multicolumn{7}{c}{BSM CP-even}\\
\sgline
0-jet & 0.271 &0.310 & 0.287 & 0.907 & 0.027 & 70.8\%\\
\dbline
\end{tabular}

\end{center}

\label{tab:results}
\end{table}

\begin{figure}
\centering
\subfloat{\includegraphics[width=0.49\textwidth]{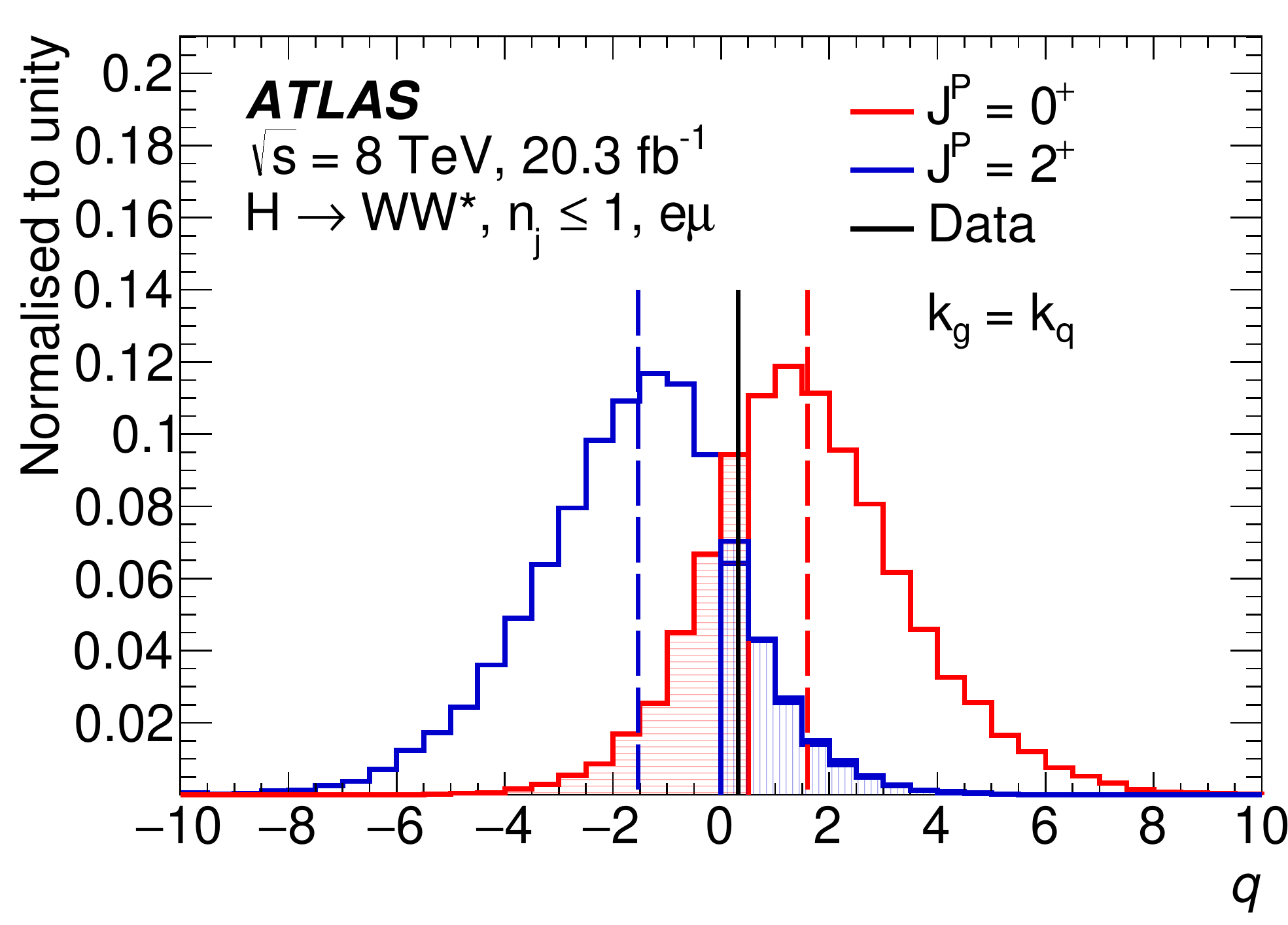}}
\caption{Test-statistic distribution for the spin-2 benchmark with universal couplings $(\kg = \kq)$ including all systematic uncertainties, with 0- and 1-jet categories combined. The median of the expected distributions for the SM (dashed red line) and the spin-2 Higgs-boson signal (dashed blue line) is also shown, together with the observed result (solid black line) from the fit to the data. The shaded areas are used to compute the observed $p$-values.}
\label{fig:spin2teststat_unblinded}
\end{figure}

 \begin{figure}
\centering

\subfloat{\includegraphics[width=0.49\textwidth]{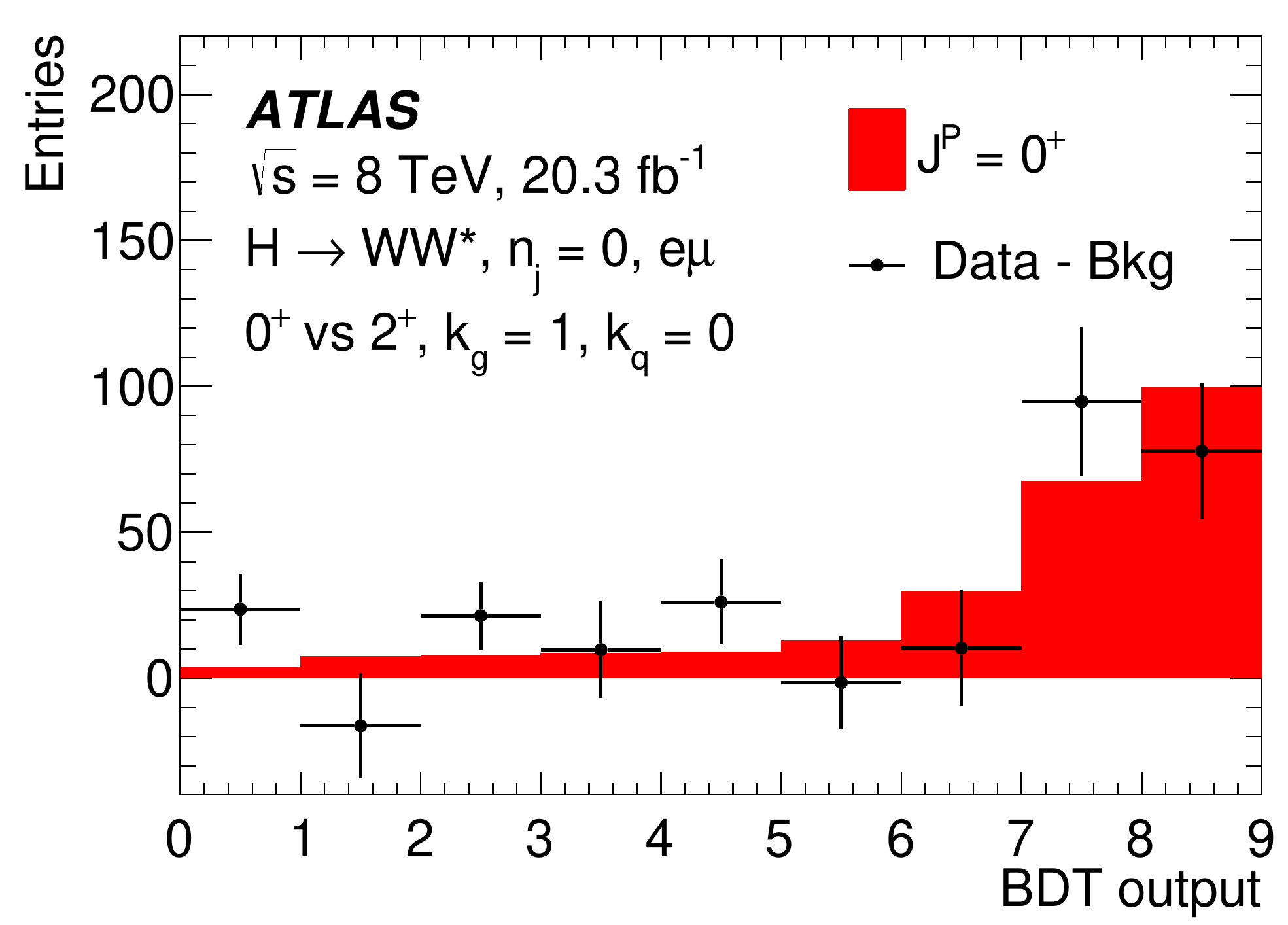}}
\subfloat{\includegraphics[width=0.49\textwidth]{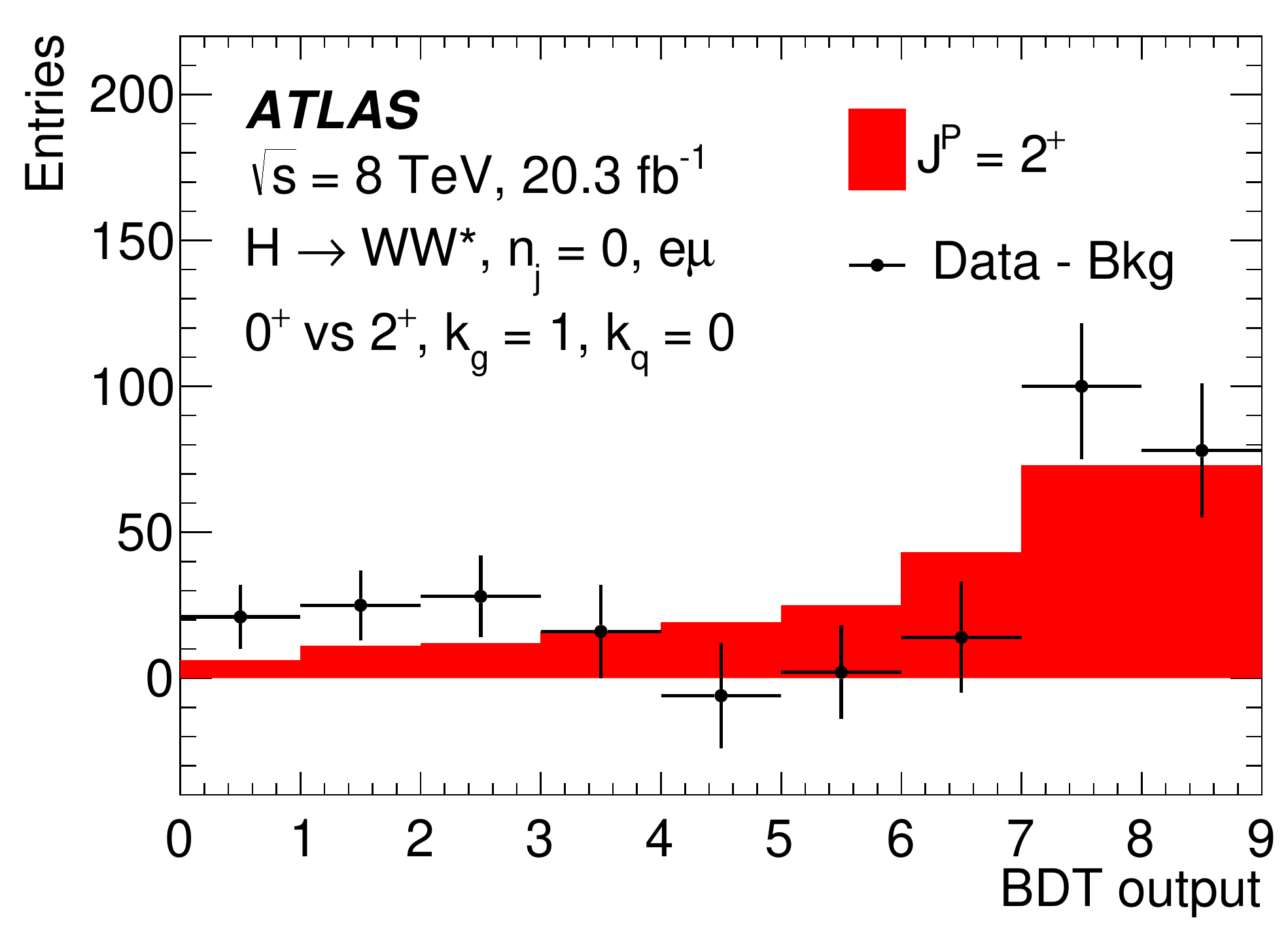}}

\caption{The unrolled one-dimensional BDT output after background subtraction and using post-fit normalisations, in the case of the spin-2 benchmark with non-universal couplings ($\kg =1, \kq = 0$), requiring the Higgs-boson \pT\ to be below 125~\GeV. The background yields are taken from the fit results, assuming the SM signal hypothesis in the left-hand plot, and the alternative spin-2
hypothesis in the right-hand plot.}
\label{fig:spin2BDTbkgsub125_0j}
\end{figure}

\subsection{Spin-0 and CP-mixing results}
\label{sec:CPunbresults}

Similar to the spin-2 fixed-hypothesis tests, the CP-even BSM Higgs and the CP-odd BSM Higgs-boson hypotheses
are tested against the SM Higgs-boson hypothesis. The expected distributions of the test statistic $q$, derived from 
pseudo-experiments for the SM versus BSM CP-odd and CP-even pure states, are shown in Fig.~\ref{fig:cpteststat_unblinded}. 
The distributions are symmetric and have no overflow or underflow bins. The overlap of the test-statistic distributions 
for the SM hypothesis and the alternative hypothesis indicates the sensitivity of the analysis to distinguish them. 
The expected sensitivity is higher for the CP-odd hypothesis than for the CP-even hypothesis. The expected and observed 
significances and \CLs\ values are summarised in Table~\ref{tab:results}. The expected significances \pexpsm\ and \pexpmusm are similar, 
because the observed and the expected SM yields are similar for the spin-0 fixed hypothesis test.  The SM hypothesis is favoured in 
both tests and the alternative hypothesis can be excluded at 96.5\% CL for the CP-odd Higgs boson and disfavoured at 70.8\% CL for 
the CP-even BSM Higgs boson.

\begin{figure}
\centering
\subfloat{\includegraphics[width=0.49\textwidth]{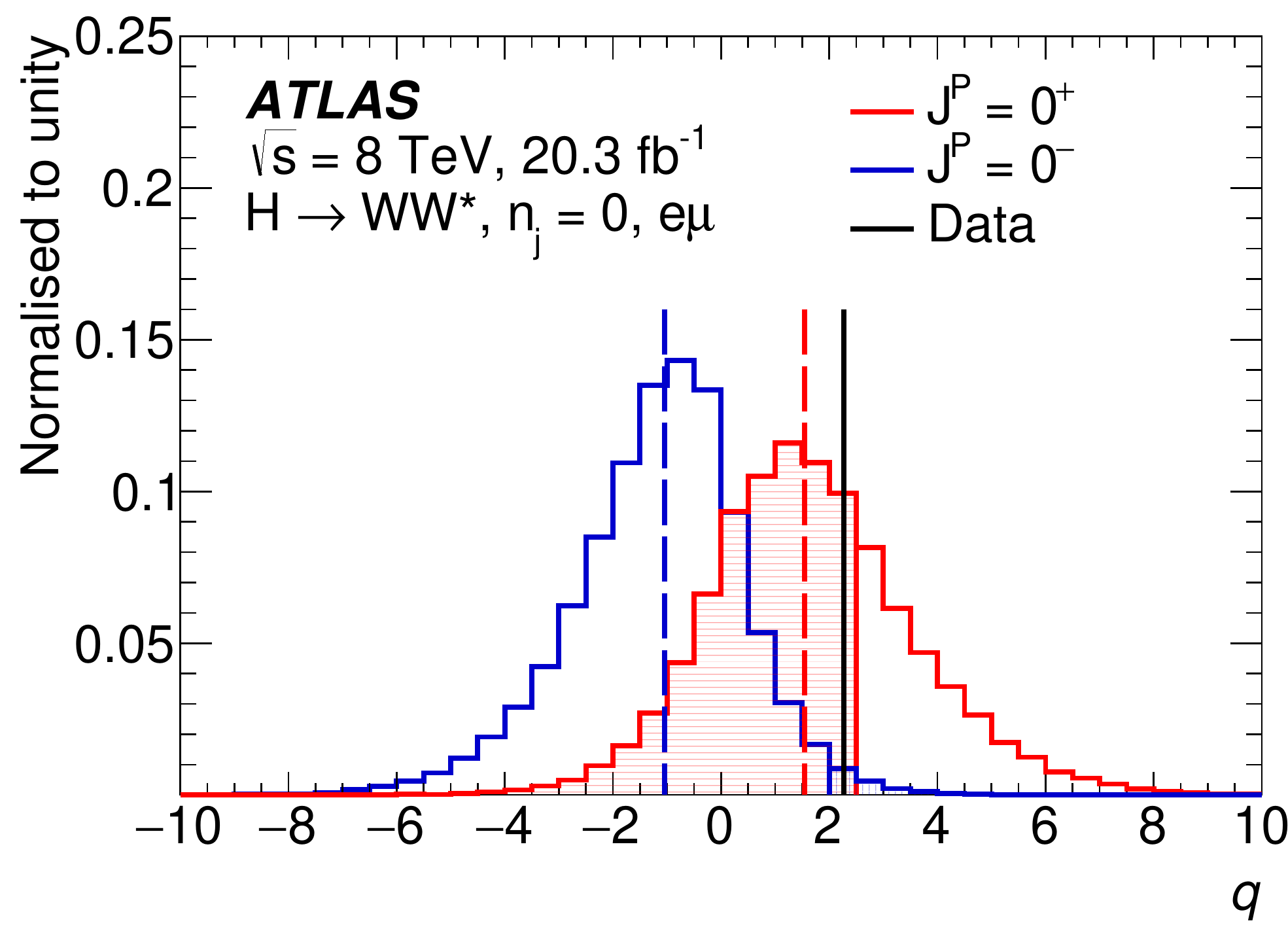}}
\subfloat{\includegraphics[width=0.49\textwidth]{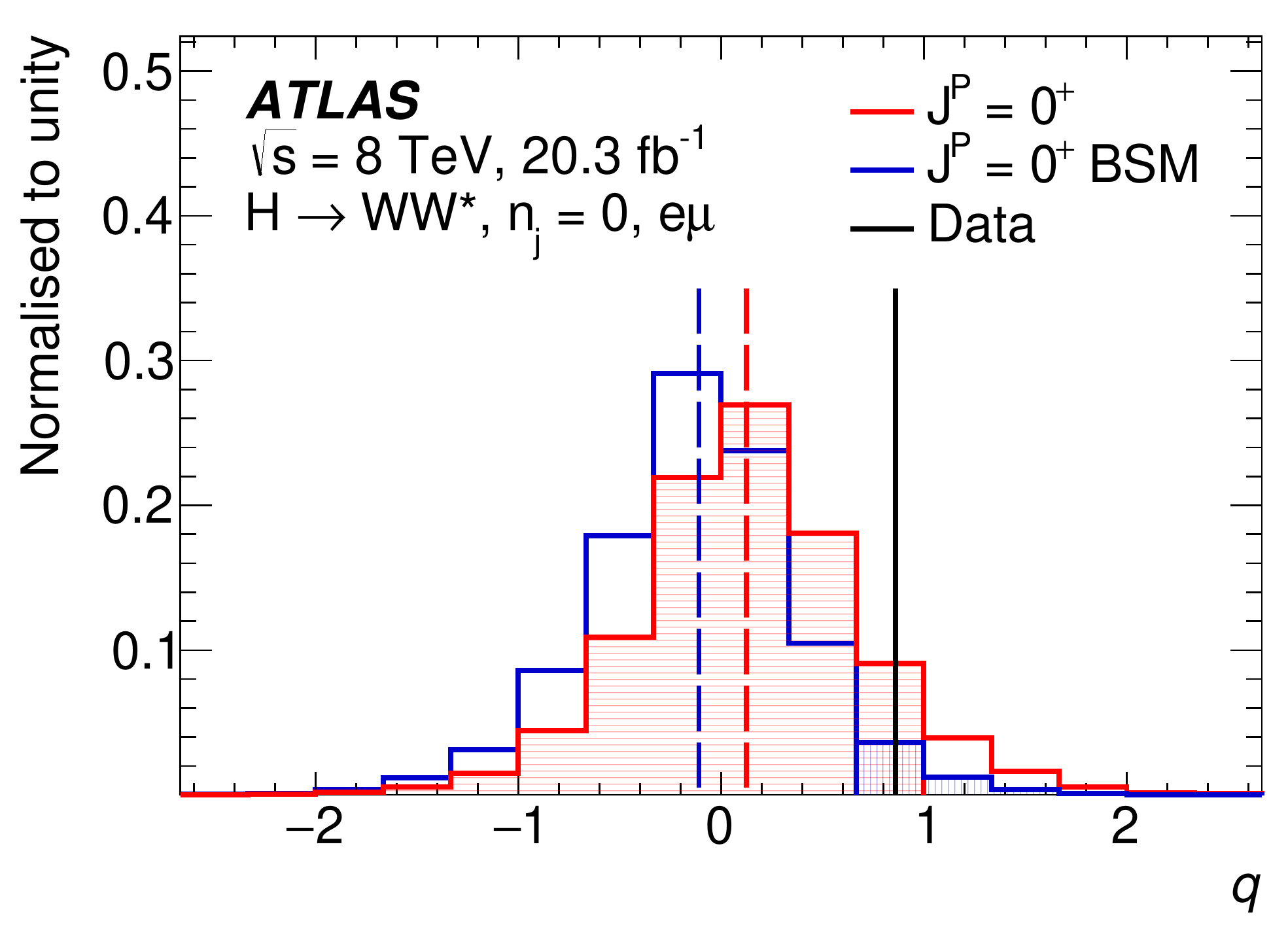}}   

\caption{Test-statistic distribution for the pure BSM CP-odd (left) and BSM CP-even (right) benchmarks, including all systematic uncertainties.
The median of the expected distributions for the SM (dashed red line) and the BSM Higgs-boson signal (dashed blue line) is also shown, together with the observed result (solid black line) from the fit to the data.
The shaded areas are used to compute the observed $p$-values.}
\label{fig:cpteststat_unblinded}
\end{figure}

\begin{figure}
\centering
\subfloat{\includegraphics[width=0.49\textwidth]{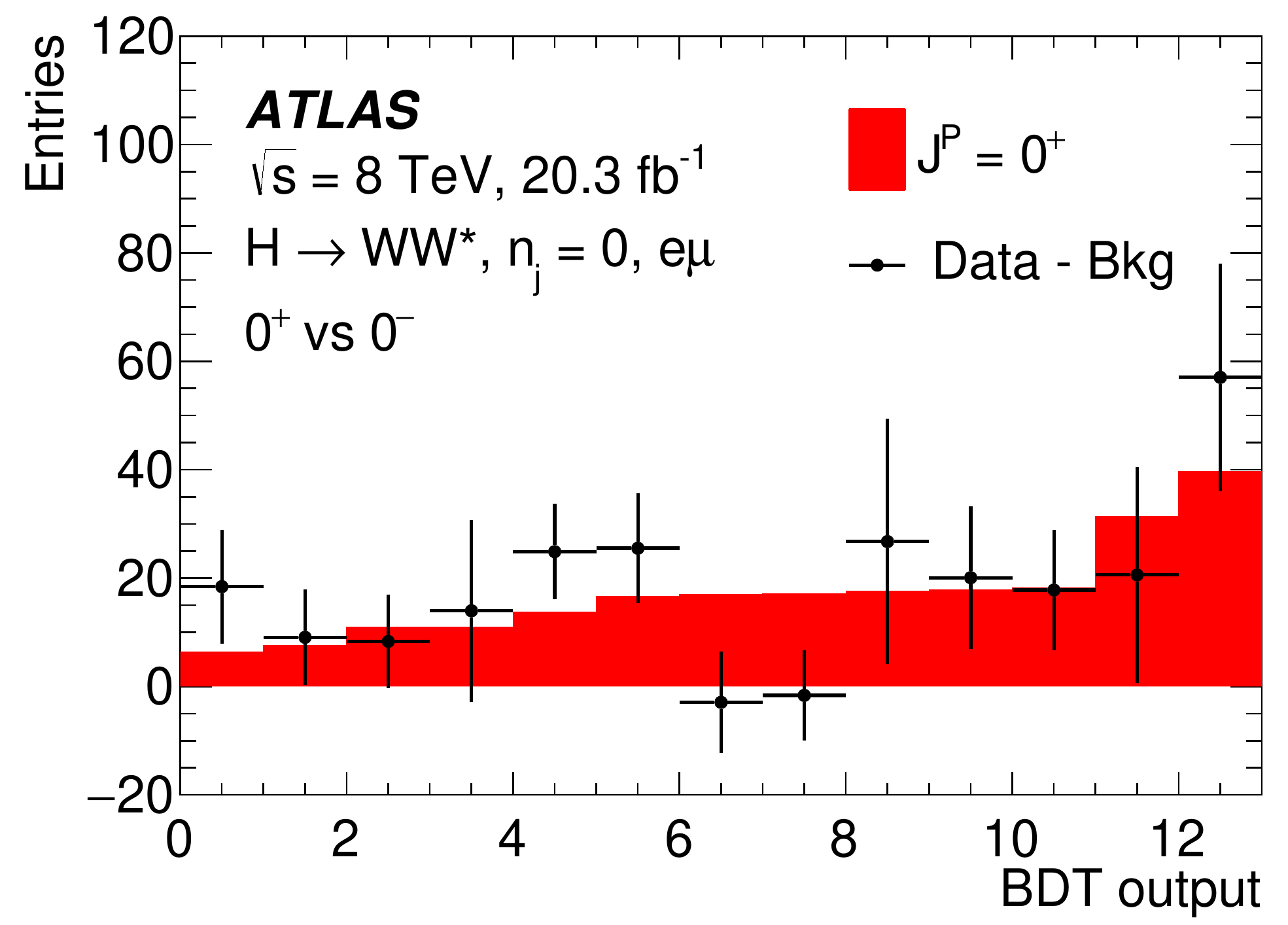}}
\subfloat{\includegraphics[width=0.49\textwidth]{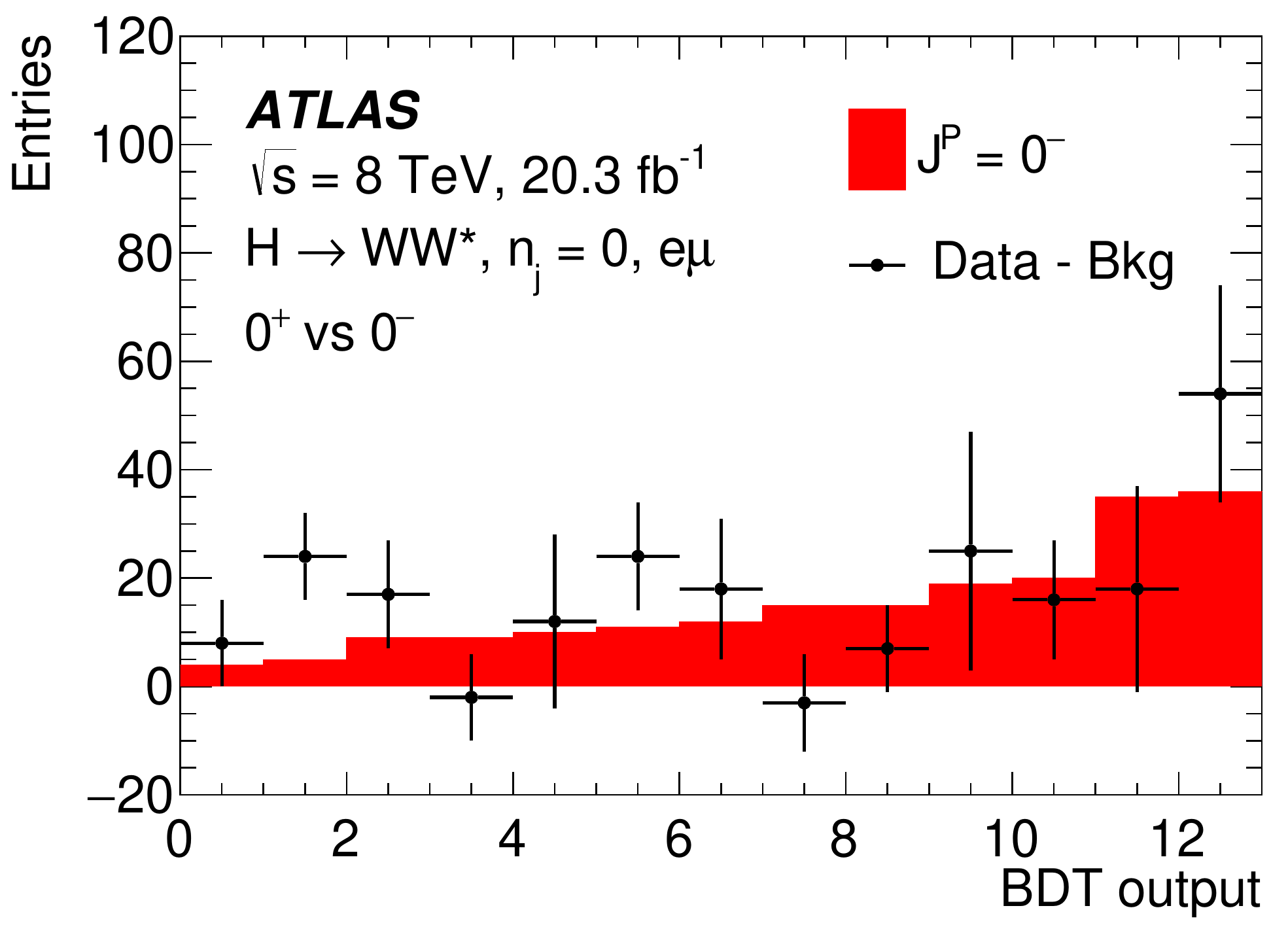}}

\subfloat{\includegraphics[width=0.49\textwidth]{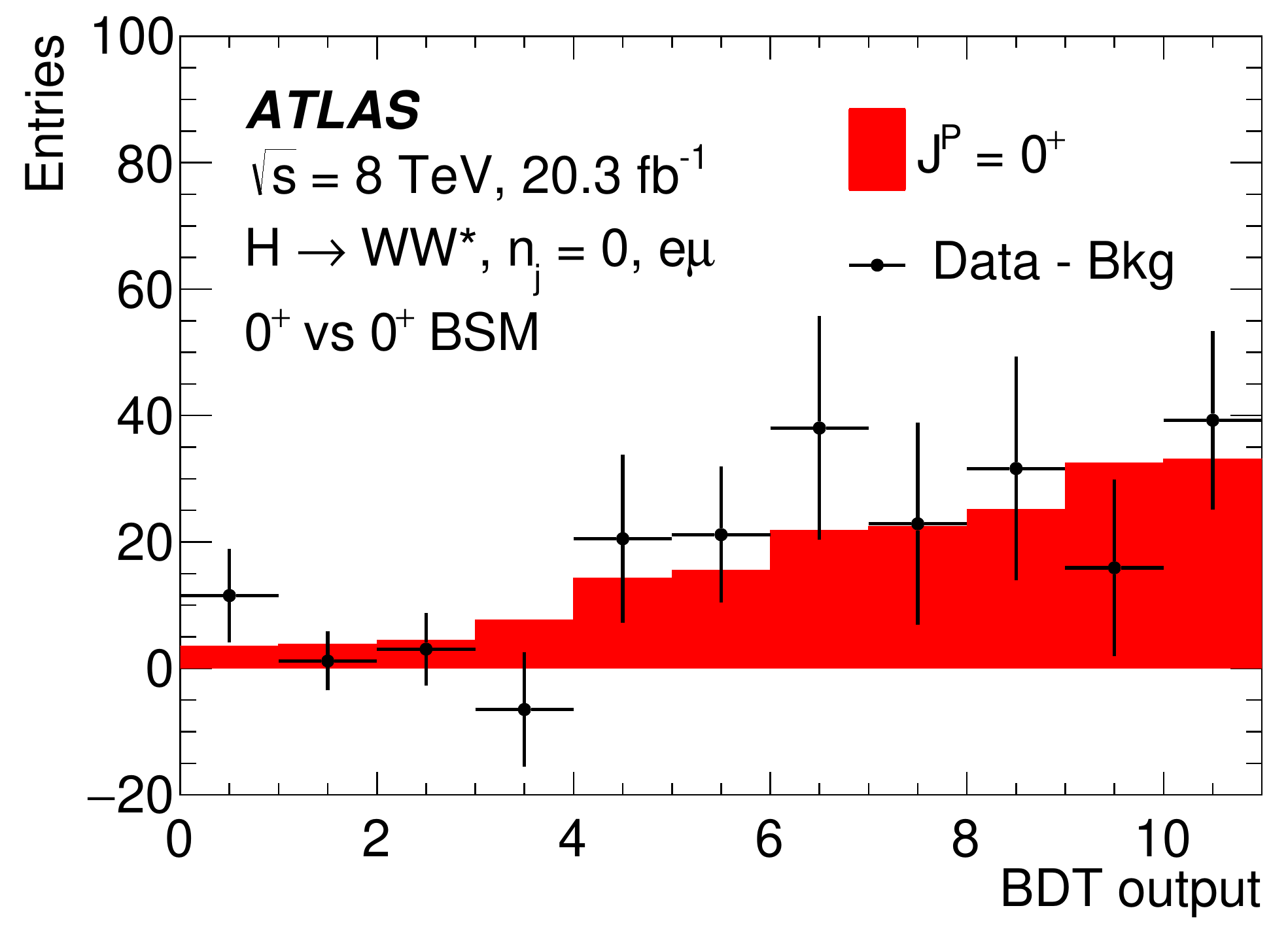}}
\subfloat{\includegraphics[width=0.49\textwidth]{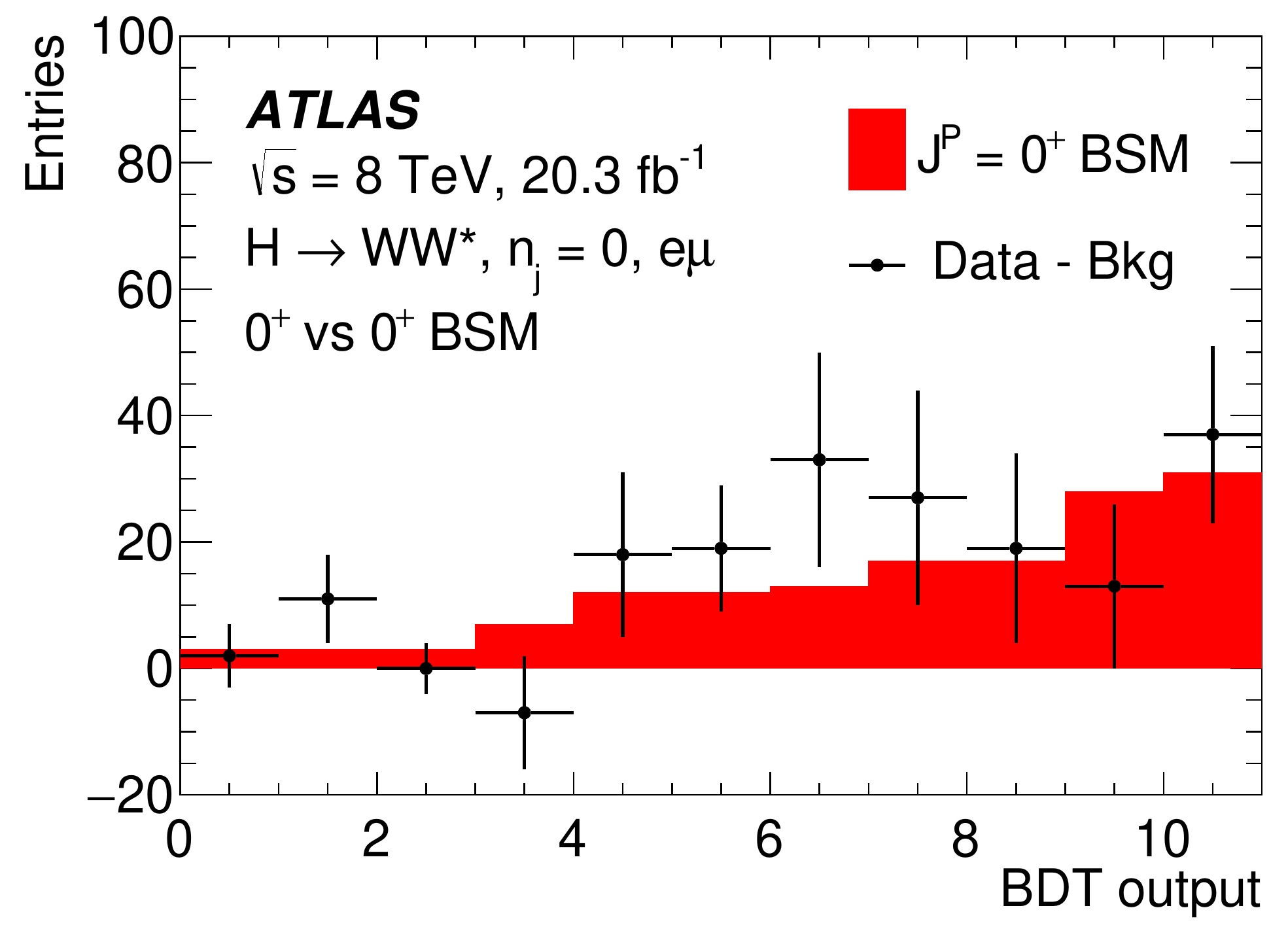}}

\caption{The unrolled one-dimensional BDT output after background subtraction in the case of the pure BSM CP-odd (top) and BSM CP-even (bottom) benchmarks.
The background yields are taken from the fit results, assuming the SM
signal hypothesis in the left-hand plots, and the alternative hypothesis
in the right-hand plots.
}
\label{fig:CPbkgsub}
\end{figure}

The unrolled BDT output distributions normalised to the post-fit values are shown in Fig.~\ref{fig:CPbkgsub}. 
These distributions show the one-dimensional unrolled BDT output for the SM and alternative signal hypotheses separately 
and compare them with the data after background subtraction. Both the signals and the background 
yields are normalised to the post-fit values. The distributions are ordered by increasing signal, and they contain bins that 
have at least three signal events and are above a signal-to-background threshold (S/B) of 0.035.  As already mentioned above,
these plots are intended for illustrative purposes only. 
The figure shows that the SM Higgs-boson hypothesis is preferred over the pure BSM CP-even or CP-odd cases. The S/B ratio 
used for the CP analysis is higher than the one used  for the spin-2 analysis because on average the bins with the highest significance 
have a higher S/B in the CP-mixing than in the spin-2 BDT output.

\begin{figure}
\centering
\subfloat{\includegraphics[width=0.49\textwidth]{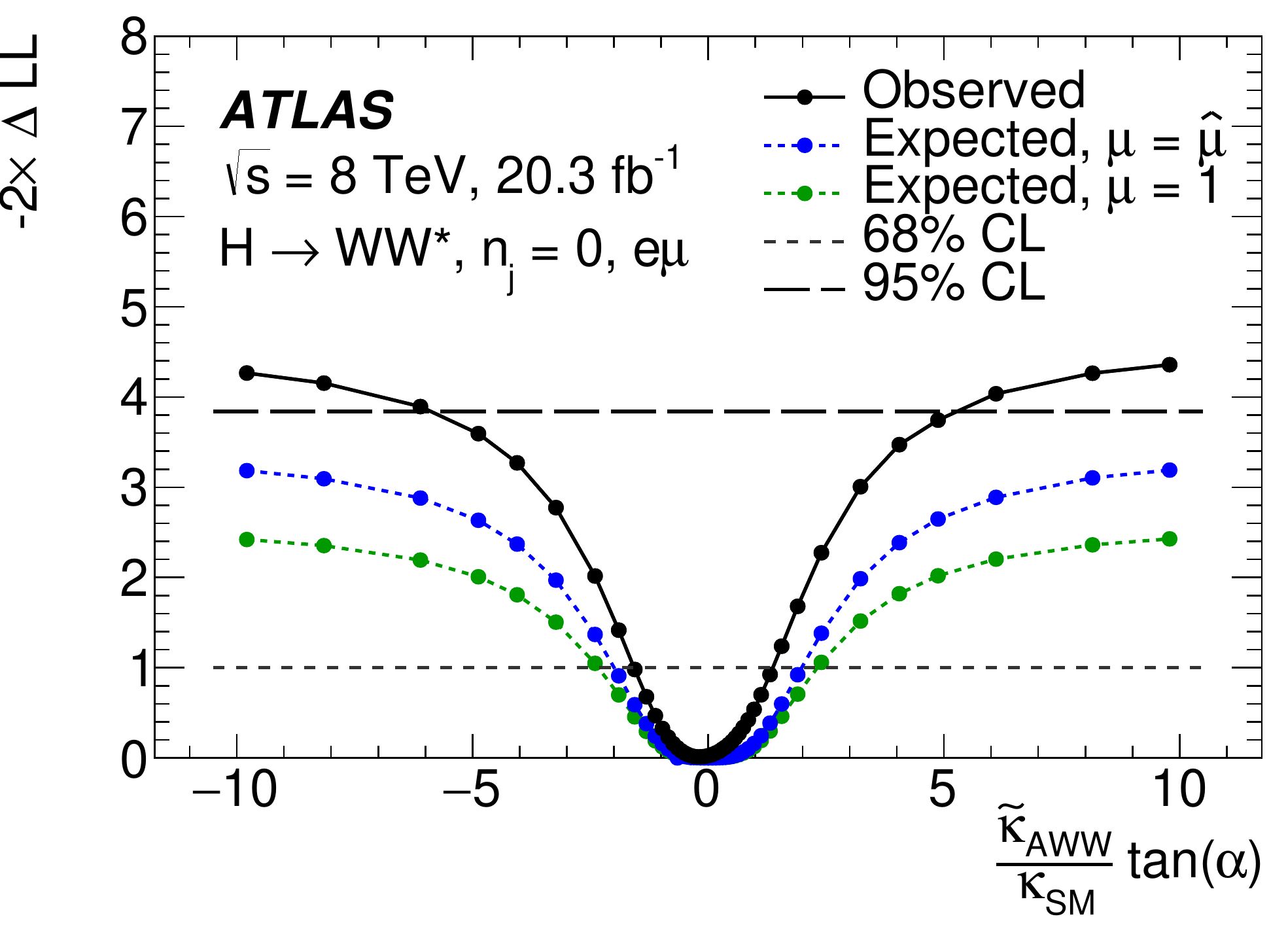}}
\subfloat{\includegraphics[width=0.49\textwidth]{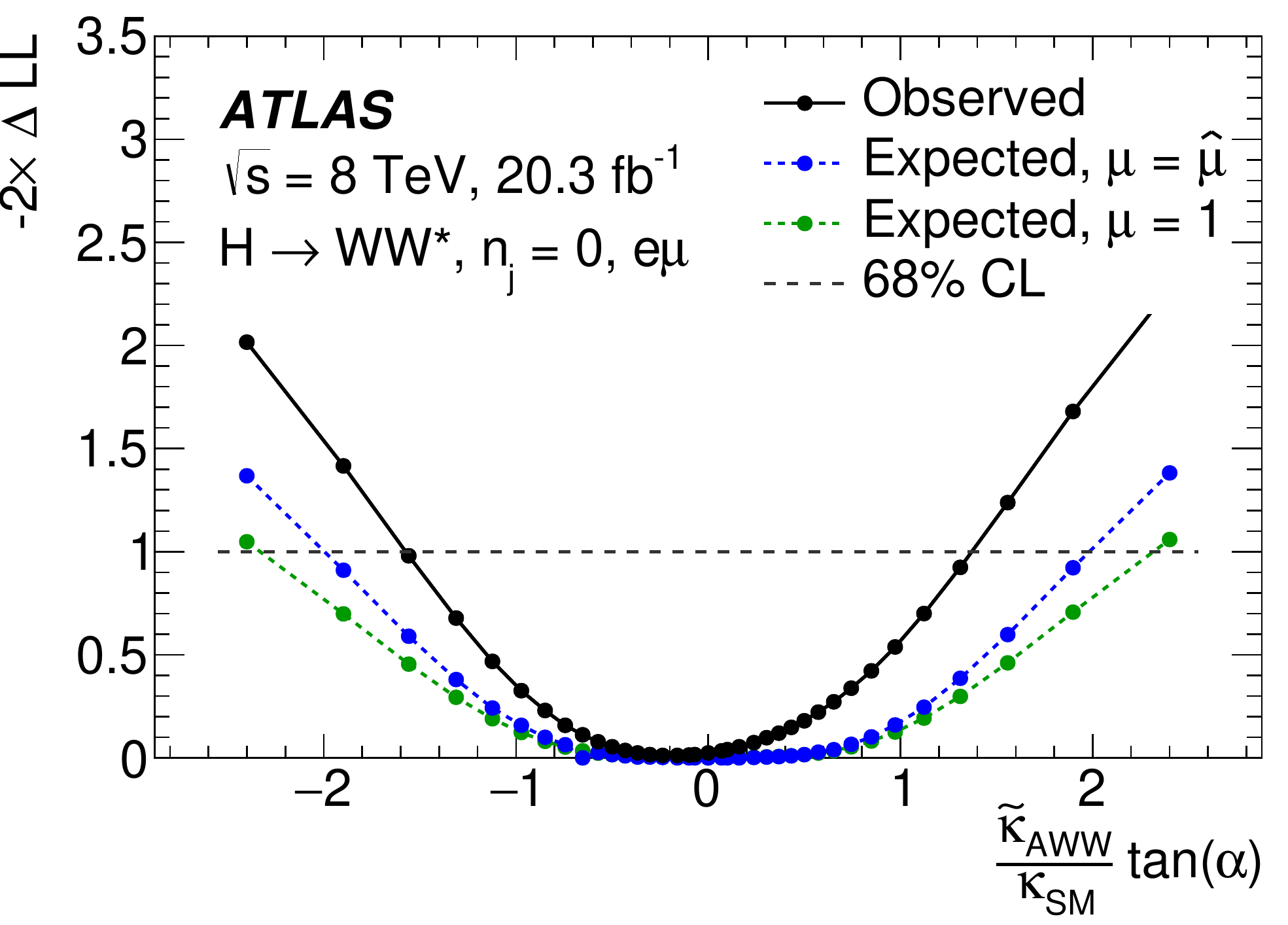}} \\
\subfloat{\includegraphics[width=0.49\textwidth]{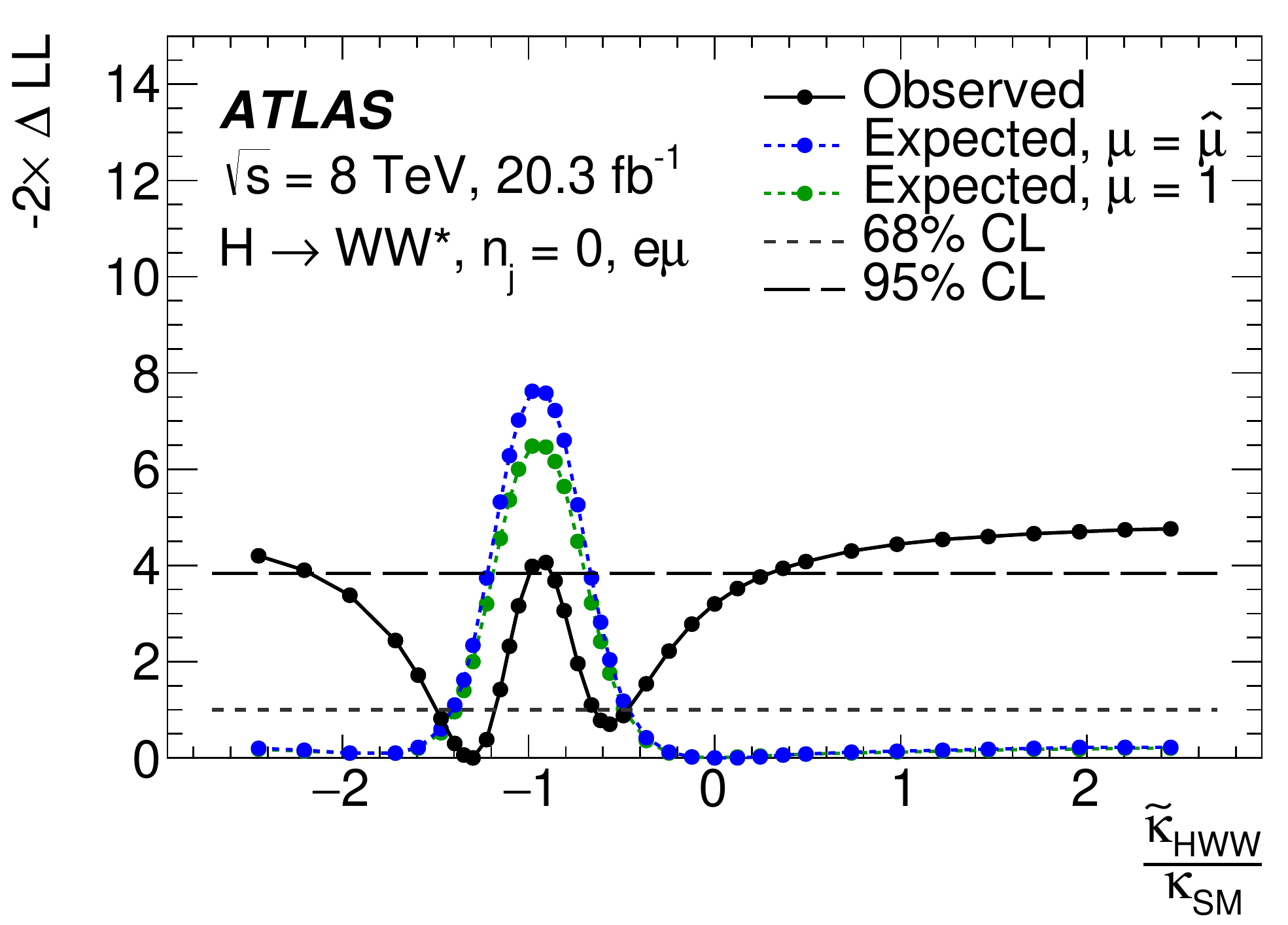}}   

\caption{The BSM CP-odd (top) and BSM CP-even (bottom) mixing scan results. The top row shows the full CP-odd scan (left) and the region around the minimum enlarged (right). The 68\% and 95\% CL exclusion regions are indicated as lying above the corresponding horizontal lines. }
\label{fig:CPscans}
\end{figure}

The compatibility of the CP-mixed signal plus background with the observed data is calculated following the prescription explained in 
Sect.~\ref{sec:stat_cp} for the two different scans (mixing of an SM Higgs boson with a BSM CP-even or CP-odd boson) 
as discussed in Sect.~\ref{sec:CPtheory}. The scan results are presented in Fig.~\ref{fig:CPscans}. 

In the case of the BSM CP-odd mixing scan (top row of Fig.~\ref{fig:CPscans}), the expected and observed curves are slightly asymmetric, 
but the sensitivity to the sign of the scan parameter is small. Due to higher observed yields for the SM hypothesis, the expected curve using the 
observed yields ($\mu=\hat{\mu}$) is above the expected curve for the yields fixed to the SM expectation ($\mu=1$). The 
minimum of the $-2\Delta\mbox{LL}$ curve is very broad and lies at $-0.2$. The value at 0 corresponds to the SM hypothesis. The values 
of $\left(\tilde{\kappa}_{\rm \sss AWW}/\kappa_{\sss \rm SM}\right) \cdot \tan \alpha$  below $-6$ and above 5 can be excluded at 95\% CL, while values below $-1.6$ and above 1.3 at 68\% CL.
The fitted signal yields and their relative uncertainties, for the SM and alternative signal hypotheses, are very stable throughout the scan.
They are given in Table~\ref{tab:postfitNorm} for the fixed-hypothesis case.

The plot on the bottom of Fig.~\ref{fig:CPscans} shows the result of the BSM CP-even scan
as a function of $\tilde{\kappa}_{\rm \sss HWW}/{\kappa}_{\sss \rm SM}$. The separation power between the SM Higgs-boson hypothesis and the BSM CP-even mixed hypothesis is enhanced in the region around --1, the observed minimum of the $-2\Delta\mbox{LL}$ distribution, 
because of the interference effect explained in Sect.~\ref{sec:variables}.  
The fitted signal yield, both for the SM and alternative signal hypotheses, is stable for values outside the observed minimum region
and similar
to the values given in Table~\ref{tab:postfitNorm} for the fixed-hypothesis case. In the region around the minimum,
the fitted BSM signal yield is higher, reaching about 370 events.
These variations are expected from the significant shape differences of the input variable distributions in this region of the parameter scan, as described in 
Sect.~\ref{sec:variables}. The relative uncertainty is stable throughout the scan, with values around 30\%.

The observed minimum of the $-2\Delta\mbox{LL}$ curve is at --1.3 and is compatible with the SM hypothesis within $1.9 \sigma$.
To further study the compatibility of the SM signal hypothesis with the observed result, several scans are performed, by fitting, instead of the real data, pseudo-data generated around the expected signal-plus-background post-fit 
BDT distribution. This means that the nuisance parameters from this test  are obtained from the fit of the SM signal to the data.
Distributions similar to the one observed in the data are reproduced by pseudo-data.
Furthermore, a fixed-hypothesis test is also performed, where the compatibility of the observed data with the SM Higgs boson versus the CP-even mixed signal corresponding to
$\tilde{\kappa}_{\rm \sss HWW}/{\kappa}_{\sss \rm SM}= -1.3$ is studied,  resulting in a $\mathrm{1-CL_\text{s}}$ of 43\% in favour of the SM and of 93\% in favour of the alternative hypothesis.

Values of the mixing parameter, $\tilde{\kappa}_{\sss \rm HWW}/{\kappa}_{\sss \rm SM}$, above 0.4 and below $-2.2$ can 
be excluded at 95\% CL, as well as in the region between $-0.85$ and $-1$. Values above $-0.5$ and below $-1.5$, as well as between $-1.2$ and $-0.65$, 
can be excluded at 68\% CL.

\section{Conclusions} 
\label{sec:conclusion}

The Standard Model $J^{P} =0^{+}$ hypothesis for the Higgs boson is compared to alternative spin/parity hypotheses using 20.3 fb$^{-1}$ of the proton--proton collision data collected by the ATLAS experiment at the LHC at $\sqrt{s}=$8 TeV and corresponding to 
the full data set of 2012. The Higgs-boson decay $WW^*\rightarrow e \nu\mu \nu$ is used to test several alternative models, including BSM CP-even and CP-odd Higgs bosons, and a graviton-inspired $J^P = 2^+$ model with 
minimal couplings to the Standard Model particles. In addition to the tests of pure $J^{P}$ states, two scenarios are considered where all the 
CP mixtures of the SM Higgs boson and a BSM CP-even or CP-odd Higgs boson are tested.

For the spin-2 benchmarks, the SM hypothesis is favoured in all tests in data and the alternative model is 
disfavoured at 84.5\% CL for the model with universal couplings and excluded at 92.5\% to 99.4\% CL for the benchmark models with non-universal couplings. 

The SM Higgs-boson hypothesis is tested against a pure BSM CP-even or CP-odd Higgs-boson hypothesis: the results prefer the SM 
Higgs-boson hypothesis, excluding the alternative hypothesis at the 70.8\% and 96.5\% levels, respectively.

The data favour the Standard Model quantum numbers in all cases apart from the scan of a CP-mixed state with a BSM CP-even 
Higgs boson, where the data prefer a mixed state with $\tilde{\kappa}_{\rm \sss HWW}/{\kappa}_{\rm \sss SM}=-1.3$, which is compatible with
the SM hypothesis within $1.9\sigma$. The $\tilde{\kappa}_{\rm \sss HWW}/{\kappa}_{\rm \sss SM}$ values can 
be excluded at 95\% CL above 0.4 and below $-2.2$, as well in the region between $-0.85$ and $-1$. 
For the mixing with a BSM CP-odd Higgs boson, the $\left(\tilde{\kappa}_{\rm \sss AWW}/\kappa_{\rm \sss SM}\right) \cdot \tan \alpha$ values above 5 and 
below $-6$ can be excluded at 95\% CL.  The preferred value corresponds to 
$\left(\tilde{\kappa}_{\rm \sss AWW}/\kappa_{\rm \sss SM}\right) \cdot \tan \alpha = -0.2$, which is compatible with the SM to within $0.5\sigma$.

\section*{Acknowledgements} 

We thank CERN for the very successful operation of the LHC, as well as the
support staff from our institutions without whom ATLAS could not be
operated efficiently.

We acknowledge the support of ANPCyT, Argentina; YerPhI, Armenia; ARC,
Australia; BMWFW and FWF, Austria; ANAS, Azerbaijan; SSTC, Belarus; CNPq and FAPESP,
Brazil; NSERC, NRC and CFI, Canada; CERN; CONICYT, Chile; CAS, MOST and NSFC,
China; COLCIENCIAS, Colombia; MSMT CR, MPO CR and VSC CR, Czech Republic;
DNRF, DNSRC and Lundbeck Foundation, Denmark; EPLANET, ERC and NSRF, European Union;
IN2P3-CNRS, CEA-DSM/IRFU, France; GNSF, Georgia; BMBF, DFG, HGF, MPG and AvH
Foundation, Germany; GSRT and NSRF, Greece; RGC, Hong Kong SAR, China; ISF, MINERVA, GIF, I-CORE and Benoziyo Center, Israel; INFN, Italy; MEXT and JSPS, Japan; CNRST, Morocco; FOM and NWO, Netherlands; BRF and RCN, Norway; MNiSW and NCN, Poland; GRICES and FCT, Portugal; MNE/IFA, Romania; MES of Russia and NRC KI, Russian Federation; JINR; MSTD,
Serbia; MSSR, Slovakia; ARRS and MIZ\v{S}, Slovenia; DST/NRF, South Africa;
MINECO, Spain; SRC and Wallenberg Foundation, Sweden; SER, SNSF and Cantons of
Bern and Geneva, Switzerland; NSC, Taiwan; TAEK, Turkey; STFC, the Royal
Society and Leverhulme Trust, United Kingdom; DOE and NSF, United States of
America.

The crucial computing support from all WLCG partners is acknowledged
gratefully, in particular from CERN and the ATLAS Tier-1 facilities at
TRIUMF (Canada), NDGF (Denmark, Norway, Sweden), CC-IN2P3 (France),
KIT/GridKA (Germany), INFN-CNAF (Italy), NL-T1 (Netherlands), PIC (Spain),
ASGC (Taiwan), RAL (UK) and BNL (USA) and in the Tier-2 facilities
worldwide.

\newpage
 
\printbibliography

\onecolumn
\clearpage
\begin{flushleft}
{\Large The ATLAS Collaboration}

\bigskip

G.~Aad$^{\rm 85}$,
B.~Abbott$^{\rm 113}$,
J.~Abdallah$^{\rm 152}$,
O.~Abdinov$^{\rm 11}$,
R.~Aben$^{\rm 107}$,
M.~Abolins$^{\rm 90}$,
O.S.~AbouZeid$^{\rm 159}$,
H.~Abramowicz$^{\rm 154}$,
H.~Abreu$^{\rm 153}$,
R.~Abreu$^{\rm 30}$
% ...
\bigskip
\\
% ...
$^{11}$ Institute of Physics, Azerbaijan Academy of Sciences, Baku, Azerbaijan\\
% ...
$^{30}$ CERN, Geneva, Switzerland\\
% ...
$^{85}$ CPPM, Aix-Marseille Universit{\'e} and CNRS/IN2P3, Marseille, France\\
% ...
$^{90}$ Department of Physics and Astronomy, Michigan State University, East Lansing MI, United States of America\\
% ...
$^{107}$ Nikhef National Institute for Subatomic Physics and University of Amsterdam, Amsterdam, Netherlands\\
% ...
$^{113}$ Homer L. Dodge Department of Physics and Astronomy, University of Oklahoma, Norman OK, United States of America\\
% ...
$^{152}$ Institute of Physics, Academia Sinica, Taipei, Taiwan\\
$^{153}$ Department of Physics, Technion: Israel Institute of Technology, Haifa, Israel\\
$^{154}$ Raymond and Beverly Sackler School of Physics and Astronomy, Tel Aviv University, Tel Aviv, Israel\\
% ...
$^{159}$ Department of Physics, University of Toronto, Toronto ON, Canada\\
% ...
\end{flushleft}

\end{document}